\documentclass[12pt]{article}
\usepackage{amssymb}
\usepackage{amsfonts}
\usepackage{amsmath}
\usepackage{mathtools}  
\usepackage{bm}
\usepackage{comment}
\usepackage{latexsym}
\usepackage{epsfig}
\usepackage{bbm}
\usepackage{enumerate}
\usepackage{float}
\usepackage{color}
\usepackage{gensymb}
\usepackage{multicol}
\usepackage{siunitx}
\usepackage{color}
\usepackage{hyperref}
\usepackage{array,multirow}
\usepackage[round,authoryear]{natbib}
\usepackage[T1]{fontenc}
\usepackage[utf8]{inputenc}
\usepackage{authblk}
\usepackage{appendix}
\usepackage{blindtext}
\usepackage{lineno}
\usepackage{nth}
\usepackage{bbm}
\usepackage{soul}
\usepackage[table,xcdraw]{xcolor}
\usepackage{setspace}
\usepackage{subcaption}
\usepackage{algorithm}
\usepackage{algpseudocode}
\usepackage{graphicx}
\usepackage{booktabs, tabularx} 
\usepackage[export]{adjustbox} 
\usepackage{setspace} 
\usepackage{ulem}
\DeclarePairedDelimiter{\abs}{\lvert}{\rvert}
\newcommand*\bigcdot{\mathpalette\bigcdot@{.5}}
\providecommand{\keywords}[1]
{
  \small	
  \textbf{\textit{Keywords---}} #1
}

\usepackage[
        a4paper,
        left=2.5cm,
        right=2.5cm,
        top=3cm,
        bottom=3cm]{geometry}
        
\begin{document}

\title{Joint modeling of wind speed and wind direction through a conditional approach}

\author{Eva Murphy\footnote{Wake Forest University. E-mail: \href{murphye@wfu.edu}{\nolinkurl{nagy@clemson.edu}}}, Whitney Huang\footnote{Clemson University. E-mail: \href{wkhuang@clemson.edu}{\nolinkurl{wkhuang@clemson.edu}}}, Julie Bessac\footnote{National Renewable Energy Laboratory. E-mail: \href{ julie.bessac@nrel.gov}{\nolinkurl{ julie.bessac@nrel.gov}}}, Jiali Wang\footnote{Argonne National Laboratory. E-mail: \href{jialiwang@anl.gov}{\nolinkurl{jialiwang@anl.gov}}}, Rao Kotamarthi\footnote{Argonne National Laboratory. E-mail: \href{vrkotamarthi@anl.gov}{\nolinkurl{vrkotamarthi@anl.gov}}}}

\date{}

\maketitle

\begin{abstract} 
Atmospheric near surface wind speed and wind direction play an important role in many applications, ranging from air quality modeling, building design, wind turbine placement to climate change research. It is therefore crucial to accurately estimate the joint probability distribution of wind speed and direction. In this work we develop a conditional approach to model these two variables, where the joint distribution is decomposed into the product of the marginal distribution of wind direction and the conditional distribution of wind speed given wind direction. To accommodate the circular nature of wind direction a von Mises mixture model is used; the conditional wind speed distribution is modeled as a directional dependent Weibull distribution via a two-stage estimation procedure, consisting of a directional binned Weibull parameter estimation, followed by a harmonic regression to estimate the dependence of the Weibull parameters on wind direction. A Monte Carlo simulation study indicates that our method outperforms two other approaches in estimation efficiency: one that utilizes periodic spline quantile regression and another that generates data from the commonly used Abe-Ley distribution for cylindrical data. We illustrate our method by using the output from a regional climate model to investigate how the joint distribution of wind speed and direction may change under some future climate scenarios. Our method indicates significant changes in the variation of wind speed with respect to some directions. \\

\end{abstract}

\doublespacing
\section{Introduction} \label{intro}
    
The atmospheric wind vector plays a significant role in many fields: air-quality monitoring \citep{airpolution}; wind building engineering \citep{building3, building2, building1}; wildfires \citep{wildfire}, just to name a few. In clean energy production, wind speed and wind direction play an important role in designing the layout of the wind farm such that the power the wind farm produces is optimal \citep{windenergy, stwf, cleanenergy}. Consequently, accurate modeling of the probability distribution of wind vector is critical in characterizing its variation. 

\begin{figure}[H]
    \centering
    \begin{subfigure}[t]{0.4\textwidth}
         \includegraphics[width = \linewidth]{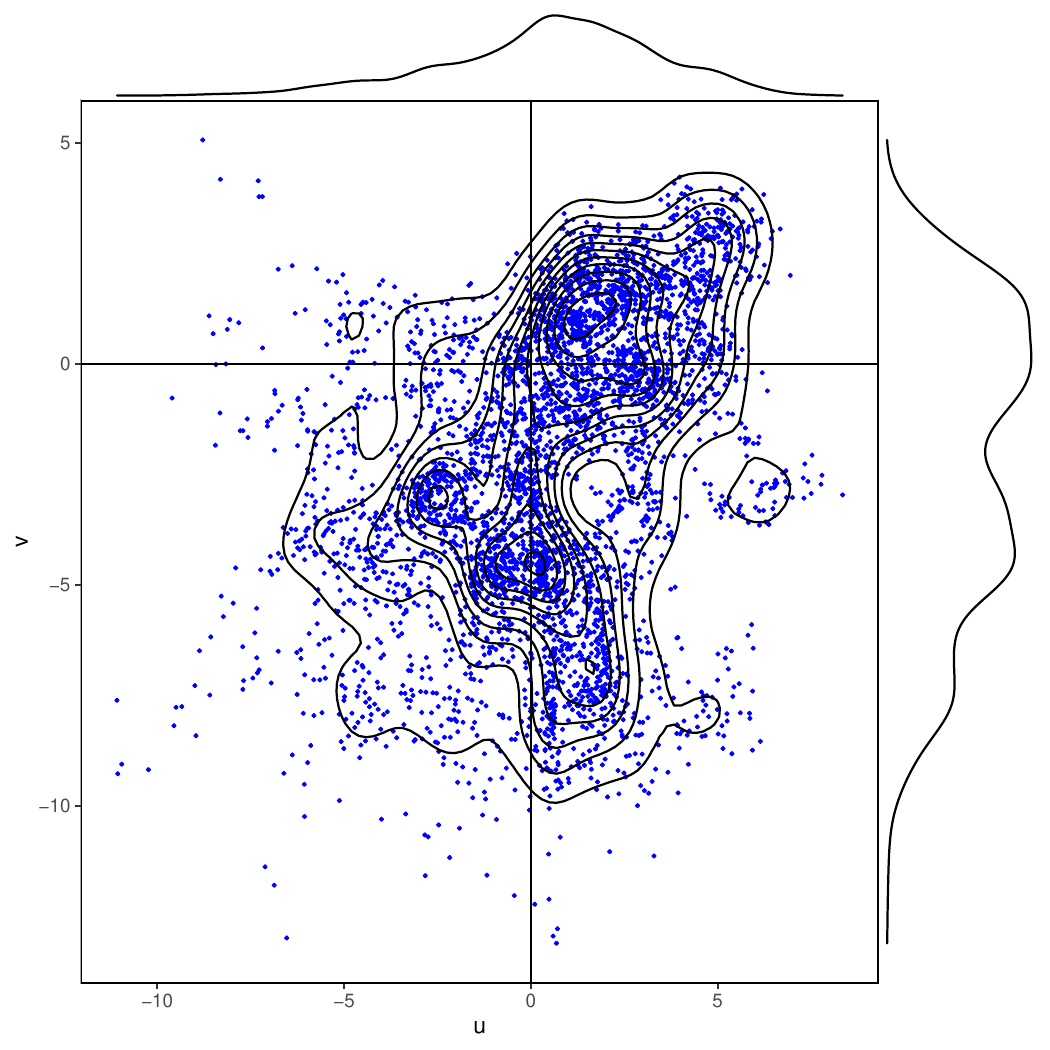}
         \caption{$(u,v)$ representation} \label{orthog}
     \end{subfigure}
     \begin{subfigure}[t]{0.4\textwidth}
         \includegraphics[width = \linewidth]{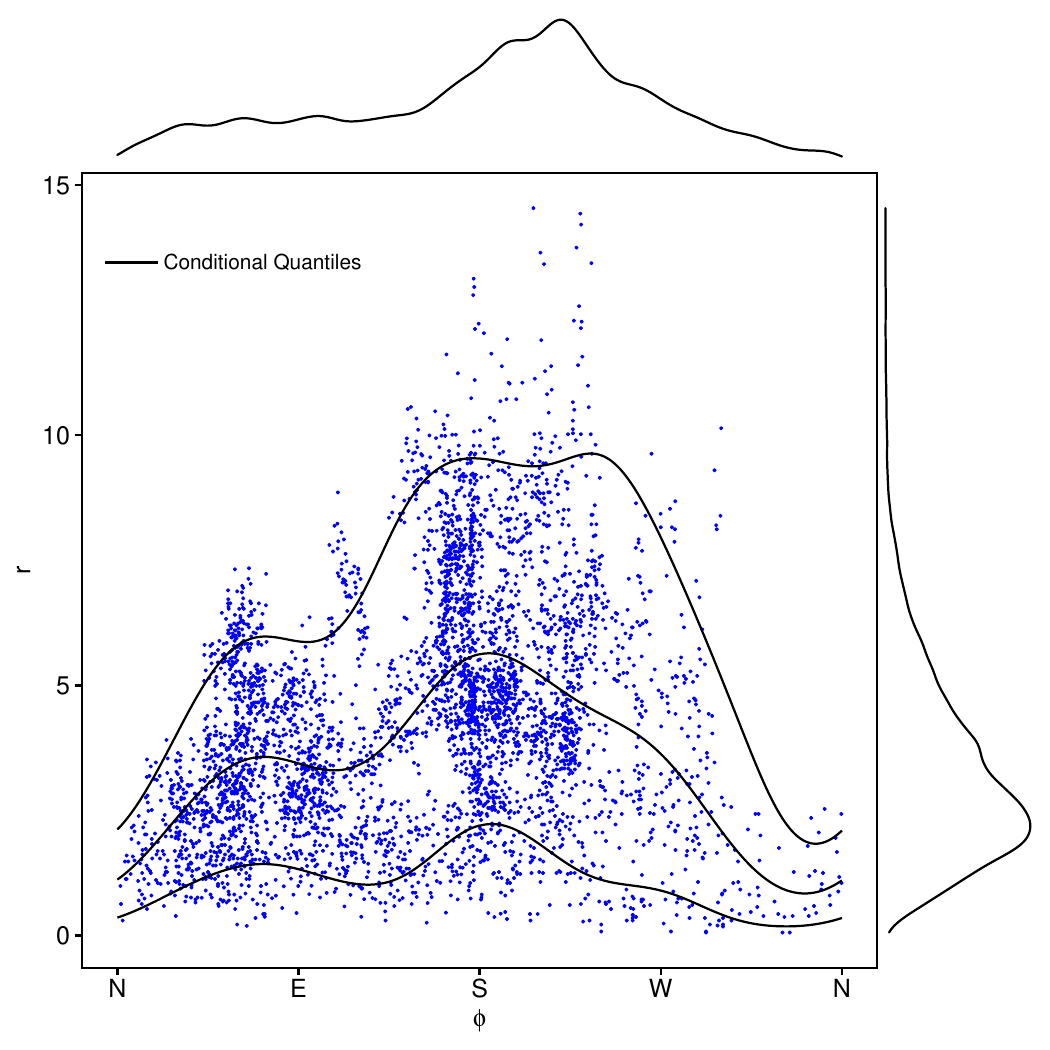}
          \caption{$(r, \phi)$ representation} 
          \label{polar}
    \end{subfigure}
    \caption{\small Representation of the wind vector in terms of Cartesian components with the corresponding marginal distributions and probability density contour curves \textbf{(Left (a))} and polar coordinates with corresponding marginal distributions and $0.05$, $0.5$ and $0.95$ conditional quantile curves \textbf{(Right (b))}.}
    \label{repres}
\end{figure}

 Wind vector can be represented in either the Cartesian coordinates $(u,v)$ (also known as zonal and meridional velocities) or the polar coordinates $(r, \phi)$ (also known as cylindrical variables), i.e. wind speed and wind direction. Transforming from the  Cartesian coordinates to polar coordinates (or vice versa) is achieved using the formula: $(u,v) = (r \sin \phi, \, r \cos \phi)$ (or $(r, \phi) = (\sqrt{u^2 + v^2},\arctan\left(\frac{u}{v}\right))$; due to meteorological convention, the $\arctan$ function is often replaced by the $\text{atan}2$ function). Fig. \ref{repres} depicts these two representations of the wind vector using 10-minute wind data measured at 10 meters above ground at the Cabauw Experimental Site for Atmospheric Research (CESAR) tower \citep{cesar} over the course of one year during the month of January. 
    
An approach for the empirical modeling of the wind vector is through the joint distribution of $(u,v)$, denoted by brackets $[U, V]$ \citep{uvmodel1,uvmodel2,uvmodel3,uvrep,uvhidden}, where, for example, a bivariate normal distribution can be used \citep{uvmodel1,uvmodel2,uvmodel3}. However, as indicated in Fig. \ref{orthog}, modeling $[V]$ (and $[U]$) using a normal distribution may not be appropriate. Copula approach \citep{ copulaJoe, copulaNelsen}, while providing a flexible alternative for modeling the marginal and dependence structures separately, can still be challenging to accommodate potentially complicated dependence structure via either parametric or non-parametric copula models. Mixture modeling approach \citep{titterington1985,lindsay1995,mclachlan2019} can also be employed to model $[U, V]$, such as mixture of bivariate distributions. Nevertheless, one needs to decide on the number of components needed and the distribution to be used, to account for outliers, and to have enough data in each cluster to adequately estimate the covariance. 
   
Under polar coordinates different statistical methods have been employed to model wind vector variations. Most existing works only consider the wind speed component of the wind vector, where there is a need to preserve the non-negativity and skewness properties. For a comprehensive review of commonly used distributions for modeling wind speed we refer the reader to \citep{wsreview}.

Wind direction has received less attention in the literature because of its circular nature \citep{berck}, however, it can be important in many applications (e.g. coastal wind direction \citep{irish}, animal behavior \citep{windchar}, fire spread \citep{fire}). A commonly used distribution for modeling wind direction is the von Mises distribution, or mixtures of von Mises distributions \citep{vonmis2}. Wind direction can also be included in the modeling of the wind speed as a regime switching guidance \citep{stwf,regimBess,Ding} in which case the regime-switching can be represented as an observed or latent variable requiring the presence of a prevalent wind direction and a clear boundary for when to switch to a new regime.  
To eliminate the regime identification \cite{stwf} propose that wind direction is used as a covariate in the modeling of the mean of the wind speed's distribution. Specifically, the authors assume that the hourly wind speed follow a truncated normal distribution $N^+(\mu, \sigma^2)$ and use sine and cosine functions to incorporate wind direction in modeling the mean, $\mu$. 
    
 Wind speed and wind direction typically exhibit an interdependent behavior: Fig. \ref{polar} shows that the (estimated) conditional quantile curves are directional dependent. Therefore, jointly modeling the distribution of wind speed and wind direction is needed. Copula approach can also be used here with a consideration that wind direction is a circular variable \citep{copula2,copula1,copula3}. Under the copula modeling framework, the joint distribution of wind speed and wind direction is separated into the marginal distributions of wind speed and wind direction, and a copula function that encodes the dependence structure of wind speed and wind direction under uniform marginals. Due to the circular nature of wind direction, there are fewer copula families that can be applied, which can be limited in modeling the wide various dependence structures of wind speed and wind direction. Recently, Wang et al. (2021) modeled the joint distribution of wind direction, wind speed, and air temperature using a vine copula framework. However, while vine copulas are designed to handle high-dimensional dependence modeling, they do not, by themselves, provide the necessary flexibility for the lower-dimensional linear-circular case we are addressing.

\begin{figure}[H]
    \centering
    \includegraphics[width = 0.33\textwidth]{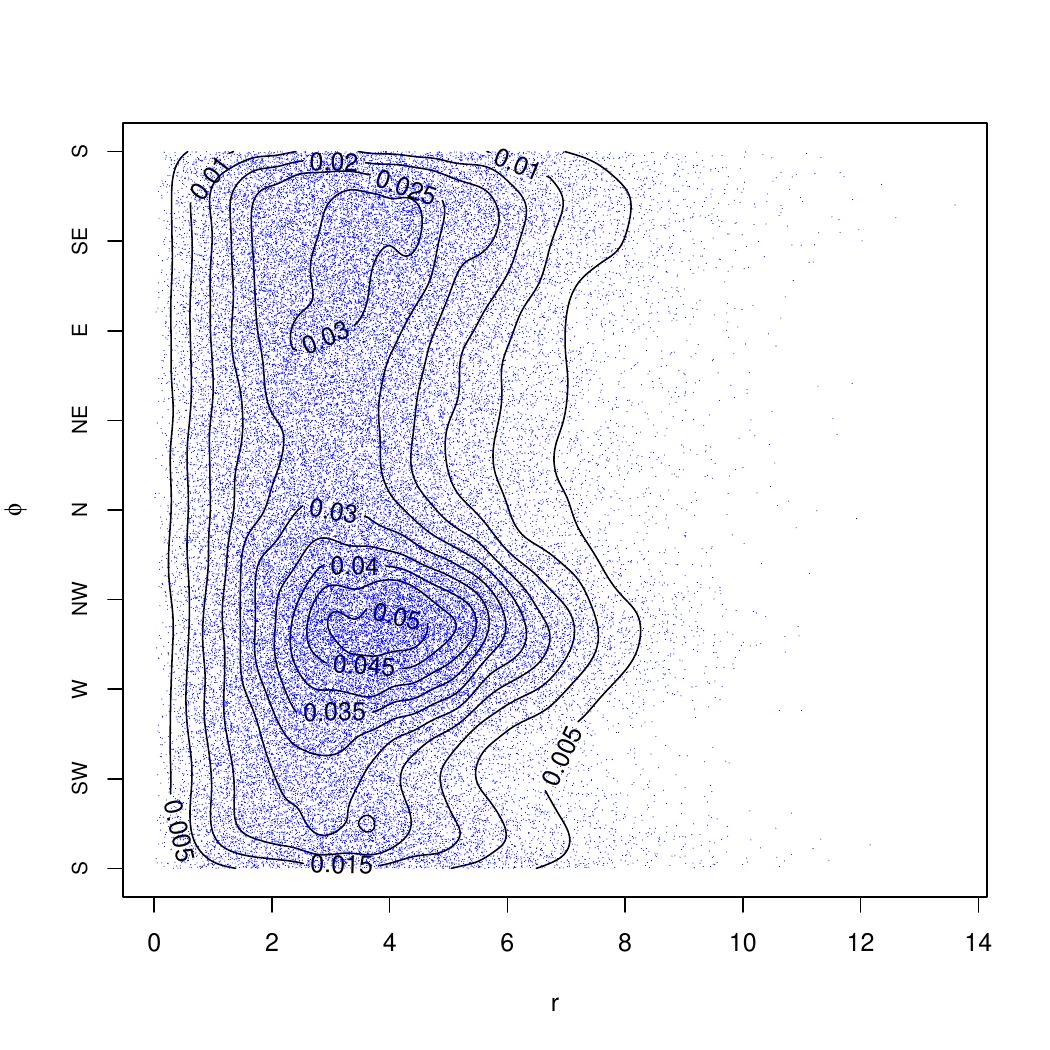}%
    \includegraphics[width = 0.33\textwidth]{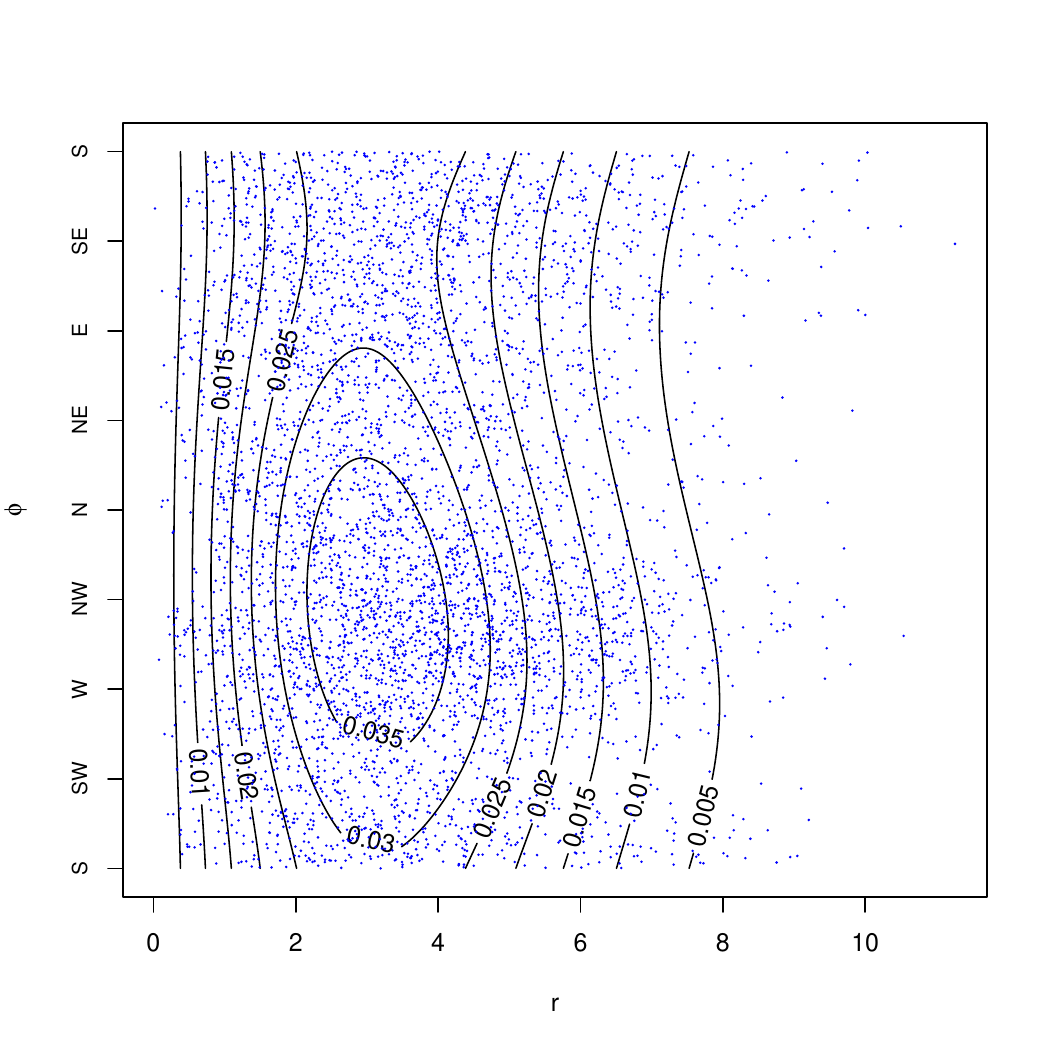}%
    \includegraphics[width = 0.33\textwidth]{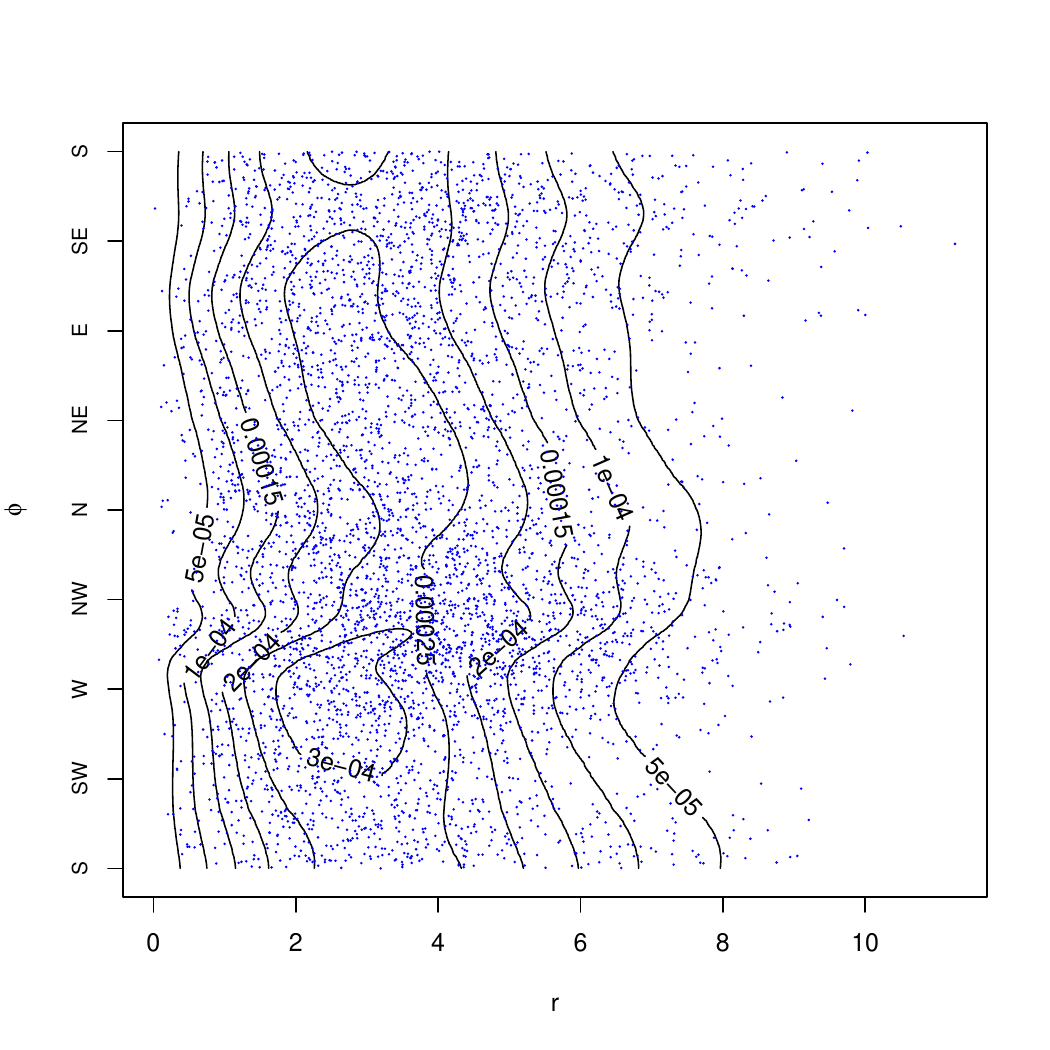}
    \caption{Density contours of wind speed and wind direction for simulated data, where $(u,v)$ are simulated from a bivariate normal distribution and then transformed to polar coordinates $(r, \phi)$, The left panel shows the `true' contour, where a two-dimensional kernel density estimation is applied to a much larger simulated dataset (62,500 points). The middle and right panels show the density contours estimated by the Abe-Ley distribution and our method, respectively, on a smaller dataset (5,000 points).}
    \label{ContourAL}
\end{figure}

Johnson and Wehrly’s distribution has been extended by \cite{abe2017tractable} by invoking a power transformation to the linear part and a perturbation to the circular part to allow for asymmetric ``sine-skewing''. This distribution combines a linear Weibull component with a circular sine-skewed von Mises component, making it widely used for cylindrical data modeling. In Fig.~\ref{ContourAL}, we present the density contours of wind speed and wind direction for the simulated data. The $(u, v)$ pairs were generated from a bivariate normal distribution with a known mean and covariance structure and then transformed into polar coordinates. The left panel displays the 'true' contour — a two-dimensional kernel density estimate based on a larger dataset of 62,500 data points. For the middle and right panels, we used a smaller dataset of 5,000 data points—intended to mimic a typical data size—to estimate the contours. The middle panel shows the density contours estimated by the Abe-Ley distribution, while the right panel shows the contours from our proposed method. The plot highlights that the Abe-Ley distribution isn’t flexible enough to fully capture the joint distribution of wind speed and direction for our study, mainly because only the shape parameter of the Weibull component depends on wind direction. While several extensions to the Abe-Ley model have been proposed, including those by \cite{Mastrantonio2018}, \cite{Lagona2015}, \cite{Sadeghianpourhamami2019}, and \cite{Eidsvik2021}, these approaches are often limited by identifiability challenges, which necessitate additional constraints. Many of these extensions also rely on highly parameterized copulas, imposing restrictions on the dependence structure and reducing the flexibility needed to fully capture the complexities of linear-circular data.

In our work, we take an alternative approach to jointly modeling wind speed and wind direction by decomposing their joint distribution into the product of the marginal distribution of wind direction and the conditional distribution of wind speed given wind direction, i.e. $[R, \Phi] = [\Phi]\cdot [ R|\Phi]$ \cite[e.g.,][]{colsaw, solari,wu2022}. These aforementioned works model the conditional distribution by binning the data by wind direction to estimate the parameters of the chosen wind speed distribution separately, then use a set number of pairs of Fourier series to model the dependence of the wind speed distribution parameters on wind direction. While proving a useful modeling framework, to the best of our knowledge, there is no systematic study on various modeling choices; furthermore, no comparison study has been done to compare with some alternative methods in terms of estimation performance. Motivated by our recent work \citep{wu2022}, this study aims to fill the gap by investigating these aspects. 
 
 In particular, we present an estimation and inference procedure, originally proposed in \citep{wu2022}, to achieve flexible modeling of wind speed and wind direction that preserves the intrinsic characteristics of the two variables, namely non-negativity and circularity. This is done by using a mixture of von Mises distributions to perform the marginal modeling of wind direction, and the Weibull distribution for the conditional modelling of wind speed, where the parameters of the Weibull distribution are modeled by means of periodic functions on wind directions, which are fitted by a two-stage estimation procedure, where their estimation uncertainties are quantified via a version of block bootstrap \citep{blockboot1, blockboot2}. 
 
 A simulation study is conducted to assess the performance of estimating the conditional distribution of wind speed using the proposed method and to compare it to two different approaches: a non-parametric quantile regression (QR) approach \citep{Koenker2005}, where the quantile curves are represented by a periodic B-spline as a function of wind direction, and the conditional distribution of wind speed given wind direction proposed by \cite{abe2017tractable} (AL). We used a custom metric, which we refer to as the mean integrated relative error (\texttt{MIRE}), to assess and compare the estimation performance of these methods when applied to cylindrical data. Our results suggest that the proposed method generally outperforms both the non-parametric QR and the AL model. We illustrate our proposed method in a climate application where changes in the present and future wind speed and wind direction distributions are estimated using the output of a regional climate model. An advantage of the proposed methodology is that it allows for the detection of changes in the distribution of wind speed with respect to some wind directions giving us additional understanding about the distribution of wind speed.

The paper is structured as follows: Section \ref{method} describes the proposed model and methods for estimation and inference. A simulation study investigating finite sample properties is described and the results are reported in Section \ref{sim}. The proposed method is applied to a regional climate model data in Section \ref{application} to study potential change of wind vector distribution under a future climate scenario. The paper concludes with a discussion and outlines of some future extensions in Section \ref{discussion}.

\section{Joint Modeling of Wind Speed and Wind Direction} \label{method}

In this section we first present the models for estimating the wind direction distribution $[\Phi]$ and the conditional distribution of wind speed $[R|\Phi=\phi]$, respectively, which together determine the joint distribution $[R, \Phi]$. The model fitting procedure for each component will then be described with more emphasize on the estimation of $[R|\Phi=\phi]$, the main contribution of this work. We make use of a block bootstrap procedure, which is capable to preserve some aspect of the temporal dependence structure without explicitly modeling it, to quantify the estimation uncertainty. The non-parametric periodic spline QR, the benchmark method for estimating $[R|\Phi=\phi]$ in this study and the AL model, will also be introduced.

\subsection{Models}

The joint distribution of wind speed and wind direction, $[R, \Phi]$, is decomposed into a product of the marginal distribution of the conditioning variable and the corresponding conditional distribution, i.e. 
\begin{equation}
\left[R, \Phi \right] =\left[ \Phi \right] \left[ R|\Phi \right],
\end{equation}
where $[\Phi]$ denotes the distribution of wind direction and $[R|\Phi]$ denotes the conditional distribution of wind speed given wind direction.

\begin{figure}[H]
    \centering
    \includegraphics[width = 0.2\textwidth]{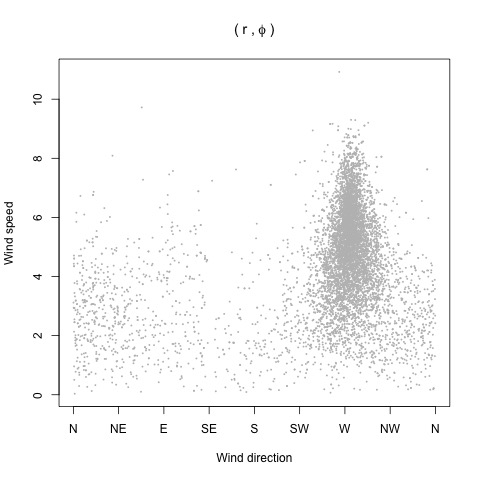}
    \includegraphics[width = 0.2\textwidth]{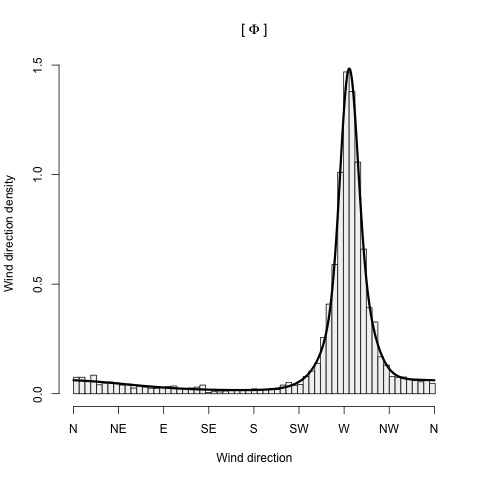}%
    \includegraphics[width = 0.2\textwidth]{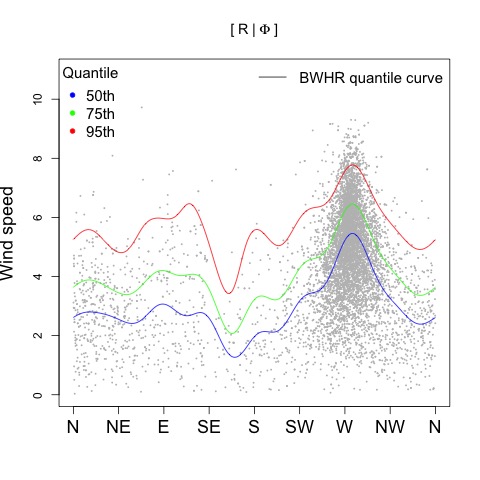}%
    \includegraphics[width = 0.2\textwidth]{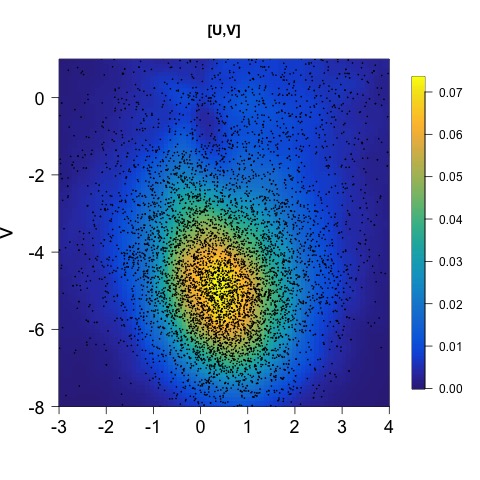}\\
    \includegraphics[width = 0.2\textwidth]{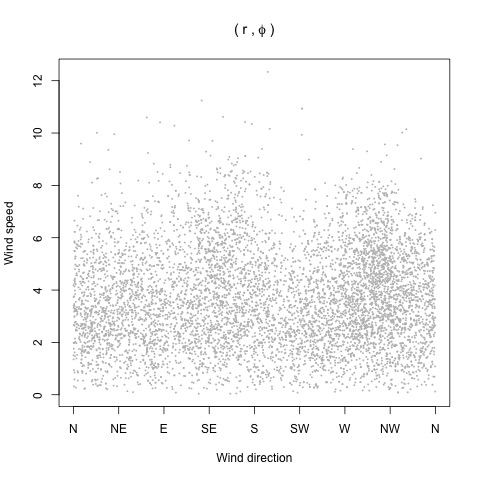}
    \includegraphics[width = 0.2\textwidth]{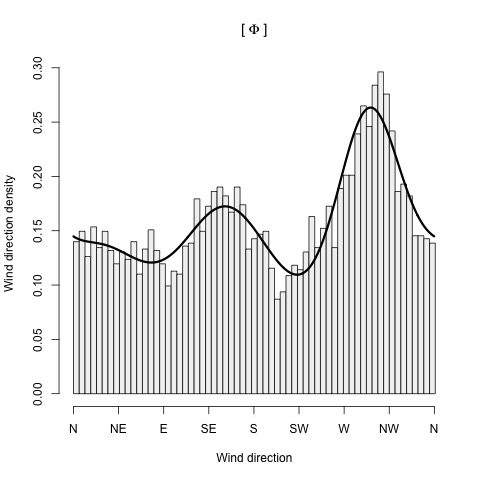}
    \includegraphics[width = 0.2\textwidth]{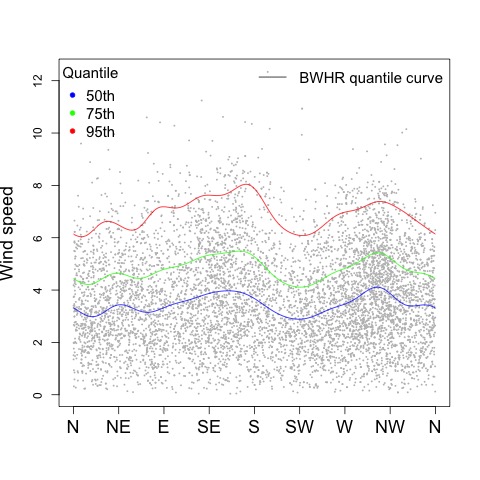}%
    \includegraphics[width = 0.2\textwidth]{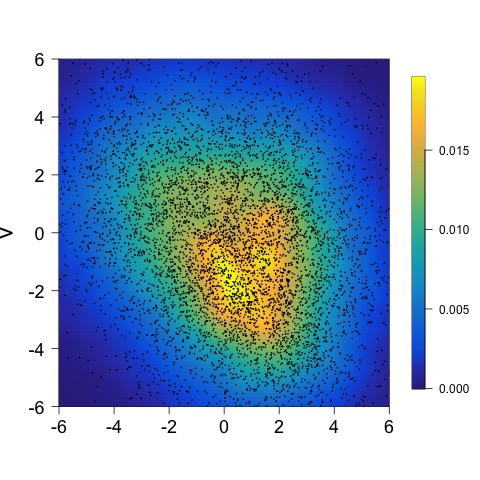}\\
    \includegraphics[width = 0.2\textwidth]{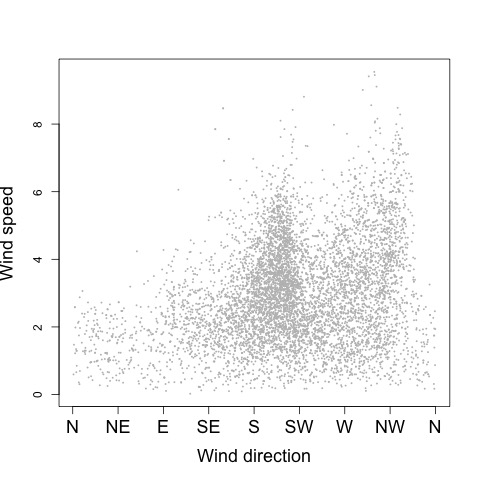}
    \includegraphics[width = 0.2\textwidth]{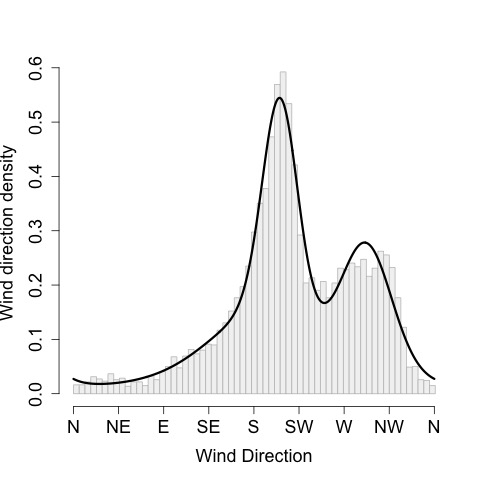}%
    \includegraphics[width = 0.2\textwidth]{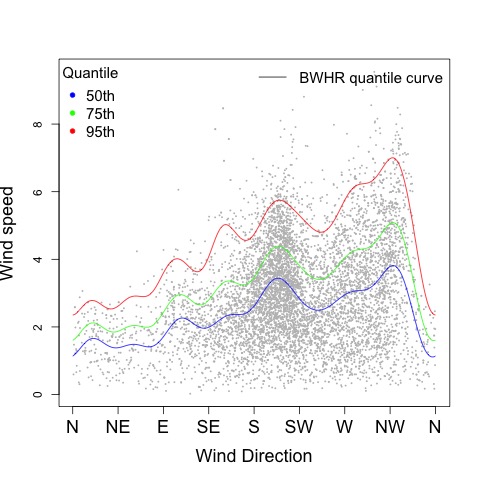}%
    \includegraphics[width = 0.2\textwidth]{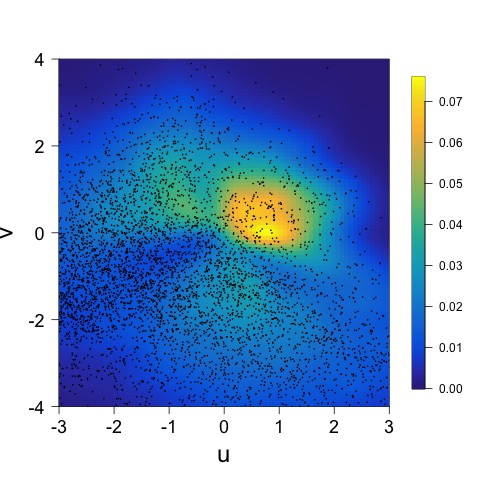}\\
    \caption{\small The wind data \textbf{(first column)} and the estimates of $[\Phi]$ \textbf{(second column)}, $[R|\Phi= \phi]$ \textbf{(third column)}, and the implied estimates of $[U,V]$ \textbf{(fourth column)} from a Texas great plain model grid cell (TX\_GP, \textbf{(top)}), a North Dakota great plain grid cell (ND\_GP, \textbf{(middle)}), and a North Carolina mountain grid cell (NC\_mtn, \textbf{(bottom)}). More information regarding these data can be found in Section 4. To estimate the density of $[U,V]$ we simulated a large data set of wind directions from the $[\Phi]$ distribution and for each wind direction we simulated wind speeds from the $[R|\Phi = \phi]$ distribution. We transformed this data to the $(u,v)$ components and a two-dimensional kernel density estimation is applied to the simulated data.} 
    \label{wswd_uv}
\end{figure}

Such a decomposition allows to estimate the joint distribution using the ``divide and conquer'' strategy, where each one-dimensional estimation problem in the right-hand side can be modeled separately and flexibly via mixture modeling \citep{titterington1985,mixture, mclachlan2019} and distributional regression \citep{distrreg} while avoiding the direct modeling of the potentially complex bivariate distribution in one step. This conditional decomposition approach is illustrated in Fig.~\ref{wswd_uv}, where flexible (yet relatively simple) models for $[\Phi]$ (second column) and $[R|\Phi]$ (third column) can produce quite complicated models for $[R, \Phi]$ and hence $[U, V]$ (fourth column).

We note that the decomposition can be done in the other way, where the conditioning variable is wind speed, i.e. in terms of $[R]$ and $[\Phi|R]$. However, it is arguably more natural to consider the conditional distribution of wind speed given wind direction than the conditional distribution of wind direction  given wind speed. Furthermore, it is our view that it is easier to perform a distributional regression with a typical scalar response than a circular response. Finally, it is worth pointing out that the conditional distribution of wind direction given wind speed can be obtained by applying Bayes' theorem, i.e., $[\Phi|R] = \frac{[\Phi][R|\Phi]}{[R]}$.

To address the complex nature of wind direction data, which often includes mixed distributions and periodic patterns, more advanced methods are necessary, such as the projected normal distribution \citep{mastrantonio2015}, the wrapped normal distribution \citep{Greco2021} and model-based clustering methods \citep{ranalli2020model}. These methods provide the flexibility required to accurately model such intricate data structures. In our study, we employ a mixture of $N_{\Phi}$ von Mises distributions \citep{vonMises,banerjee} to model the distribution of wind direction,  $[\Phi]$. The von Mises distribution is a distribution commonly used for modeling circular data and has the following probability density function (pdf): 
\begin{equation}
    f_{vM}(\phi| \mu, \kappa) = \frac{1}{2\pi I_0(\kappa)} \exp\left[ \kappa \cos(\phi - \mu)  \right],
    \label{vM}
\end{equation}
where $\mu \in [0, 2\pi)$ and $\kappa \geq 0$ are the parameters of the distribution representing the directional mean and concentration of the distribution, respectively, and $I_0(\kappa)$ is the modified Bessel function of the first kind and order zero \citep{bessel}. However, the von Mises distribution is unimodal and therefore, may not produce a satisfactory fitting for wind direction data that is multimodal such as shown in the middle column of Fig.~\ref{wswd_uv}. Hence, our choice of using a finite mixture of von Mises distribution.  Here, the density of the wind direction is modeled as follows:
\begin{equation}
    f_{\Phi}(\phi| \mathbf{\mu,} \mathbf{\kappa}, \mathbf{\omega})= \sum_{j= 1}^{N_{\Phi}} \omega_j f_{vM}(\phi; \mu_j, \kappa_j),
    \label{mvM}
\end{equation}
where $\omega_j$, $j = 1, \ldots, N_{\Phi}$, are weights such that $\sum_{j = 1}^{N_{\Phi}} \omega_j = 1$.

To model $[R|\Phi]$, the conditional distribution of wind speed given wind direction (directional wind speed distribution hereafter), a parametric conditional distribution is assumed while allowing the parameters of the distribution be smooth yet flexible periodic functions of the wind direction. To this end, a Weibull distribution \citep{weibull, monahan1} is considered where the parameters are modeled via Fourier series. The resulting pdf is given as follows:  
\begin{eqnarray}
    f_{R|\Phi}(r|\phi, \alpha, \beta) & = &\left( \frac{\alpha(\phi)}{\beta(\phi)} \right) \left( \frac{r}{\beta(\phi)} \right)^{\alpha(\phi)-1} \exp\left[\left( -\frac{r}{\beta(\phi)}\right)^{\alpha(\phi)}\right],\\
     \alpha(\phi) & =& b_{\alpha, 0} + \sum_{k=1}^{K_{\alpha}}\Big[a_{\alpha,k}\cos(k\phi) + b_{\alpha,k}\sin( k\phi)\Big],\\
\beta(\phi) & =& b_{\beta, 0} + \sum_{k=1}^{K_{\beta}}\Big[a_{\beta,k}\cos(k\phi) + b_{\beta,k}\sin( k\phi)\Big].
    \label{Weibull}
\end{eqnarray}

\subsection{Estimation Procedure}
\label{sec:BWHR}

The parameters pertaining to the von Mises mixture distribution, $\mu_j,\, \kappa_j,\, \omega_j$, $j = 1 \ldots N_\Phi$, are estimated using the Expectation Maximization (EM) algorithm \citep{em, banerjee}. The number of von Mises components ($N_{\Phi}$) can be determined using the Bayesian information criterion (BIC) (\cite{bic}, Chapter 6).
Estimating the directional wind speed distribution $[R|\Phi=\phi]$ involves a pair of harmonic regression where the number of the harmonic terms is needed to be determined. Ideally, one could perform a likelihood-based estimation by fitting the directional dependent Weibull distribution to wind speed and direction data $\{r_{i}, \phi_{i}\}_{i=1}^{n}$ with given $K_{\alpha}$ and $K_{\beta}$. However, it is our experience that doing so tends to be numerically unstable (see supplementary materials Sec.~SM~1) and may lead to spurious fitted curves. The approach we take is a two-step procedure where the data points $\{r_{i}, \phi_{i}\}_{i=1}^{n}$ are divided into $N$ bins  in the direction domain, a Weibull distribution is fitted to wind speeds within each bin via maximum likelihood (ML) method. In the second step a pair of harmonic regression are conducted (equations (5) and (6) in the previous section, respectively) where the ML estimates across all bins (i.e., $\{\hat{\alpha}_{j}\}_{j=1}^{N}$, $\{\hat{\beta}_{j}\}_{j=1}^{N}$) are treated as the ``responses'', where $N$ is the number of direction bins in the first step. Additionally, the estimated standard errors $\{\hat{\text{se}}(\hat{\alpha}_{j})\}_{j=1}^{N}$ and $\{\hat{\text{se}}(\hat{\beta}_{j})\}_{j=1}^{N}$ are incorporated to enable a weighted version of harmonic regression to be performed to estimate the parameters in equations (5) and (6).

Our estimation procedure involves several ``tunning'' parameters, namely, $N$ (number of bins), $K_{\alpha}$ and $K_{\beta}$ (number of pairs of trigonometric functions). We make the following suggestions for determining these values: 
\begin{itemize}
  
    \item The first suggestion pertains to binning the wind direction data using equal width or equal frequency (i.e., each bin contains the same number of data points) bins. Note that equal frequency binning approach leads to unequal bin sizes due to variation of wind direction distribution, hence resulting in an ``uneven'' modeling of $\alpha(\phi)$ and $\beta(\phi)$ in that of very little variation in the directions with sparse data while potentially too much variation in the directions that the data are dense. Therefore, we recommend using equal width directional bins, which in general gives better overall estimation performance (see supplementary material Sec.~SM~2).
    
    \item Next, regarding the choice of the number of bins, $N$.
    Our sensitivity analysis (see supplementary material Sec.~SM~2) suggests that choosing $N=36$ bins is enough to get a good estimation of the directional wind speed distribution as long as the size of the data set is sufficiently large (e.g., $n>3000$); additionally, there is no clear need to let $N$ grow with $n$ where $n$ is ``large'' because the ``support'' of $\Phi$ is fixed regardless of the sample size.
    
    \item In determining the number of pairs of Fourier series , i.e., $K_{\alpha}$ and $K_{\beta}$, one must have that $N \geq 2K+2$ to avoid having more parameters than the sample size ($N$ in this case) when fitting the harmonic regression in Step 2. For a chosen value of $N$, $K_\alpha$ and $K_\beta$ can be determined using BIC. However, we find that this strategy often gives to a ``saturated'' model, i.e., $N \approx 2K+2$ which leads to overfitting. Therefore, we suggest to use $K_\alpha = K_\beta = 8$. This choice is compared to the choice of BIC and resulted in a smaller integrated mean relative error (MIRE, see its definition in Section~\ref{sec:Est_preformance}) as shown in the supplementary material (Sec.~SM~3).

\end{itemize}

We name the estimation method for $[R|\Phi=\phi]$ the Binned Weibull Harmonic Regression, \texttt{BWHR} hereafter, and we summarize it in \textbf{Algorithm}~\ref{algorithm1} below. 

\begin{algorithm}[H]
\caption{Binned Weibull - Harmonic Regression }\label{alg:cap}
\textbf{Input:} $\{r_{i}, \phi_{i}\}_{i=1}^{n}$\\
\textbf{First step}: bin the wind data to estimate the parameters of the Weibull distribution and to compute a summary statistics of wind direction (e.g. medians) as follows:
\begin{itemize}
    \item Bin the data by dividing the wind direction into $N$ bins, i.e. choose equal sized bins such that each bin has on average sufficient number of data points (e.g., 200 points). 
   \item In each bin, represent wind direction data by a summary statistic, $\{\tilde{\phi}_{i}\}_{j = 1}^N$, and fit a two parameter Weibull distribution to the wind speed data. The parameters of each of the Weibull distributions are estimated using Maximum Likelihood Estimation (MLE) method to obtain the estimates $\{\hat{\alpha}_j, \hat{\beta}_j\}_{j=1}^N$ and their corresponding standard errors $\{\hat{\text{se}}(\hat{\alpha}_j), \hat{\text{se}}(\hat{\beta}_i)\}_{j=1}^N$.
\end{itemize}

\textbf{Second step}: estimate the directional dependent function of Weibull parameters, $\alpha(\phi)$ and $\beta(\phi)$, using harmonic regression via WLS as follows:
\begin{itemize}
    \item Use the direction summary statistic and the MLEs of each of the Weibull distributions as data points, i.e. $\{\tilde{\phi}_{j}, \hat{\alpha}_j\}_{j = 1}^N$ and $\{\tilde{\phi}_{j}, \hat{\beta}_j\}_{j = 1}^N$ , to regress the MLEs on the direction using periodic functions (e.g.\ a fixed $K$ pairs of harmonic functions): $\hat{\alpha}(\phi) = b_{\alpha, 0} + \sum_{k=1}^{K}\Big[a_{\alpha,k}\cos(k\tilde{\phi}) + b_{\alpha,k}\sin( k\tilde{\phi})\Big]$, $\hat{\beta}(\phi) = b_{\beta, 0} + \sum_{k=1}^{K}\Big[a_{\beta,k}\cos(k\tilde{\phi}) + b_{\beta,k}\sin( k\tilde{\phi})\Big]$. The weights (of WLS) are the squares of the inverses of the standard errors (SE) of the MLEs such that MLEs with lower SE receive higher weights than those with higher SE.
   \end{itemize}
\textbf{Output:} $\alpha(\phi)$ and $\beta(\phi)$

\label{algorithm1}
\end{algorithm}

Simulation-based approach can be used to obtain the implied joint distribution of $\widehat{[R, \Phi]}$ based on $[\hat{{\Phi}}]$ and $\widehat{[R|\Phi]}$. Specifically, one can simulate a large number of random variates, $\{\phi_{\text{sim}, i}\}_{i=1}^{\ell}$ from $\hat{[\Phi]}$, and for each random variate $\phi_{\text{sim},i}$, simulate a paired random variate $r_{\text{sim}, i}$, from $\widehat{[R|\Phi = \phi_{\text{sim}, i}]}$. Transforming the simulated data $(\phi_{\text{sim}, i}, r_{\text{sim}, i})_{i=1}^{m}$ using the transformation formulas presented in Section \ref{intro} we can approximate the distribution of $\widehat{[U,V]}$ up to a Monte Carlo sampling error, which is negligible with a large $\ell$.



\subsection{Periodic Quantile Regression} 

Quantile regression (QR) is a general method for estimating conditional quantiles of the response variable $Y$. Rather than modeling the conditional mean of the response, i.e. $E[Y]$, as a function of the covariates $x$'s as commonly done in regression analysis, QR models the conditional quantile, i.e. $Q_{Y}(\tau|X) = \inf\{y: F(y|X)\geq \tau\}$, for quantile level $\tau \in (0,1)$ \citep{quantreg}. One can approximate the underlying conditional distribution $[Y|X]$ by estimating a set of conditional quantile levels (i.e. $Q_Y(\tau_k|X = x), \tau_k \in [0,1], k = 1, \ldots,K, x \in \mathbb{R} $) \citep{quantreg2}.

Building on this foundation, recent advances have been made in directional data analysis. \cite{quantdir} introduce a novel concept of quantiles tailored for directional data, addressing the challenge of defining quantiles in circular or spherical spaces by leveraging the angular Mahalanobis depth. \cite{circquant} propose a residual for circular-linear regression models that enhances diagnostic accuracy and model reliability, even with small samples. In our study, when modeling the directional wind speed distribution, wind direction is considered a covariate. To this end, we model the $\tau$-quantile of the directional wind speed distribution ($[R|\Phi]$) using QR with periodic B-splines. The use of splines in the quantile setting has been explored in prior research, including the work by \cite{Frumento2021} on quantile regression coefficient functions with longitudinal data and \cite{Frumento2016} study on quantile regression coefficient functions.  In our work, the estimator of $Q_{R|\Phi}(\tau|\phi)$ takes the following form: 
\begin{equation}
    \hat{Q}_{R|\Phi}(\tau|\phi) = B(\phi)^T \hat{\beta}(\tau).
\end{equation}
The vector $B(\phi)$ is a periodic B-spline with degree of freedom $df$. The periodic B-spline is used to preserve the circular property of the directional wind speed distribution while providing modeling flexibility in terms of quantile functions. We select the degree of freedom ($df = 18$) by conducting a sensitivity study (supplementary material Sec.~SM~4) and conclude that this choice leads to a reasonable compromise regarding bias and variance trade-off. 
The coefficient vector $\beta(\tau)$ is estimated as follows:
\begin{equation}
    \hat{\beta}(\tau) =\arg \min_{\beta} \sum_{i = 1}^n \rho_{\tau}(r_i - B(\phi_i)^T\beta(\tau)),
\end{equation}
where $\rho_\tau (y) = y(\tau - \mathbbm{1}_{(y<0)})$ is the quantile loss function. We call this periodic B-spline quantile regression model and its estimation procedure \texttt{BPQR} hereafter. While it is not part of our proposed method, we use it as a baseline for comparing the estimation performance of our approach.

\subsection{Abe-Ley distribution}
The Abe-Ley distribution, introduced by \cite{abe2017tractable}, combines the sine-skewed von Mises and Weibull distributions, making it suitable for cylindrical data, such as wind speed and direction. Cylindrical data are represented in the form $(r, \phi)$, where $r \in [0, \infty]$ denotes the magnitude (linear part) and $\phi \in (-\pi, \pi]$ represents the angle (circular component). The pdf of the Weibull sine-skewed von Mises distribution proposed by \cite{abe2017tractable} is as follows:
\begin{equation}
    f(r, \phi) = \frac{\alpha \beta^\beta}{2\pi\cosh \kappa}\left( 1+\lambda \sin(\phi - \mu) \right)r^{\alpha - 1} \exp(-(\beta r)^\alpha (1 - \tanh(\kappa) \cos(\phi - \mu))),
\end{equation}
where $\alpha > 0 $ and $\beta > 0$ are the shape and scale parameters of the linear magnitude, $\mu \in (-\pi, \pi]$ is the circular location and $\lambda \in [-1,1]$ is the circular skewness. The parameter $\kappa \geq 0 $ represents the circular concentration and the dependence between the circular and linear parts; for example, $\kappa = 0$ implies that the circular and linear parts are independent. 

One attractive property of the Abe-Ley joint distribution is that the conditional distribution $[R|\Phi]$, which corresponds to directional wind speed distribution in our application, as well as the marginal distribution $\Phi]$, which corresponds to wind direction distribution in our application, can both be derived from  the Abe-Ley joint distribution, as given by the following:
\begin{eqnarray*}
    f(r|\phi) &=& \alpha\left[\beta (1-\tanh(\kappa) \cos(\phi - \mu))^{\alpha}\right]^{\alpha}r^{\alpha -1} \exp\left[-(\beta(1 - \tanh(\kappa) \cos(\phi - \mu))^{1/\alpha}r)^\alpha \right], \label{ALcond}\\
    f(\theta) &=& \frac{1 - \tanh^2(\kappa/2)}{2\pi}\frac{1+\lambda \sin(\theta - \mu)}{1 + \tanh^2(k/2) - 2\tanh(k/2)\cos(\theta - \mu)}
    \label{ALdir}
\end{eqnarray*}
To model the $\tau$-quantile of the directional wind speed distribution $[R|\Phi]$ we use the fact that this conditional distribution is the Weibull distribution with the shape parameter $\beta(1- \tanh(\kappa) \cos(\phi - \mu))^{1/\alpha}$ , as noted by the authors. Specifically, we estimate the parameters of the Abe-Ley distribution using the likelihood function: 
\begin{eqnarray*}
    \mathcal{L}(\alpha, \beta, \mu, \kappa, \lambda) &=& (\alpha - 1) \sum_{i=1}^{n} \log x_i - \beta \alpha \sum_{i=1}^{n} x_i^\alpha \left(1 - \tanh(\kappa) \cos(\theta_i - \mu)\right) \\ &+& \sum_{i=1}^{n} \log\left(1 + \lambda \sin(\theta_i - \mu)\right) + n(\alpha \log \beta + \log \alpha - \log(2\pi \cosh(\kappa))).
\end{eqnarray*}
Due to the lack of closed-form expressions for the maximum likelihood estimates, numerical optimization methods are necessary. These estimated parameters are used to compute the $\tau$-quantile of the Weibull distribution with the specified shape parameter. In our work, we refer to  this method as the \texttt{AL} method.

\subsection{Estimation Uncertainty}
\begin{figure}[H]
       \centering
      \includegraphics[width = 0.33\textwidth]{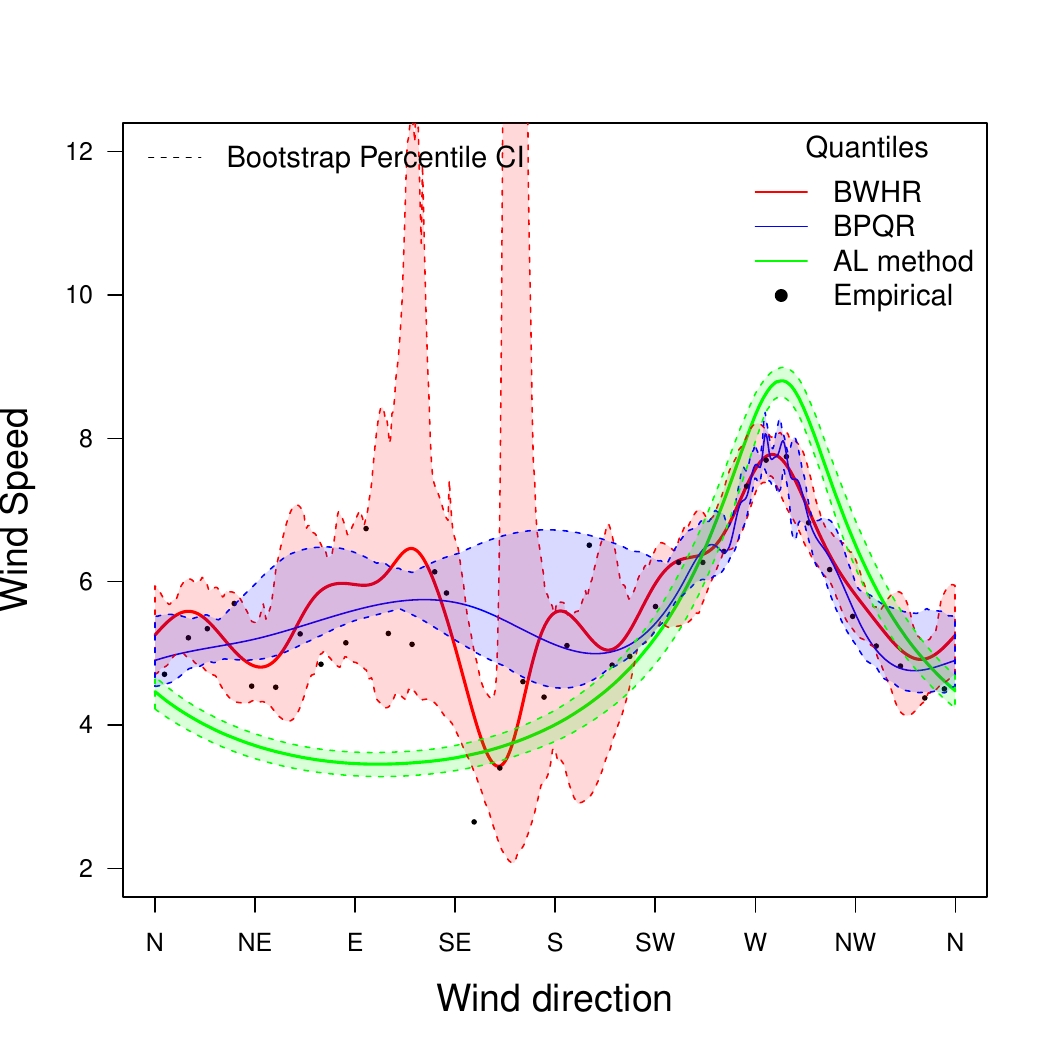}%
      \includegraphics[width = 0.33\textwidth]{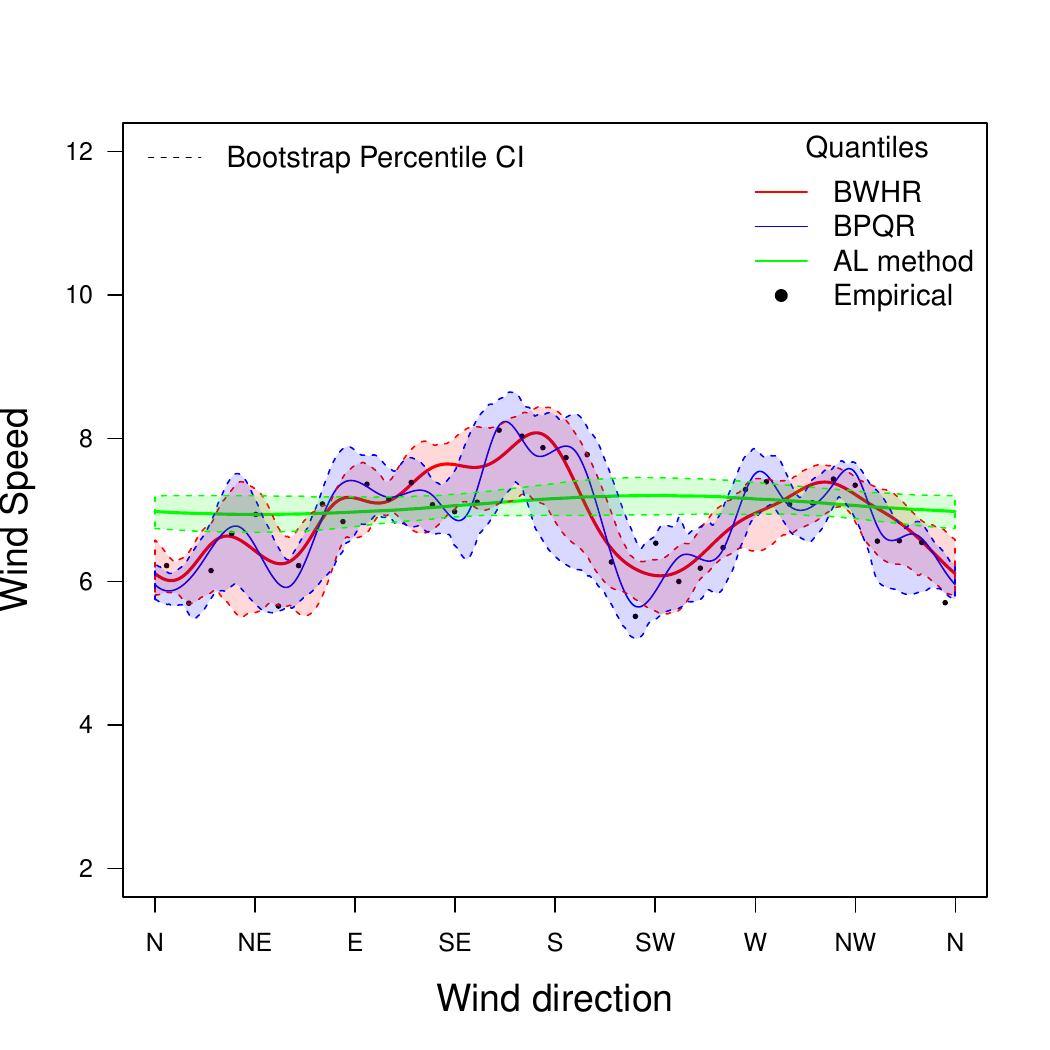}%
      \includegraphics[width = 0.33\textwidth]{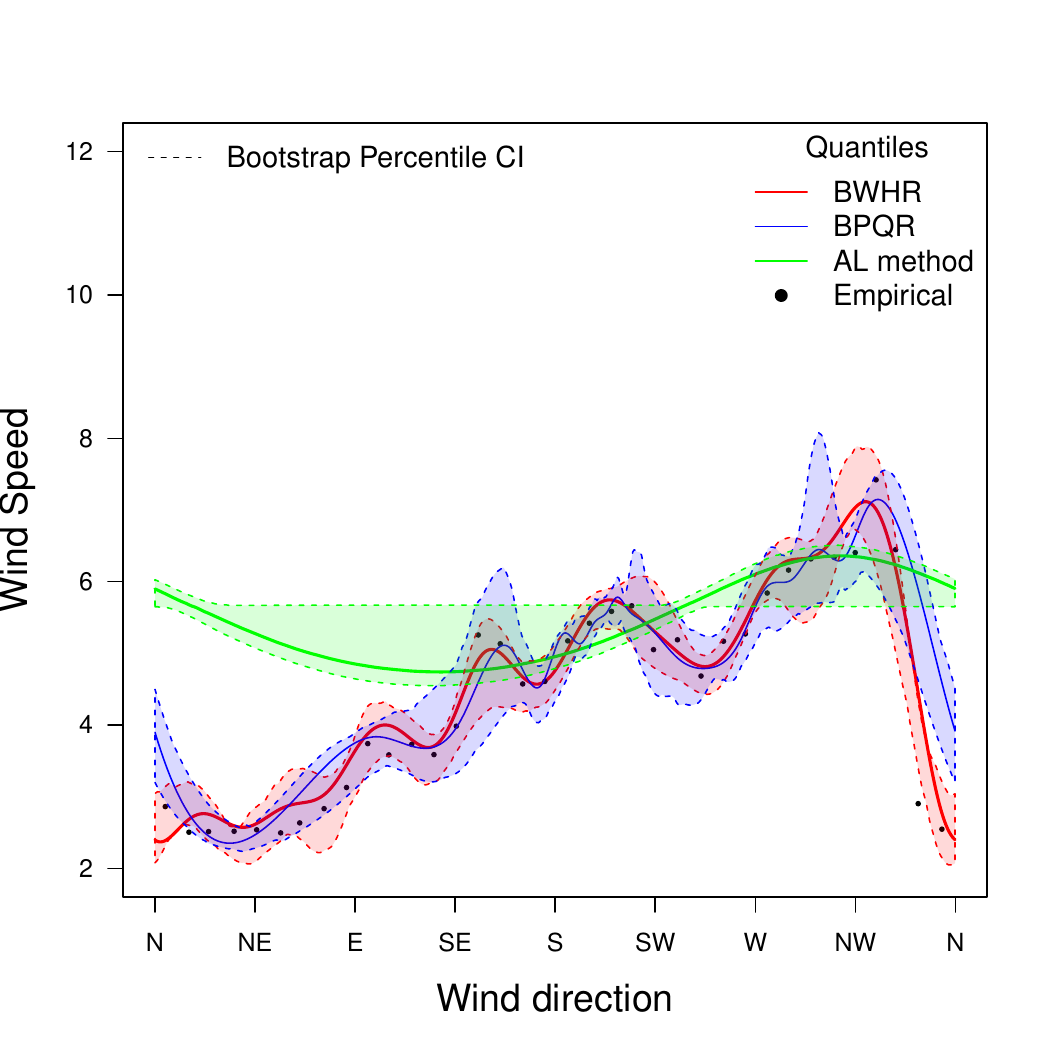}\\
      \includegraphics[width = 0.33\textwidth]{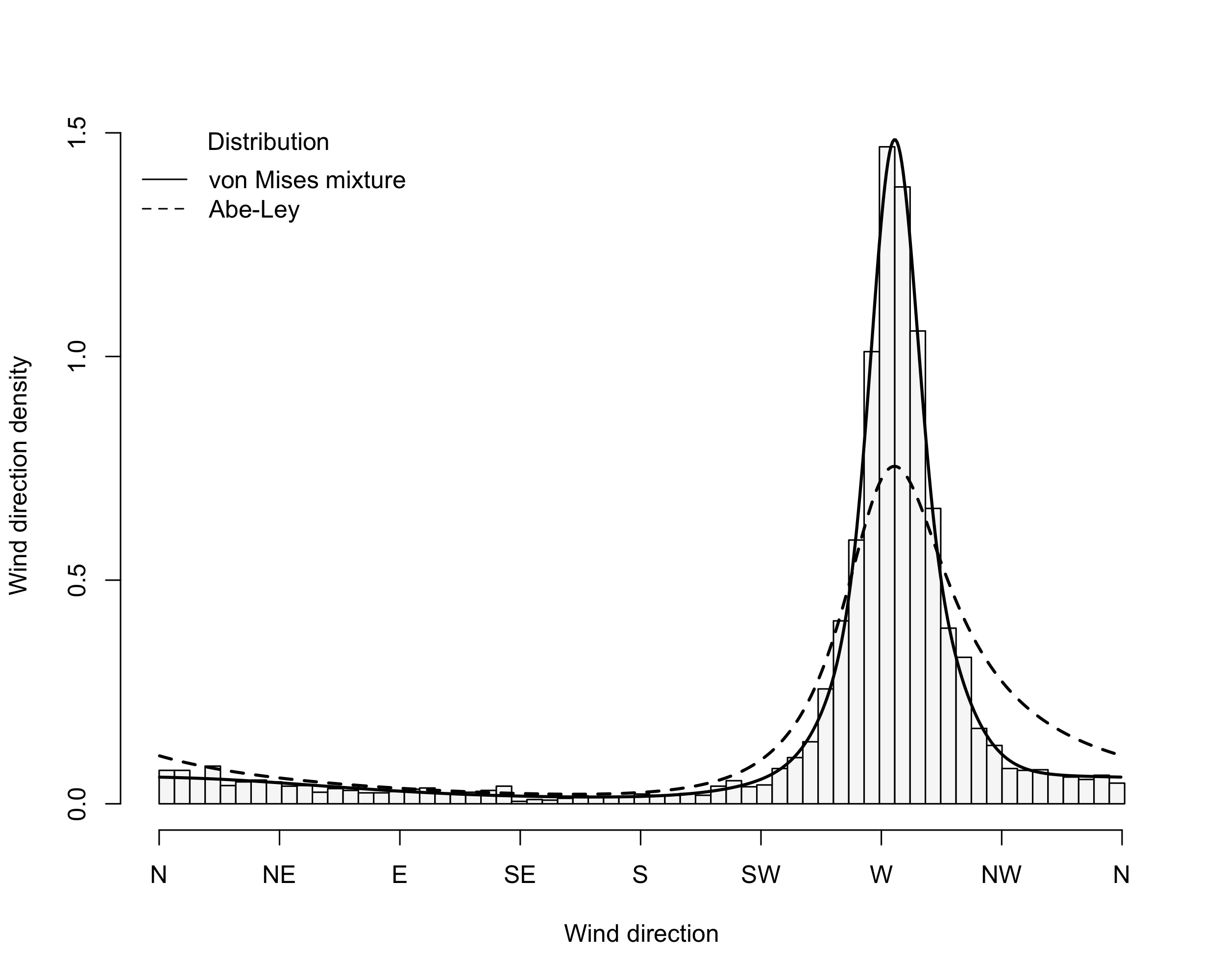}%
      \includegraphics[width = 0.33\textwidth]{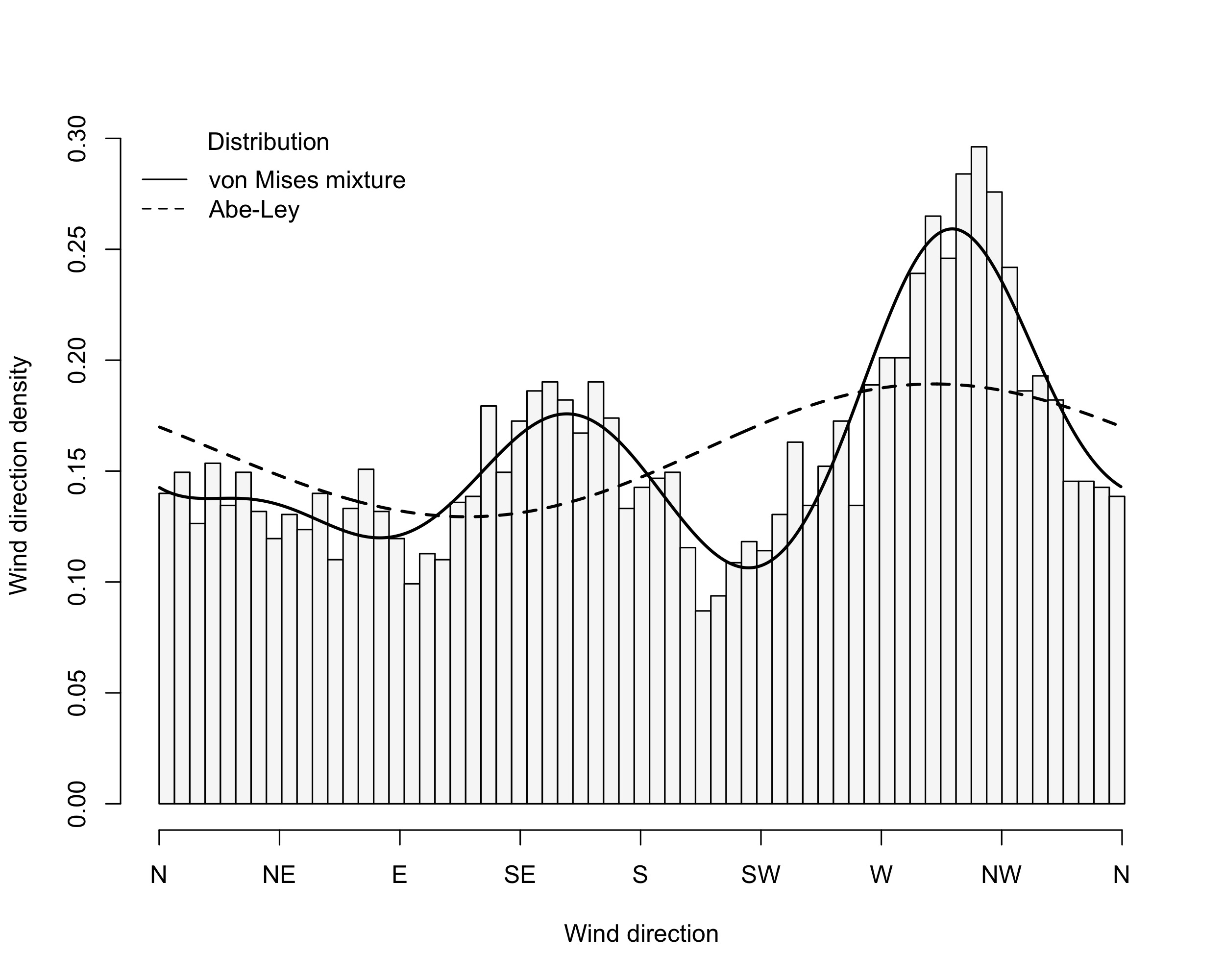}%
      \includegraphics[width = 0.33\textwidth]{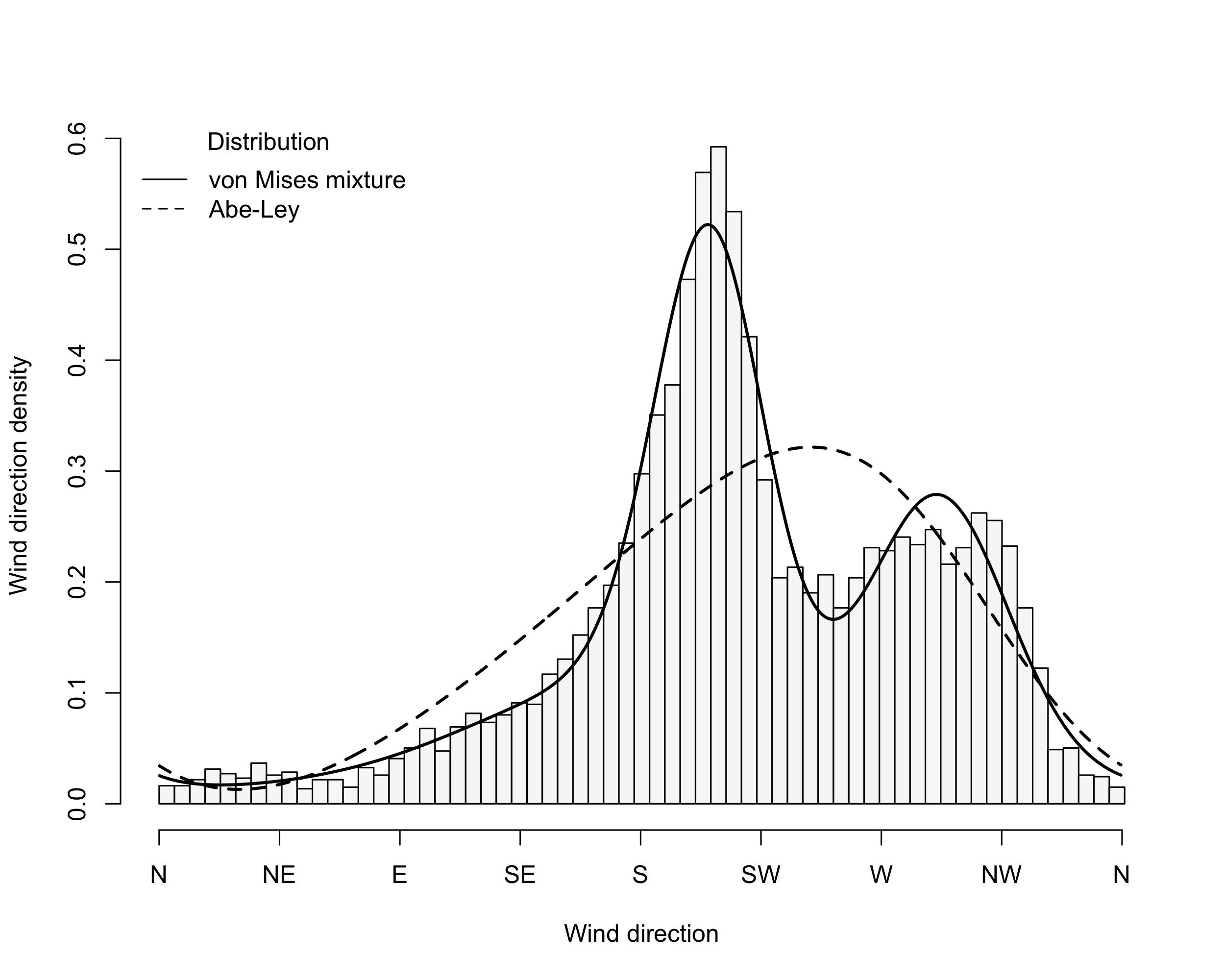}%
        \caption{\small \textbf{Top row:} $95\%$ bootstrapped percentile confidence intervals of the $95\%$ quantile curve using \texttt{BWHR} (red), \texttt{BPQR} (blue), and \texttt{AL} method (green). The solid dots represent the (binned) empirical $95\%$ quantile estimates. \textbf{Bottom row:} Estimated wind direction distribution at three different locations using von Mises mixture distribution (\textbf{solid line}) and the wind direction distribution derived from the Abe-Ley distribution (\textbf{dashed line}). }
       \label{bootstrap}
   \end{figure}

To quantify the uncertainty associated with the parameter estimation we use a version of block bootstrap \citep{kunsch1989} for all three methods, \texttt{BWHR}, \texttt{AL} method and \texttt{BPQR}. The block bootstrap is used in order to preserve the interannual temporal dependence presented in the climate application. Specifically, we draw 500 block bootstrap samples, where a block  represents one year \citep{blockboot}. In particular, since our data $\{r_{i}, \phi_{i}\}_{i=1}^{n}$ represent wind speed and wind direction measured during one season we can arrange them into blocks, where each block represents a year. The $100\times(1-\alpha)\%$ bootstrap percentile interval is constructed using the $\frac{\alpha}{2}$ upper and lower percentiles. Fig. \ref{bootstrap} shows these intervals for Texas Great Plain (TX\_GP), North Dakota Great Plain (ND\_GP) and North Carolina mountains (NC\_mtn) location along with the computed $95\%$ quantile curve and empirical quantile estimates. 
   
We conclude this section by noting that one can apply a one-step approach to estimating the directional wind speed distribution, providing a more coherent uncertainty quantification. However, in some instances, including our case, this approach may not produce robust results because of the more complex optimizations involved. In our specific context, a one-step method like estimating the directional wind speed distribution using maximum likelihood estimation (MLE) without binning can result in erratic estimates due to numerical challenges or oversmoothing, particularly when wind direction data are sparse. We provide results from this method in the supplementary material (SM1 Figure 1), showing that the unbinned MLE tends to overly smooth the distribution in these sparse sections. Our multistep procedure, which incorporates binning, mitigates this problem by focusing on local areas, offering a more accurate representation of the wind speed distribution. By implementing this multistep procedure, we can then effectively quantify uncertainty using the bootstrap method, as it offers a comprehensive view of uncertainty estimation when other methods may not be applicable. As illustrated in Fig.~\ref{bootstrap}, particularly in the first column where the wind direction distribution is sparse, our proposed method, the \texttt{BWHR}, shows higher uncertainty in areas such as the SE and S directions. This aligns with the understanding that limited data leads to greater uncertainty. In contrast, the \texttt{AL} method exhibits a lower uncertainty in these regions. The bootstrap method not only enhances our ability to capture uncertainty, acknowledging the challenges of achieving precise estimates with limited data, but also allows us to assess the impact of outliers on parameter estimation. Specifically, it reveals that in regions where the wind direction distribution has low probabilities, our proposed method appropriately reflects higher uncertainty.

\section{Simulation Study} \label{sim}

The purposes of this simulation study are fourfold: (i). to demonstrate how we implement the \texttt{BWHR} method, (ii). to assess the performance of the wind direction distribution estimation, $[\Phi]$, by a von Mises mixture distribution, (iii). to compare the \texttt{BWHR} method to the \texttt{BPQR} and AL methods in terms of $[R|\Phi=\phi]$, and (iv). to investigate the performance of the \texttt{BWHR} method in estimating the changes in the joint distribution of wind speed and direction. Simulated wind direction distributions were chosen to represent a variety of scenarios, ranging from a distribution concentrated at a dominant direction to one that is roughly uniformly spread across all directions. In our simulation study, all distributions were estimated using a von Mises mixture distribution applied to our motivated datasets, which led to at least three components. The first two scenarios incorporated four components, while the third used three.

\subsection{Design}

In this simulation study we mimic the joint distribution of wind speed and direction at three different locations presented in Section \ref{application}. 
Specifically, at each location the true data generating mechanism is estimated using the output from a regional climate model (see Section~ \ref{application} for more details) by the followings:
\begin{enumerate}
        \item Fit a bivariate Normal mixture distribution to $\{(u_{i},v_{i})\}_{i=1}^{n}$ of the model output. We purposely specify the data generating mechanism under the Cartesian coordinates to assess the flexibility of our conditional approach specified under the polar coordinates. The bivariate normal mixture fitting is done using the \texttt{mclust()} function from the \texttt{mclust} package \citep{mclust} in \texttt{R} \citep{R}. 
        \item Transform the fitted bivariate normal mixture distribution to $[R, \Phi]$ distribution and compute the distribution of wind direction $[\Phi]$ and the conditional distribution $[R|\Phi]$.
    \end{enumerate}

For each location we generate 500 replicates, each having a sample size of 7360 wind speed and wind direction data points, corresponding to 3-hourly wind data of one season (assuming 92 days per season) over the course of 10 years. To evaluate the performance in estimating \textit{changes} in the directional wind speed distribution, we apply the above steps to the model output for both historical and future scenario (see Section~\ref{application} for further details) to compute their differences. \texttt{BWHR}, \texttt{AL} method and \texttt{PBQR} are fitted to these simulated data and the estimated quantiles are subtracted to assess the estimation performance (see Section~\ref{sec:Est_preformance}).

\subsection{Illustration of BWHR}
\begin{figure}[H]
    \centering
    \includegraphics[width = 0.45\textwidth]{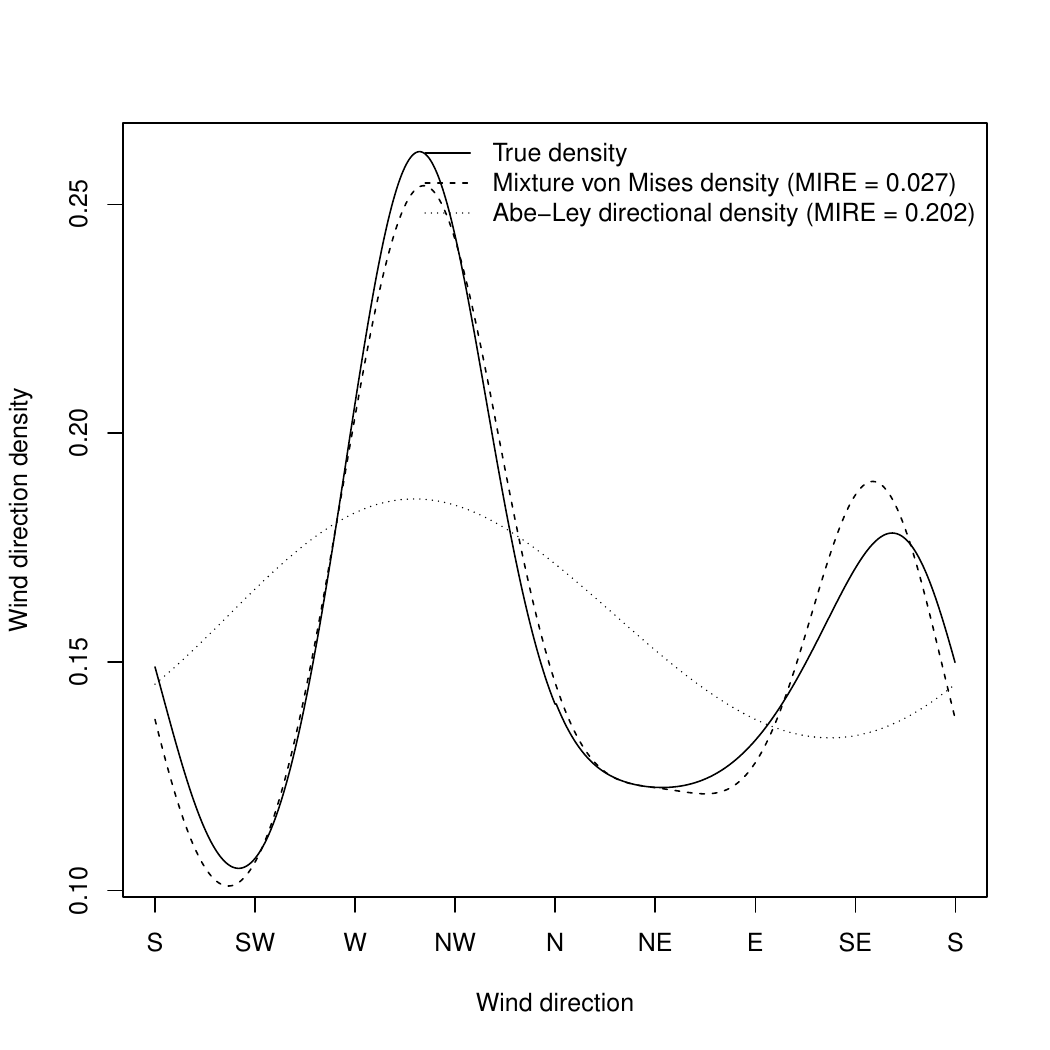}
    \includegraphics[width = 0.45\textwidth]{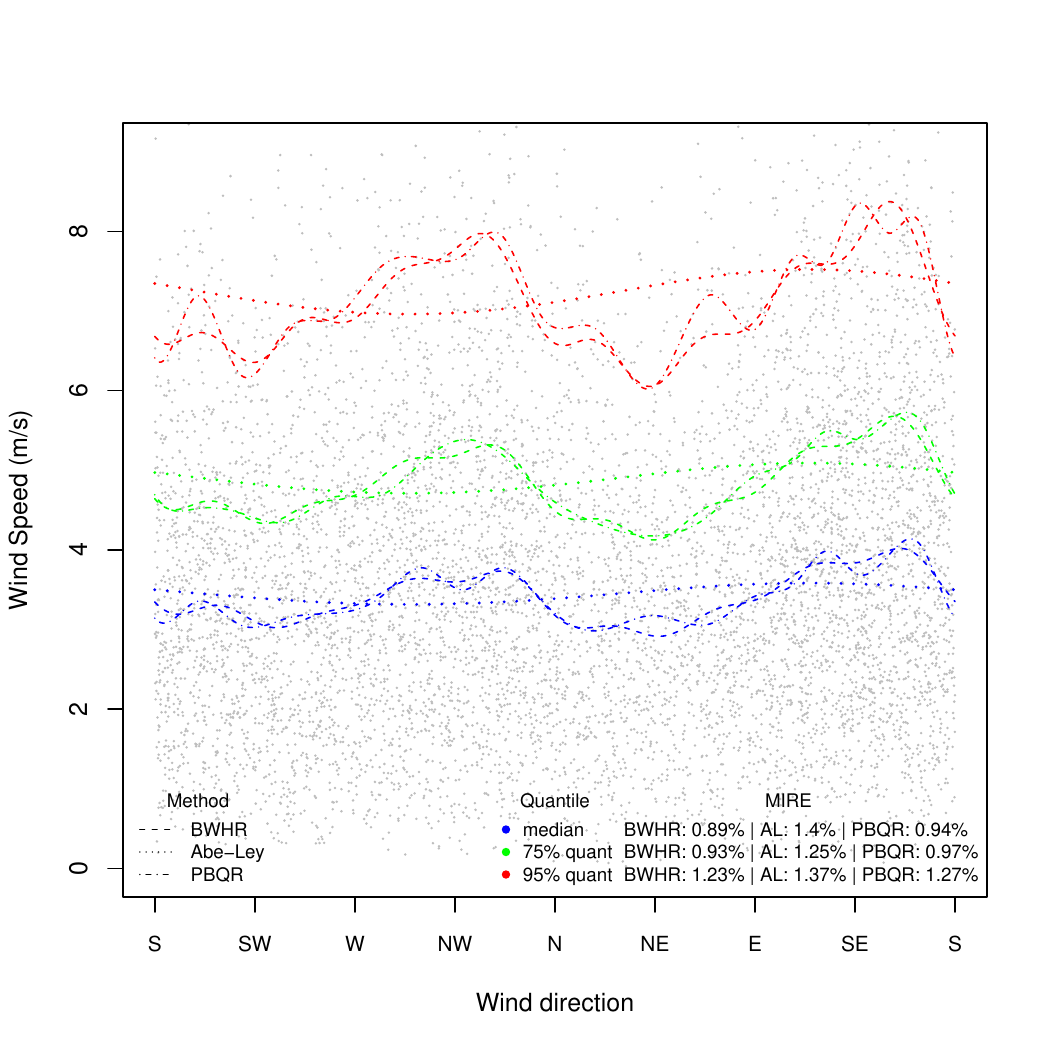}
    \caption{\small An illustration of applying the von Mises mixture distribution and the Abe-Ley directional distribution along with the \texttt{BWHR}, \texttt{BPQR} and \texttt{AL} methods to a simulated data set. (\textbf{Left}) Wind direction distribution is estimated using a 4 component von Mises mixture distribution (dashed line) and the Abe-Ley directional distribution (dotted line). (\textbf{Right}) We bin wind direction into $N=36$ equally spaced bins and we use $K_\alpha = K_\beta= 8 $ pairs of Fourier series to obtain the \texttt{BWHR} quantiles (dashed lines); the \texttt{BPQR} (dash-dot lines) are computed using 18 degrees of freedom; the dotted lines represent the quantiles of the \texttt{AL} conditional distribution}. 
    \label{methodillus}
\end{figure}

We use a simulated data series (see Fig. \ref{methodillus}) to illustrate the \texttt{BWHR}, AL method and \texttt{PBQR} as described in Section \ref{method}. To estimate wind direction we use the \texttt{movMF()} function from the \texttt{movMF} \citep{movFMR} package in \texttt{R}. The number of components of the von Mises mixture distribution is determined using BIC. 

For the \texttt{BWHR} method we choose the number of bins to be $N = 36$, so that there are, on average, about 200 data points across these bins, and we set $K_\alpha = K_\beta = 8$. These choices were made following the suggestions from Section~\ref{sec:BWHR}. A sensitivity analysis is included in the supplementary materials (Sec.~SM~2 and SM~3).  
After determining $N$, $K_{\alpha}$, and $K_{\beta}$, \textbf{Algorithm}~\ref{algorithm1} can be applied to estimate $\alpha(\phi)$ and $\beta(\phi)$. The estimation performance of $[R|\Phi=\phi]$ is evaluated by examining a few selected quantile curves ($.5, .75$ and $.95$, see Fig. ~\ref{methodillus}  for an example).

In implementing the \texttt{BPQR} method, we model each selected quantile of the directional wind speed distribution separately where a periodic B-spline is used to represent the directional quantile curve. We use the \texttt{quantreg} \citep{quantregR} and \texttt{pbs} \citep{pbsR} packages in \texttt{R} \citep{R} to carry out the estimation. In generating the periodic B-spline we need to determine the degrees of freedom ($df$). Based on a sensitivity analysis we choose $df = 18$ to balance bias and variance.

In implementing the \texttt{AL} model we optimize the parameters of the Abe-Ley full cylindrical distribution with the \texttt{DEoptim} library \citep{DEoptim}, which maximizes the log-likelihood function. After optimization, we derive the shape and scale parameters for the Weibull distribution using the model parameters and the wind direction. Finally, we utilize these parameters along with the \texttt{qweibull()} function in \texttt{R} to calculate the desired quantiles.

\subsection{Estimator Performance}
\label{sec:Est_preformance}
It is important to incorporate wind direction distribution when quantifying the estimation performance. Therefore, we define the mean integrated relative error (MIRE) with respect to wind direction density, $f(\phi)$, specifically, 
\begin{equation}
\mathrm{MIRE} = \int_0^{2\pi} \abs*{ \frac{\hat{g}(\phi) - g(\phi)}{g(\phi)}} f(\phi)d\phi \, ,
\end{equation}
where $g(\phi)$ is the estimand (the quantity of interest) as a function of direction (e.g.,\ directional $\tau$ - quantile of wind speed ($q_{\tau}(\phi)$), wind direction probability density function ($f(\phi)$), or quantile difference ($q_{\tau,\text{diff}}(\phi)$)). This metric is computed by discretizing the wind direction domain using a set of evenly spaced grid points $\{0 < \phi_1 < \phi_2< \ldots< \phi_m < 2\pi\}$ with a sufficiently large $m$ (we use $m = 629$ here) and using the following formula:

\begin{equation}
   \mathrm{MIRE} \approx \frac{\sum_{i = 1}^m f(\phi_i) \abs*{\frac{\hat{g}(\phi_i) - g(\phi_i)}{g(\phi_i)}}\Delta \phi}{\sum_{i=1}^m f(\phi_i)}.
\end{equation}

We summarize the results of the simulation study in Table ~\ref{MIREvals1} - Table~\ref{MIREvals_diff1}. In general, the wind direction distribution estimation performs reasonably well by using a mixture of von Mises distributions. At NC\_mtn location, however, we note that the mixture of von Mises distributions is not able to capture well the distribution of wind direction, compared to the other locations, likely due to the fact that this is a mountainous region with a more complex wind direction distribution that is not well modeled by a von Mises mixture distribution. In the case of estimating the directional wind speed distribution, the \texttt{BWHR} method consistently outperforms both the \texttt{AL} method and the \texttt{BPQR} method across all three locations. Finally, when estimating changes in the directional wind speed distribution, we conclude that the \texttt{BWHR}  method  performs as good as the other two methods, in some cases, may even outperform them. However, we note that caxution should be taken when interpreting the performance of the \texttt{AL} method in estimating changes in the directional wind speed distribution, as it shows a significantly higher relative error in estimating the quantiles.

We end this section by investigating the circumstances in which one approach, either the \texttt{BWHR} method, the \texttt{AL} method, or the \texttt{BPQR} method, outperforms the others. Both the \texttt{BWHR} and \texttt{BPQR} methods consistently surpass the \texttt{AL} method in performance when estimating the quantiles of the directional wind speed distribution. This is attributed to the AL method's limited flexibility in capturing the directional aspect, where only the shape parameter depends on wind direction. As noted earlier, the \texttt{BWHR} method demonstrates strong performance in estimating the directional wind speed distribution based on the MIRE metric. However, in the case of wind directions with significantly lower occurrence probabilities, the \texttt{BPQR} method performs slightly better. Moreover, representing the directional wind speed distribution using the \texttt{BWHR} allows for a data generating mechanism and, thus, we can use a simulation-based approach to examine the joint distribution of $[R, \Phi]$ (hence $[U, V]$ ) and their changes. We see this property as an advantage of \texttt{BWHR} over \texttt{BPQR} since QR method by itself does not completely determine the conditional distribution and hence the joint distribution.

\begin{table}[H]
\captionof{table}{\small{Boxplots of MIRE values of wind direction (\textbf{section (a)}), \texttt{BWHR} method, \texttt{AL} method and \texttt{BPQR} method (\textbf{section (b)}). The tables contain the MIRE values averaged over the 500 replicates along with the standard deviation values of MIRE of mixture of von Mises (\texttt{MvM}), Abe-Ley wind direction distribution (\texttt{AL}) (\textbf{section (a)}), \texttt{BWHR} method, \texttt{AL} method and \texttt{BPQR} method (\textbf{section (b)}).}}
\label{MIREvals1}
\subcaption{Wind direction}
\begin{tabularx}{\linewidth}{@{}c X @{}}
    \includegraphics[width = 0.25\textwidth, valign = c]{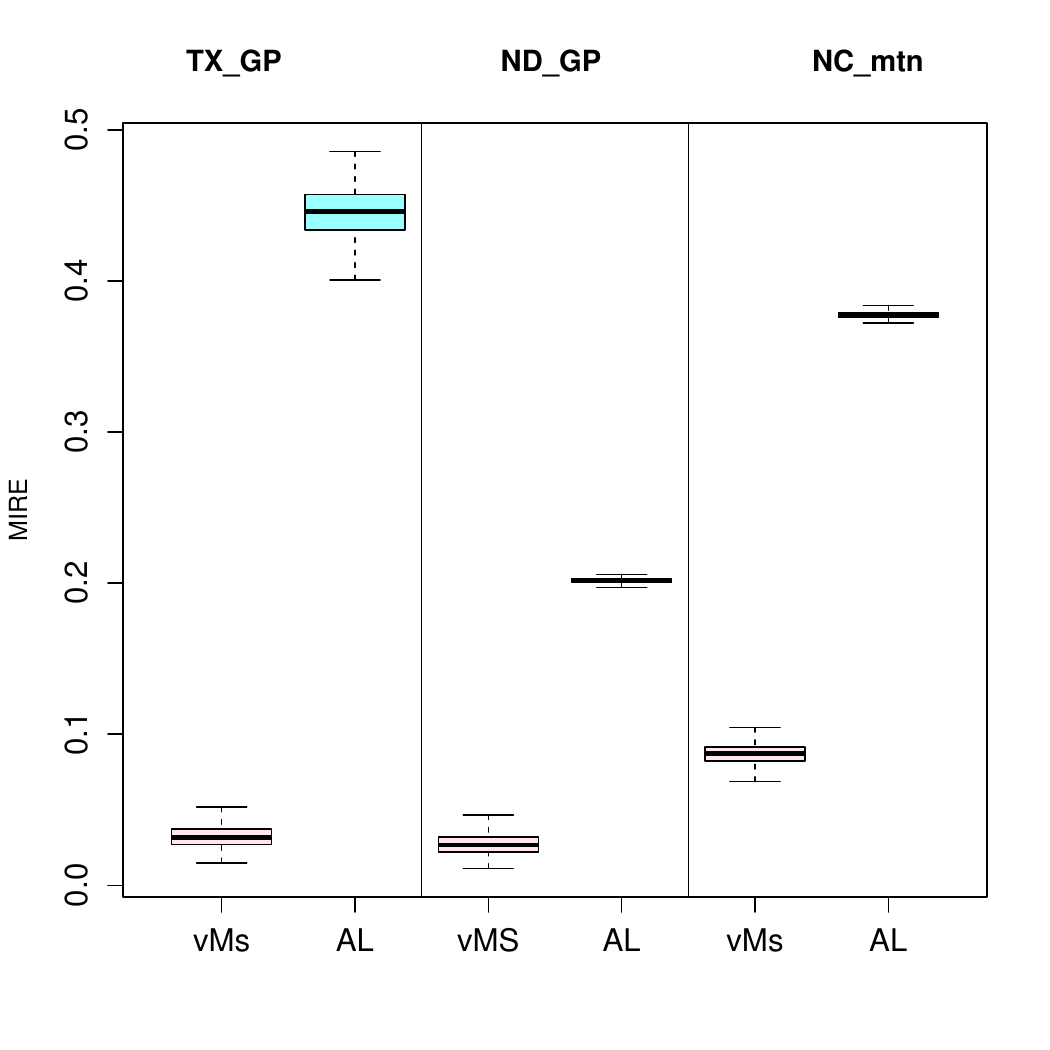}%
    &
    \small{\begin{tabular}{l|c|c}
        & Method & MIRE{\tiny($sd$)} \\
        \hline
        \multirow{2}{*}{\texttt{TX\_GP}}
        &\texttt{MvM} & 0.032 {\tiny(0.007)}  \\
        & \texttt{AL}& 0.446 {\tiny(0.017)}  \\
        \hline
        \multirow{2}{*}{\texttt{ND\_GP}}
        &\texttt{MvM}& 0.027 {\tiny(0.008)}  \\
        & \texttt{AL}& 0.202 {\tiny(0.002)}  \\
        \hline
        \multirow{2}{*}{\texttt{NC\_mtn}}
       &\texttt{MvM} & 0.087 {\tiny(0.007)}  \\
       & \texttt{AL}&  0.378 {\tiny(0.002)}  \\
        \hline
\end{tabular}}
    \label{WIMREvals1}
\end{tabularx}\\

\subcaption{Directional wind speed}
\centering
\includegraphics[width = .3\textwidth, valign = c]{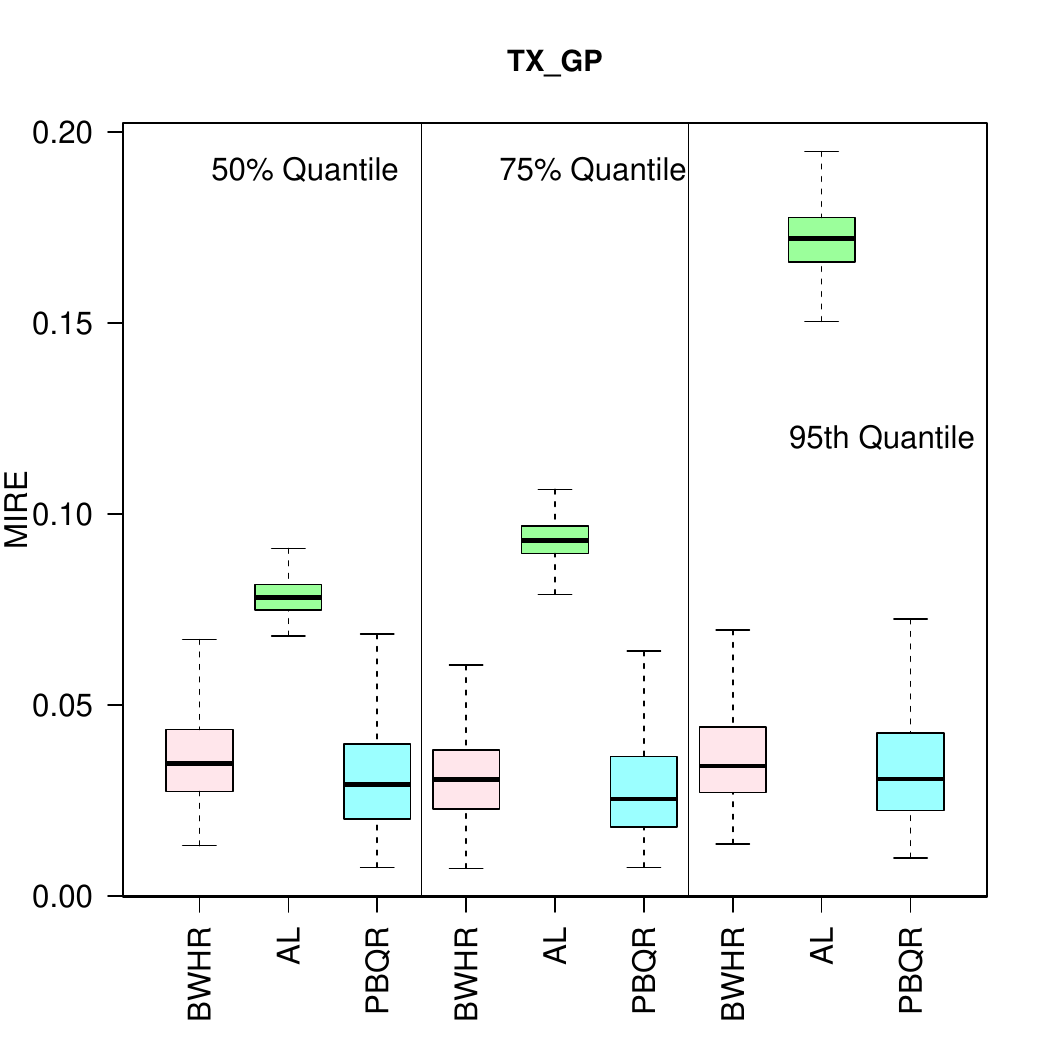}%
\includegraphics[width = .3\textwidth, valign = c]{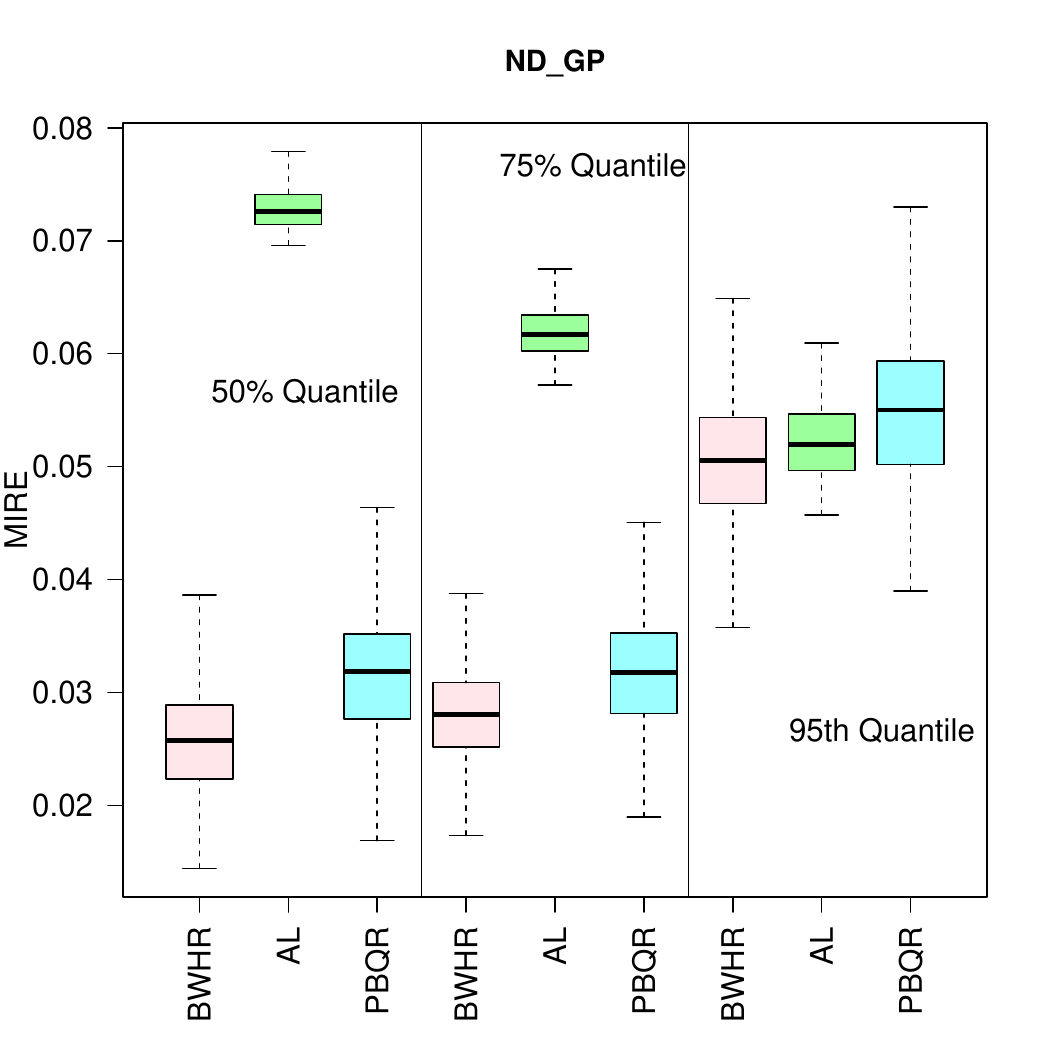}%
\includegraphics[width = .3\textwidth, valign = c]{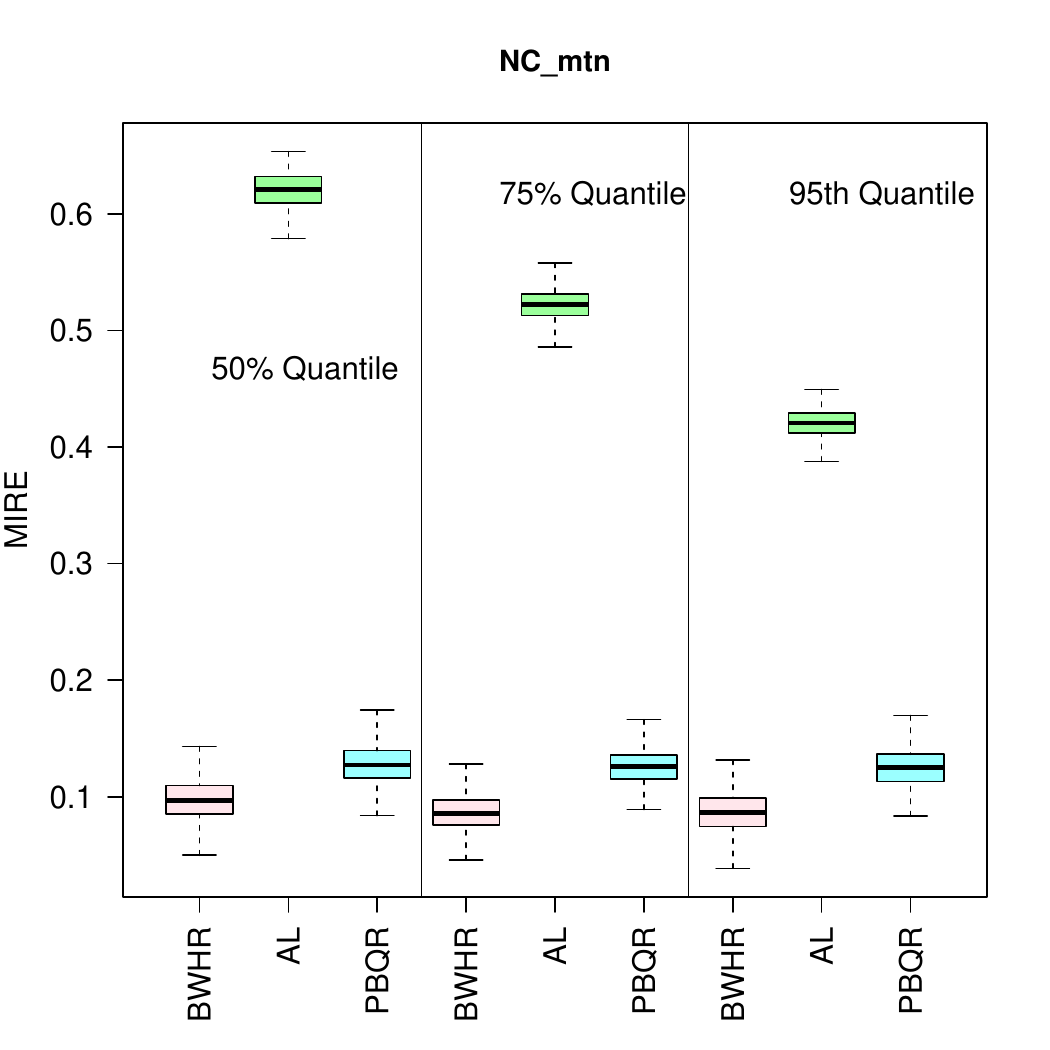}\\
\vspace{1cm}
    \centering
    \small
    \begin{tabular}{l|c|c|c|c}
        & Method & \footnotesize{MIRE$_{q_{\tau=0.95}}$} & MIRE$_{q_{\tau=0.75}}$ & MIRE$_{q_{\tau=0.50}}$ \\
        \hline
        \multirow{3}{*}{\texttt{TX\_GP}}
        & \texttt{PBQR} & \small{0.034} {\tiny(0.015)} & 0.029 {\tiny(0.014)} & 0.032 {\tiny(0.015)} \\
        & \texttt{BWHR} & \small{0.036} {\tiny(0.012)} & 0.032 {\tiny(0.012)} & 0.037 {\tiny(0.013)} \\
        & \texttt{AL method} & \small{0.172} {\tiny(0.009)} & 0.093 {\tiny(0.005)} & 0.079 {\tiny(0.005)} \\
        \hline
        \multirow{3}{*}{\texttt{ND\_GP}}
        & \texttt{PBQR} & 0.055 {\tiny(0.007)} & 0.032 {\tiny(0.005)} & 0.032 {\tiny(0.006)} \\
        & \texttt{BWHR} & 0.051 {\tiny(0.006)} & 0.028 {\tiny(0.004)} & 0.026 {\tiny(0.005)} \\
        & \texttt{AL method} & 0.052 {\tiny(0.009)} & 0.062 {\tiny(0.005)} & 0.073 {\tiny(0.005)} \\
        \hline
        \multirow{3}{*}{\texttt{NC\_mtn}}
        & \texttt{PBQR} & 0.125 {\tiny(0.017)} & 0.126 {\tiny(0.015)} & 0.128 {\tiny(0.018)} \\
        & \texttt{BWHR} & 0.088 {\tiny(0.019)} & 0.087 {\tiny(0.016)} & 0.097 {\tiny(0.018)} \\
        & \texttt{AL method} & 0.426 {\tiny(0.009)} & 0.528 {\tiny(0.006)} & 0.626 {\tiny(0.005)} \\
        \hline\\

    \end{tabular}
\end{table}

\vspace{0.35cm}

\begin{table}[H]

\captionof{table}{\small{Boxplots of MIRE values of quantile differences (\textbf{section (a)}). The tables contain the MIRE values averaged over the 500 replicates along with the standard deviation values of MIRE of wind direction (\textbf{top section}), \texttt{BWHR} method, \texttt{AL} method and \texttt{BPQR} method.}}
\label{MIREvals_diff1}
\subcaption{Directional quantile differences}
\centering
\includegraphics[width = .3\textwidth, valign = c]{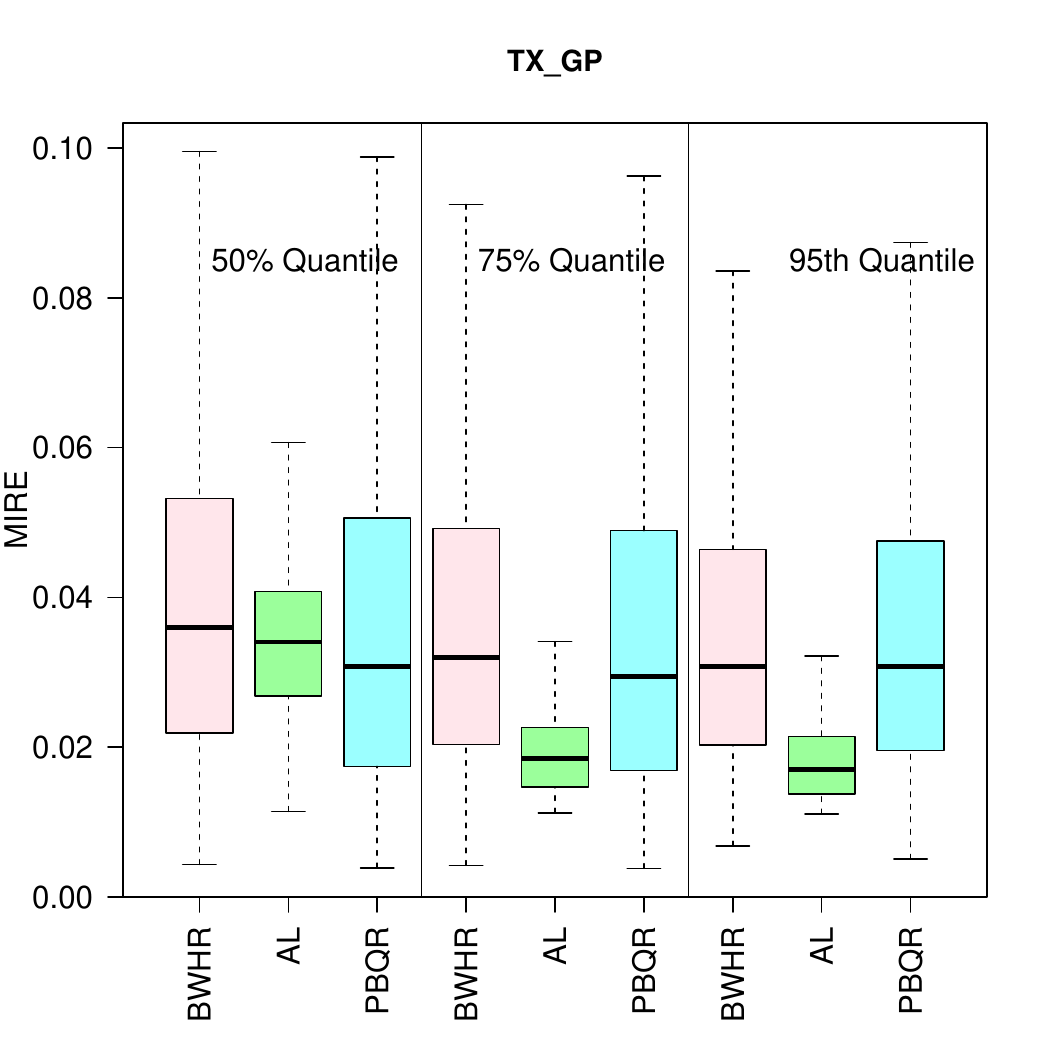}%
\includegraphics[width = .3\textwidth, valign = c]{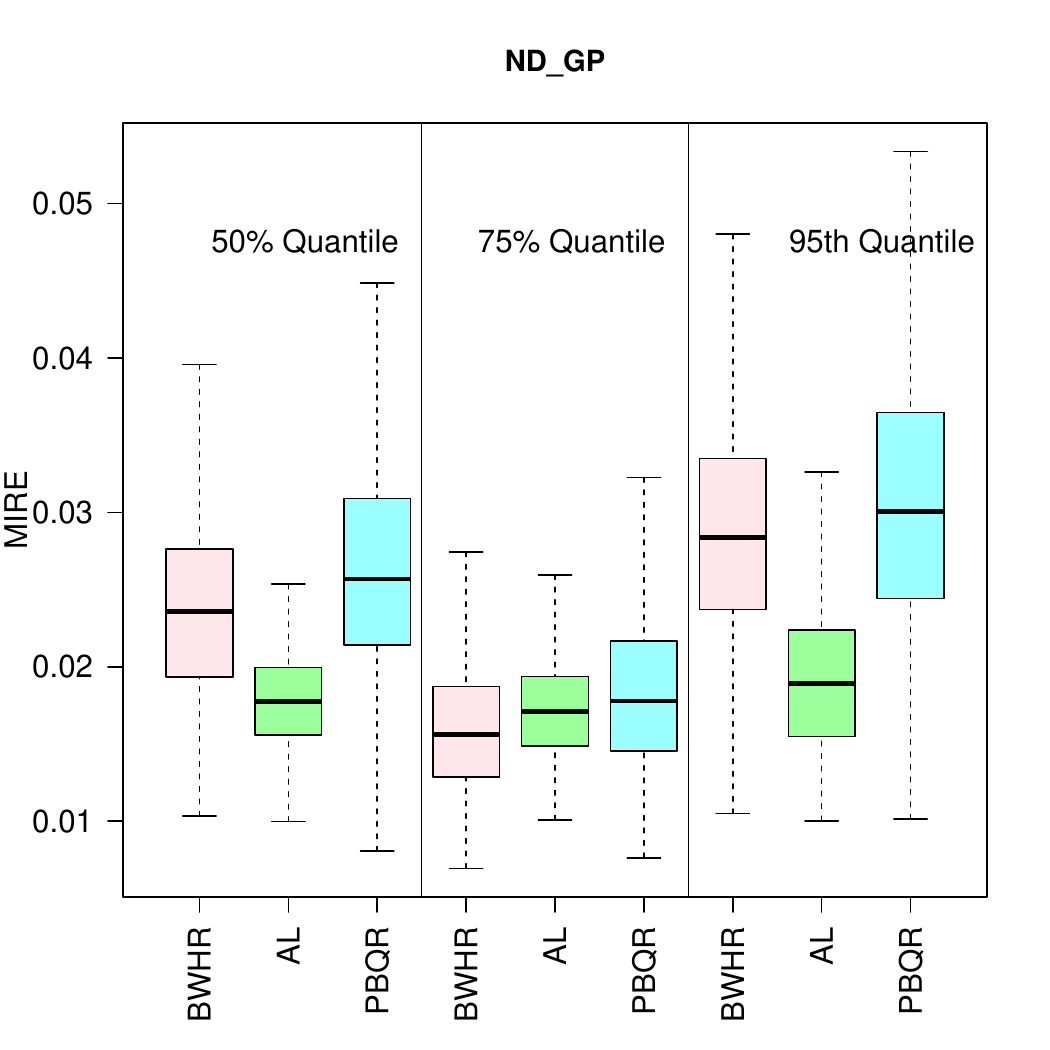}%
\includegraphics[width = .3\textwidth, valign = c]{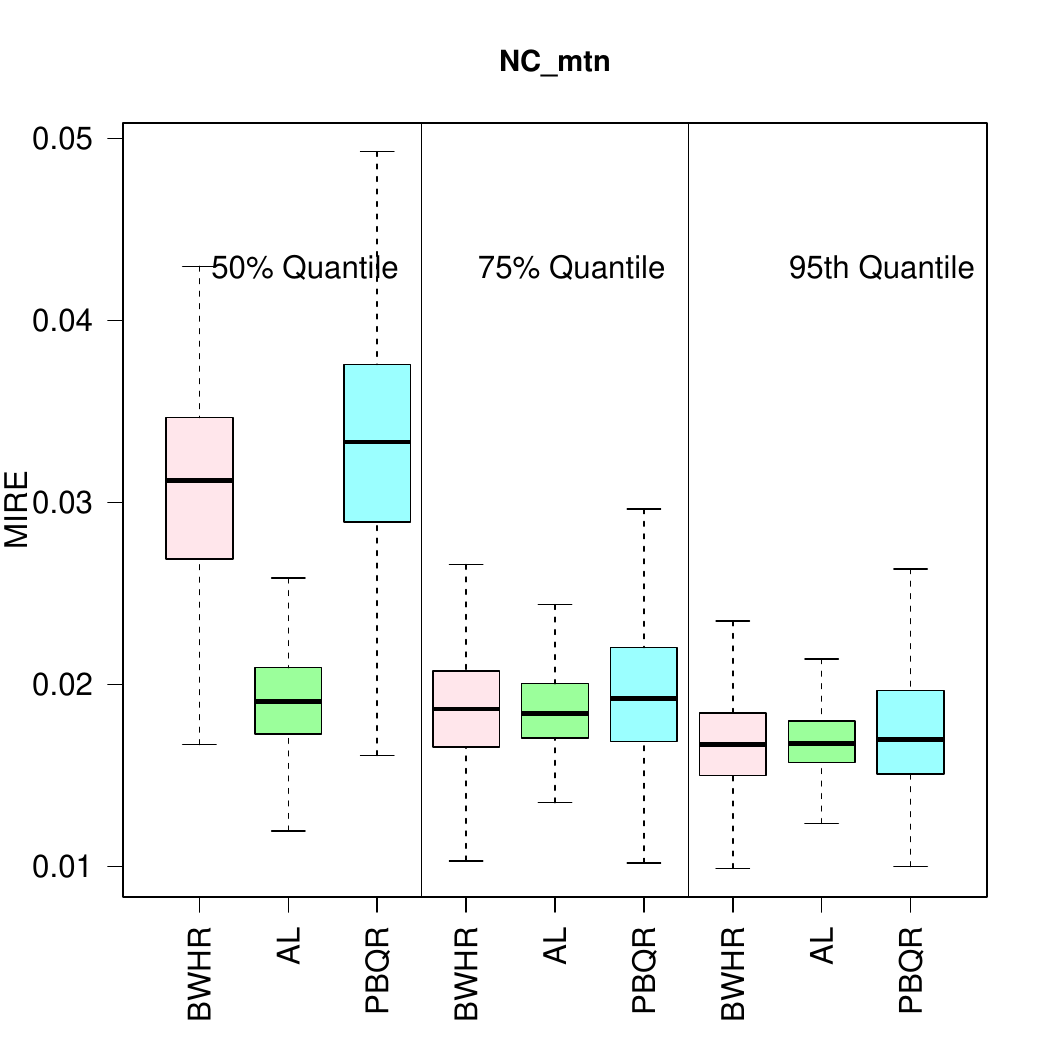}\\
\vspace{0.5cm}
\small
\begin{tabular}{l|c|c|c|c}
    & Method & \footnotesize{MIRE$_{q_{\tau=0.95}}$} & MIRE$_{q_{\tau=0.75}}$ & MIRE$_{q_{\tau=0.50}}$ \\
    \hline
    \multirow{3}{*}{\texttt{TX\_GP}}
    & \texttt{PBQR} & \small{0.037} {\tiny(0.022)} & 0.037 {\tiny(0.025)} & 0.039 {\tiny(0.025)} \\
    & \texttt{BWHR} & \small{0.037} {\tiny(0.020)} & 0.039 {\tiny(0.022)} & 0.042 {\tiny(0.024)} \\
    & \texttt{AL method} & \small{0.019} {\tiny(0.006)} & 0.020 {\tiny(0.006)} & 0.036 {\tiny(0.011)} \\
    \hline
    \multirow{3}{*}{\texttt{ND\_GP}}
    & \texttt{PBQR} & 0.031 {\tiny(0.009)} & 0.018 {\tiny(0.006)} & 0.026 {\tiny(0.007)} \\
    & \texttt{BWHR} & 0.029 {\tiny(0.008)} & 0.016 {\tiny(0.004)} & 0.024 {\tiny(0.006)} \\
    & \texttt{AL method} & 0.018 {\tiny(0.005)} & 0.017 {\tiny(0.003)} & 0.018 {\tiny(0.003)} \\
    \hline
    \multirow{3}{*}{\texttt{NC\_mtn}}
    & \texttt{PBQR} & 0.018 {\tiny(0.004)} & 0.020 {\tiny(0.004)} & 0.034 {\tiny(0.007)} \\
    & \texttt{BWHR} & 0.017 {\tiny(0.003)} & 0.019 {\tiny(0.003)} & 0.031 {\tiny(0.006)} \\
    & \texttt{AL method} & 0.017 {\tiny(0.002)} & 0.019 {\tiny(0.003)} & 0.02 {\tiny(0.004)} \\
    \hline
\end{tabular}
\end{table}

\section{Application} \label{application}
The aim of this application is to estimate how the joint distribution of wind speed and direction may change from present to a possible future climate condition. According to the 2021 climate change report published by the Intergovernmental Panel on Climate Change (IPCC) \citep{ipcc}, research on the possible impact of climate change on wind patterns is warranted as it is a resource for clean energy. However, compared to temperature and precipitation, it is still uncertain how wind speed and direction distribution might change under future climate scenarios. For example, \cite{windclimate} found that observational data exhibit a decreasing trend in the annual mean wind speeds while reanalysis data show an increasing trend in the annual mean wind speeds. Furthermore, \cite{ windclimate2} show that reanalysis data from different Climate models show inconsistent trends in wind speed distribution. We aim to supplement these findings by examining any changes in the distribution of wind speed with respect to wind direction using our method \texttt{BWHR}.

\subsection{Data Analysis}

\begin{figure}[H]
\begin{multicols}{2}
\includegraphics[width = 0.42\textwidth]{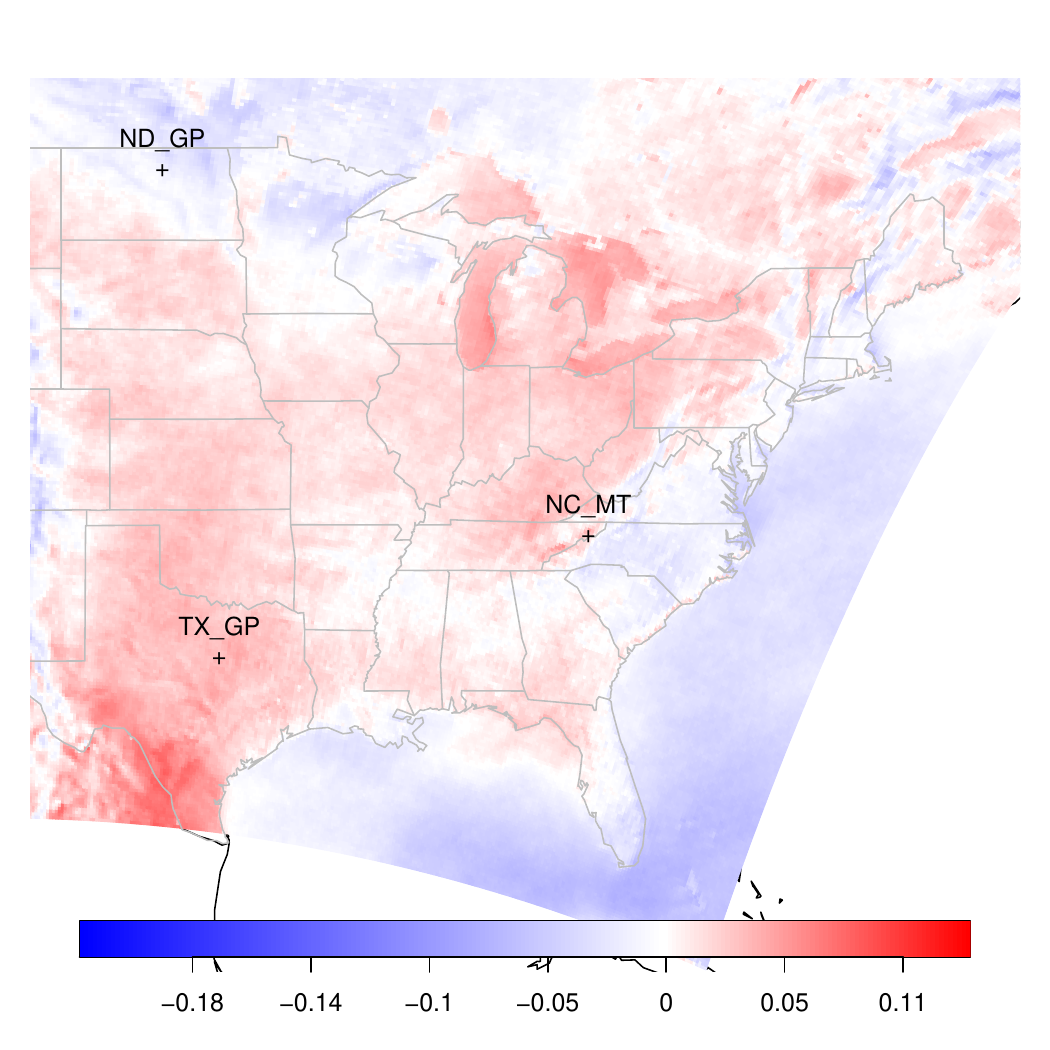}

 \includegraphics[width = 0.21\textwidth]{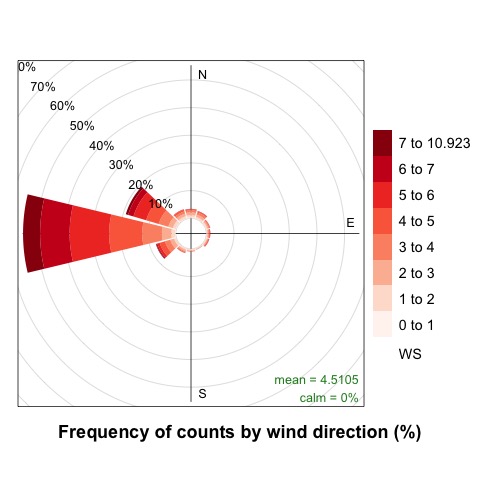}%
    \includegraphics[width = 0.21\textwidth]{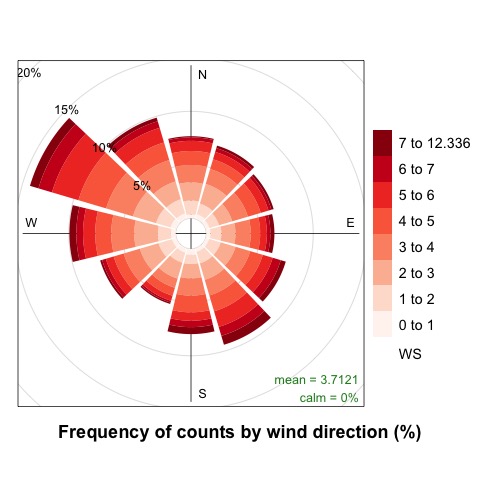}%
    \includegraphics[width = 0.21\textwidth]{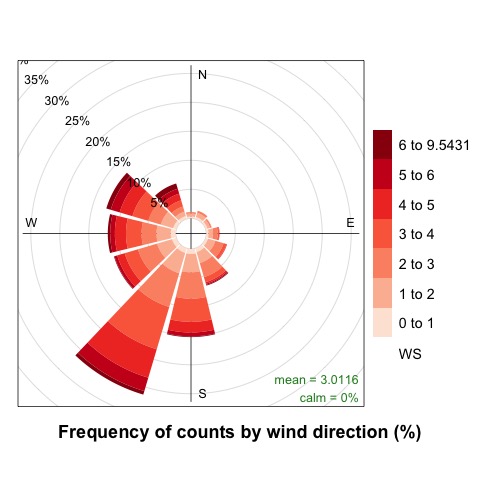}

    \includegraphics[width = 0.21\textwidth]{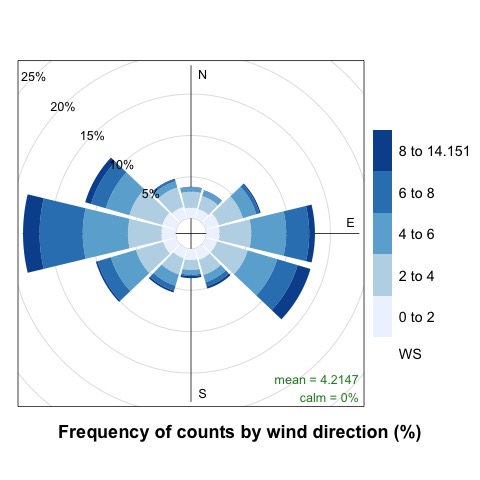}%
    \includegraphics[width = 0.21\textwidth]{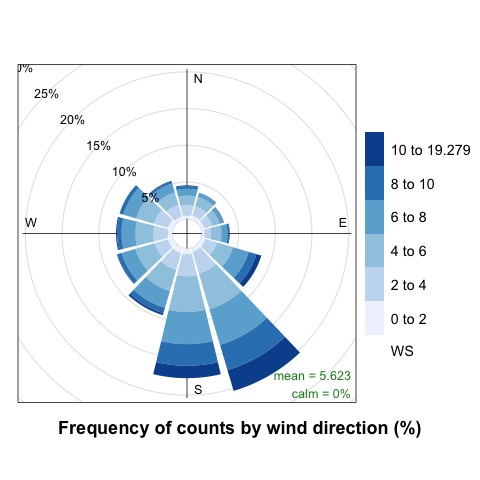}%
    \includegraphics[width = 0.21\textwidth]{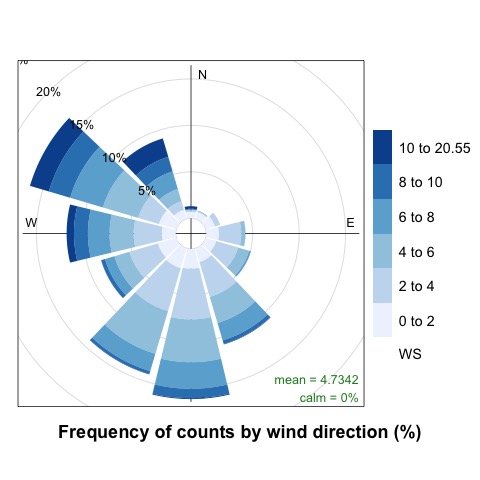}%
\end{multicols}

    \caption{\small \textbf{(Left)} studied locations: Texas Great Plain (TX\_GP), North Dakota Great Plain (ND\_GP), North Carolina mountain (NC\_mtn). We use the ``+'' symbol to denote the center of these grid cells. The map is colored by the ratio change in 10-year mean 95th-percentile of wind speed distribution projected by WRF driven by Community Climate System Model 4 (CCSM4), see data description in Section~ \ref{application}. \textbf{(Right)}: Windrose diagrams for TX\_GP (left column), ND\_GP (middle column), and NC\_mtn (right column) for summer (top row) and winter (bottom row).}
    \label{maploc}

\end{figure}

In this study we use 3-hourly output from 12km Weather Research and Forecasting (WRF) \citep{WRF} regional climate simulations driven by the Community Climate System Model 4 (CCSM4) \citep{data} under 10-year historical (1995-2004) and a future time-period (2085-2094) under a representative concentration pathway (RCP) 8.5 during the summer season (June, July, August) and winter season (December, January, February). More details on these simulations can be found in \citep{wang2015,zobel2018a,zobel2018b}. To provide some spatial diversity of wind speed and wind direction distributions we use the output from three locations: Texas Great Plain (TX\_GP), North Dakota Great Plain (ND\_GP) and North Carolina mountains (NC\_mtn) (see Fig. ~\ref{maploc}). Rose diagrams at these locations for summer and winter seasons are shown in Fig.~\ref{maploc}. The distributions of wind speed and wind direction at these locations are depicted in the top row of Fig.~\ref{diff} - \ref{diffwin}, and Fig.~\ref{wdboot}, respectively. The wind direction distributions were chosen to feature different scenarios, in particular, ranging from a distribution concentrated around one direction to a distribution that is ``evenly'' distributed along all directions.  

\subsection{Change in the Wind Direction Distribution}

\begin{figure}[H]
    \centering
    \includegraphics[width = .33\textwidth]{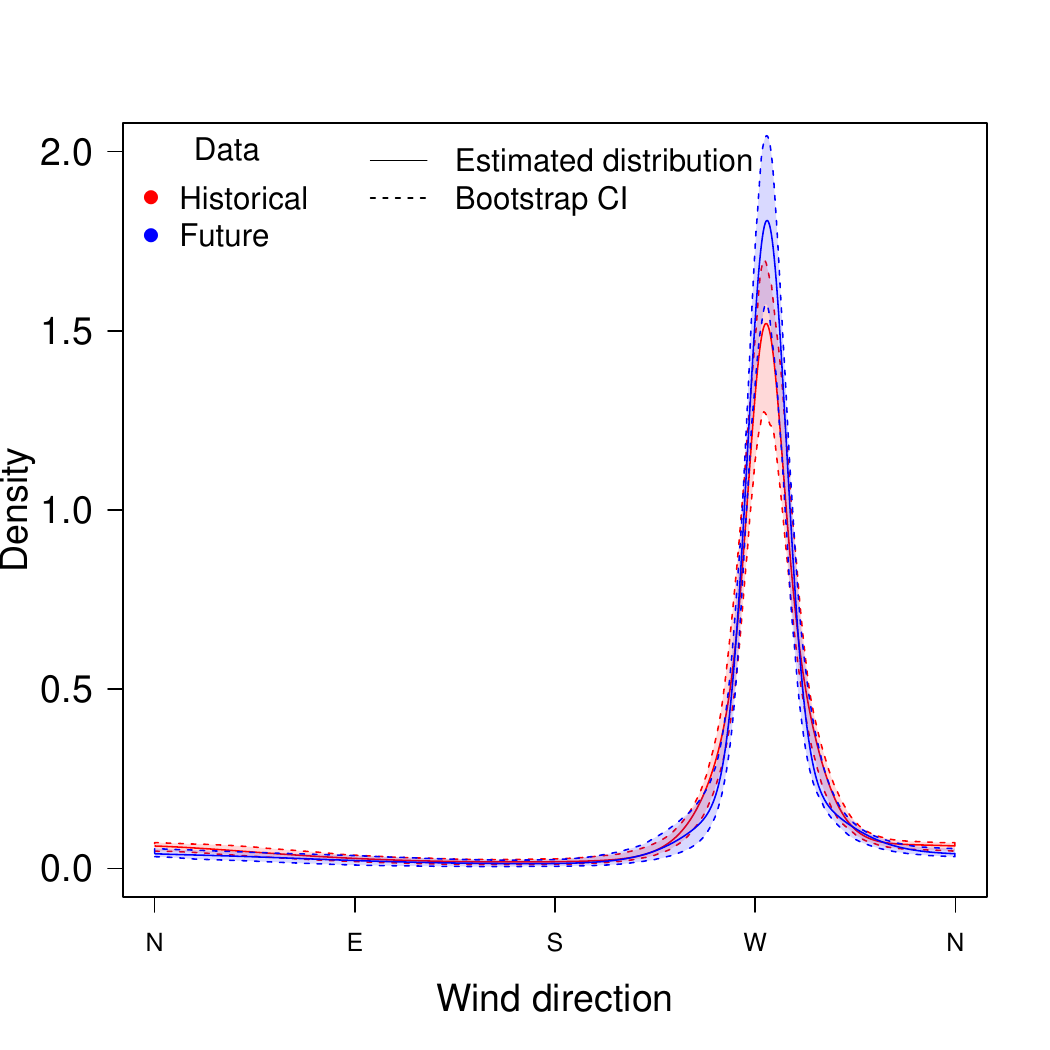}%
    \includegraphics[width = .33\textwidth]{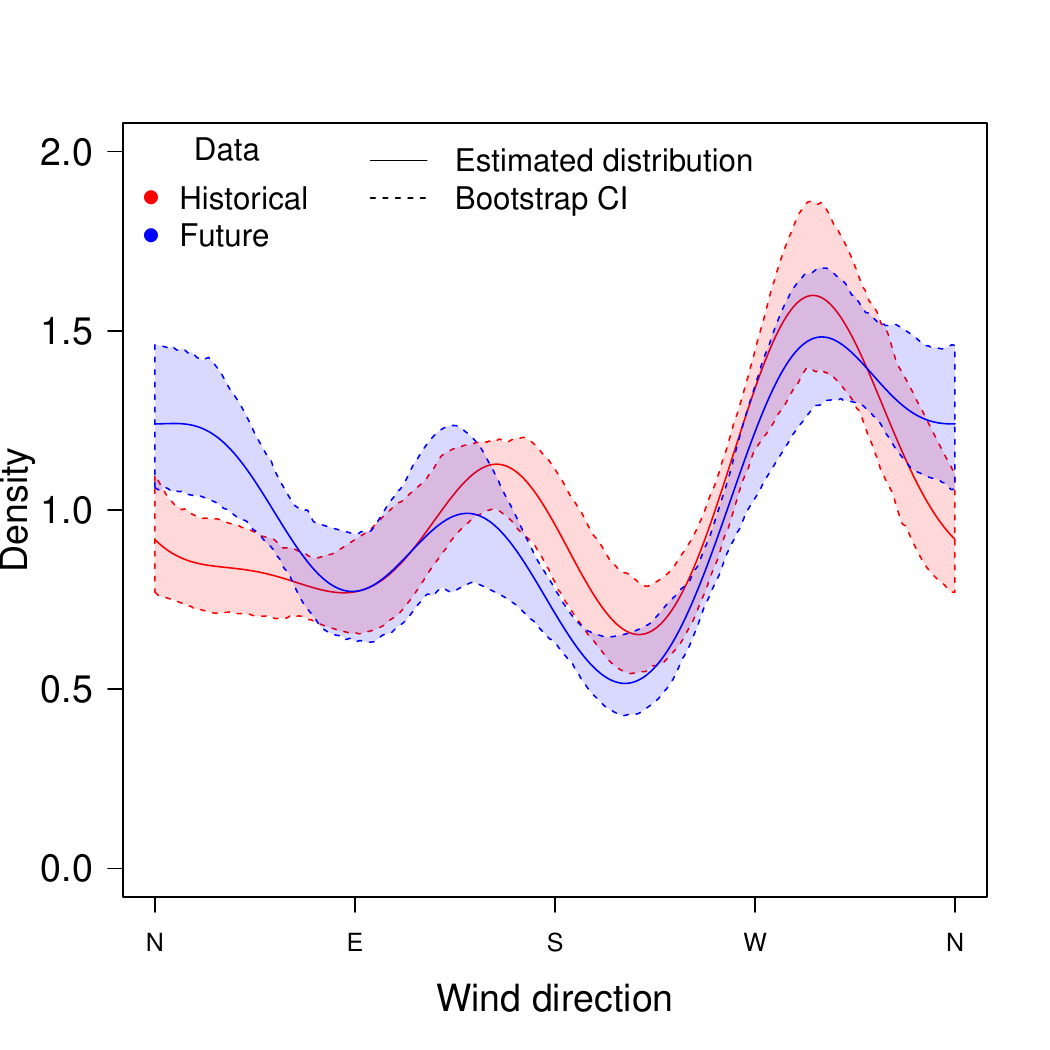}%
    \includegraphics[width = .33\textwidth]{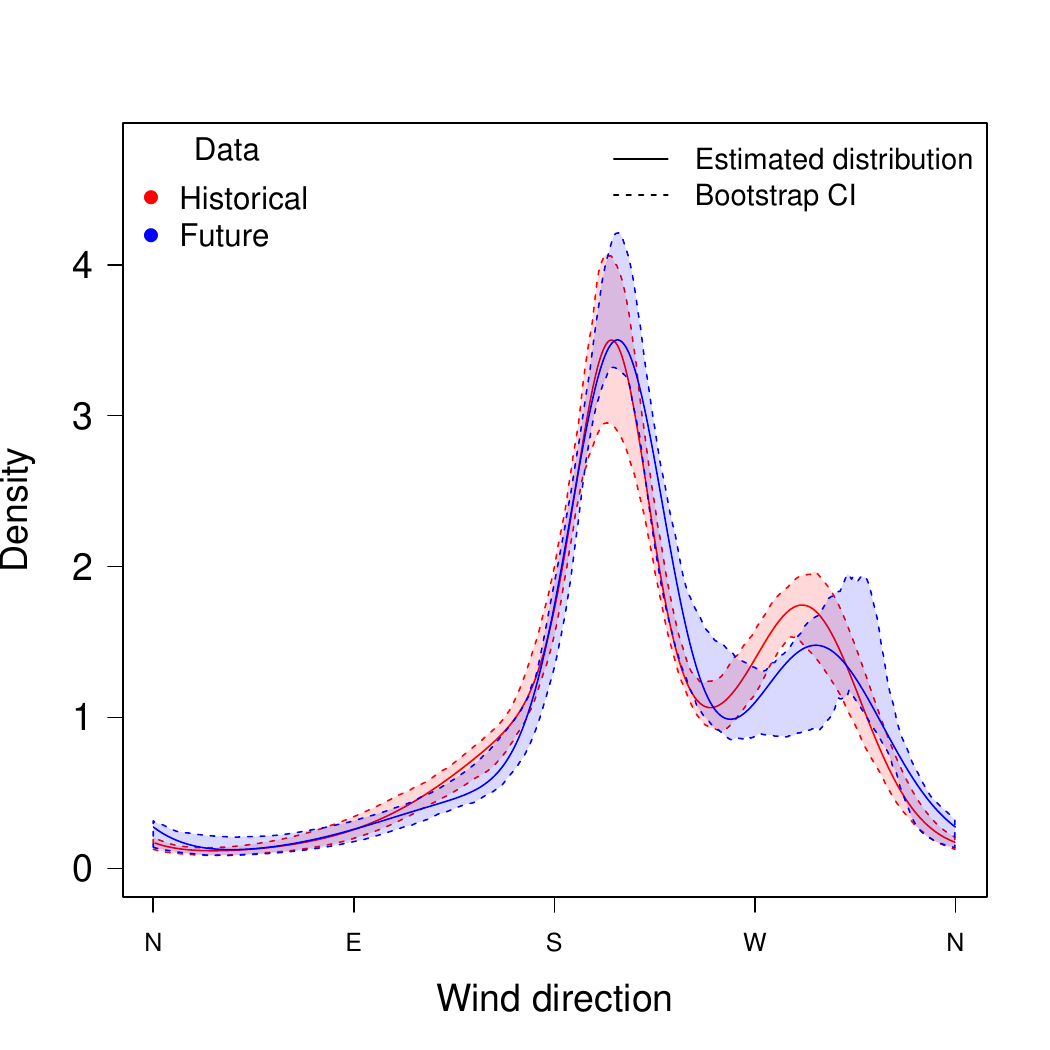}
    \includegraphics[width = .33\textwidth]{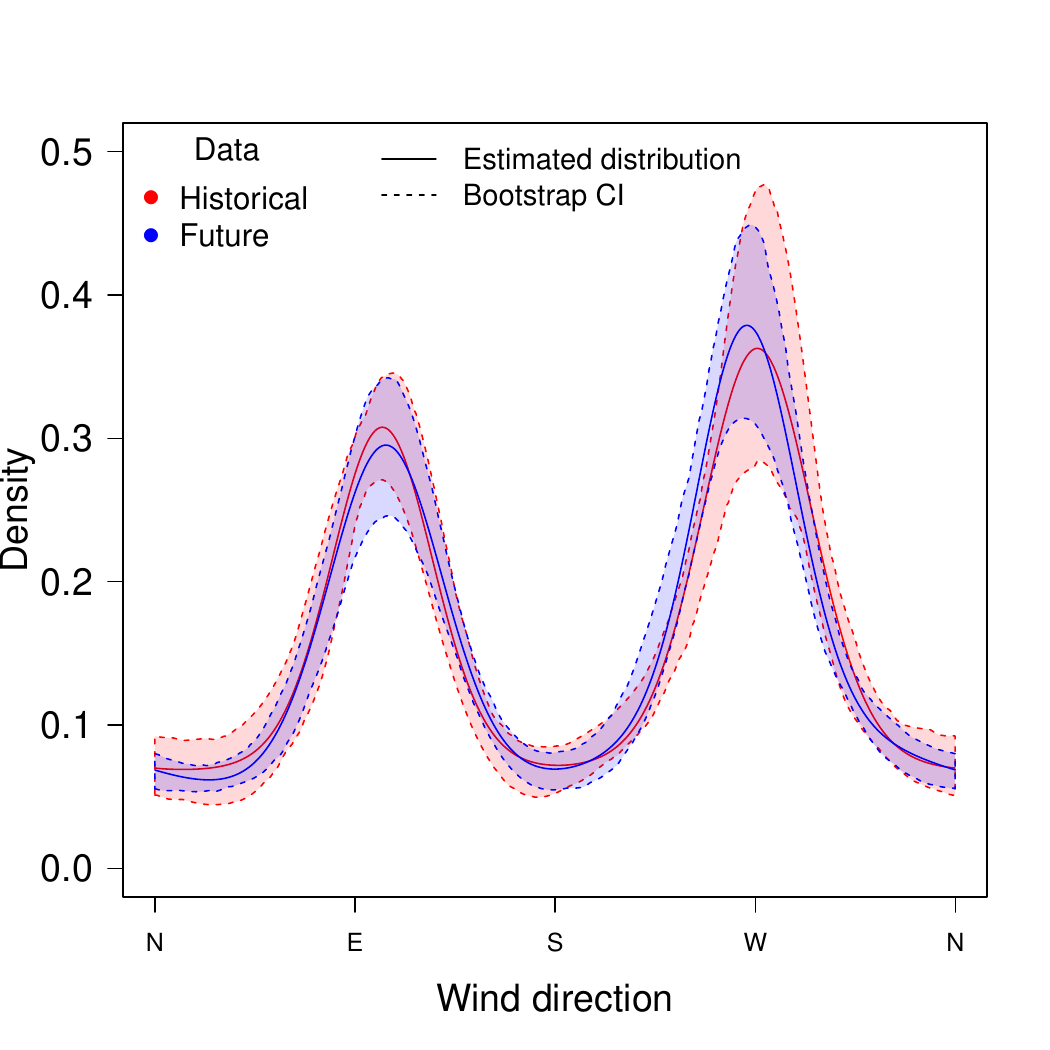}%
    \includegraphics[width = .33\textwidth]{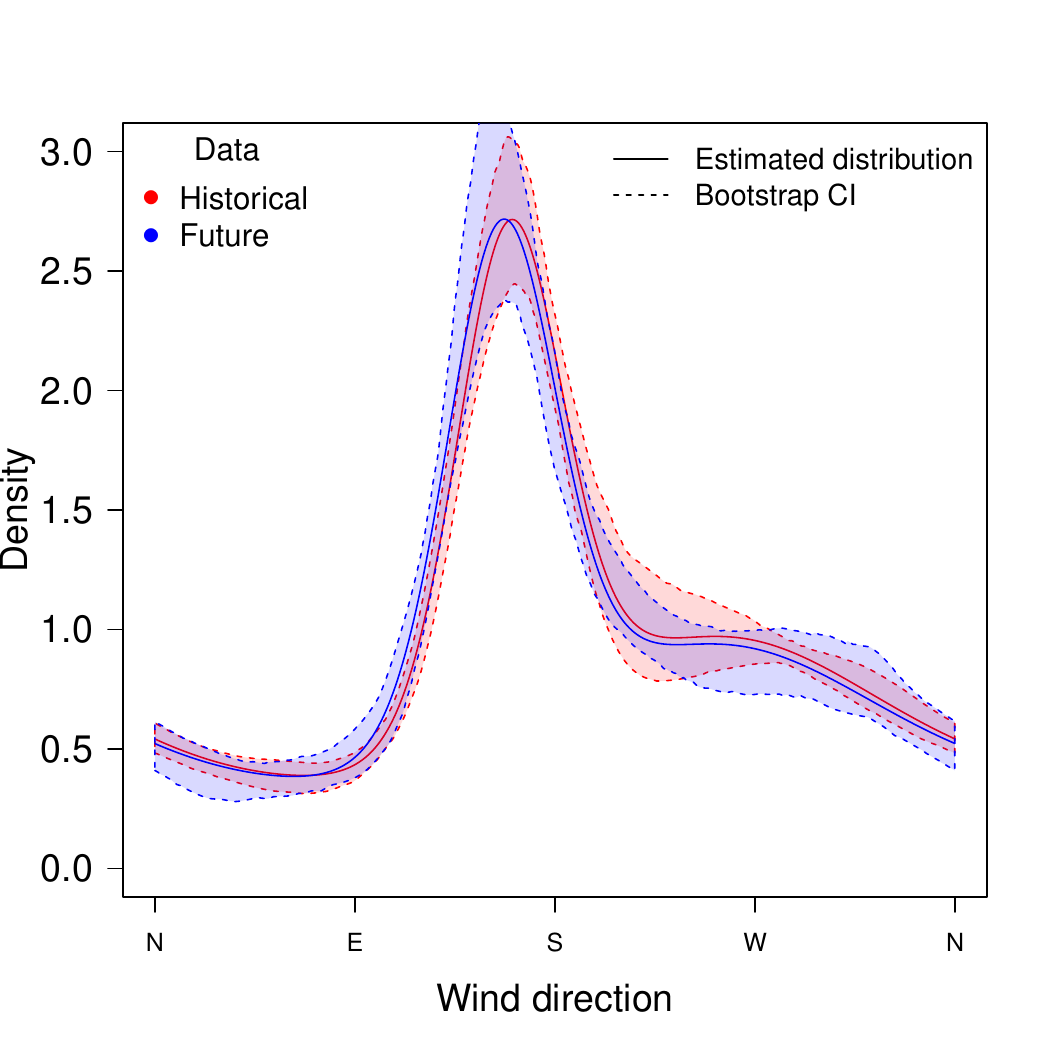}%
    \includegraphics[width = .33\textwidth]{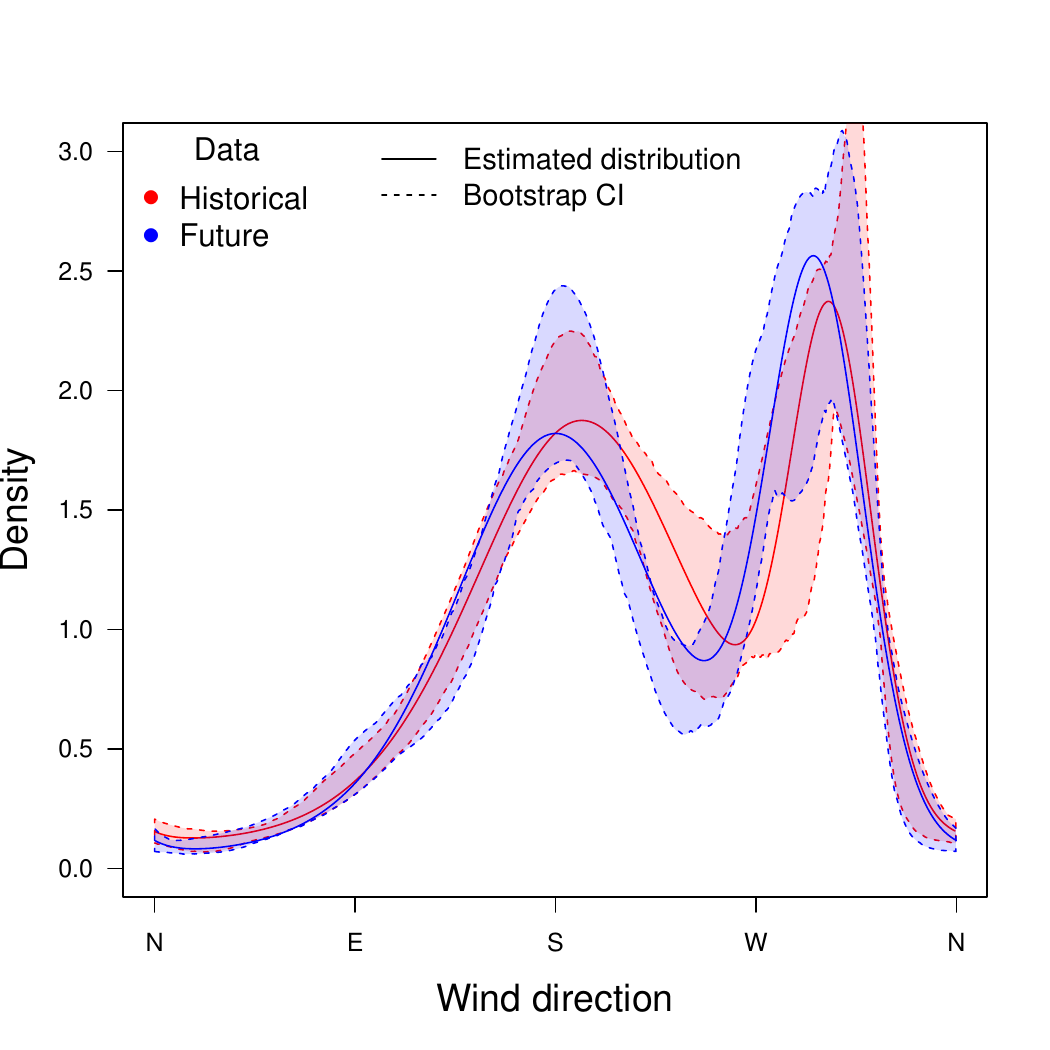}
    \caption{ Bootstrap confidence intervals of historical (red color) and future (blue color) wind direction distributions during summer (\textbf{top}) and winter (\textbf{bottom}) seasons. Each column represents a different location.}
    \label{wdboot}
\end{figure}

To investigate the changes in the wind direction we fit a von Mises mixture distribution to wind direction data at each location, where the number of components was chosen using the BIC. To quantify for the estimation uncertainty in wind direction we use the block boostrap procedure explained in Section \ref{method}. Fig.~\ref{wdboot} illustrates that the estimated wind direction distribution at TX\_GP remains relatively stable throughout both winter and summer seasons, showing minimal or negligible changes. Conversely, at ND\_GP, there are indications of variations in wind direction, particularly towards northerly and southerly directions during the summer season. Similarly, at NC\_mtn, changes in wind direction are observed specifically concerning westerly direction. 

\subsection{Change in the Directional Wind Speed Distribution}

\begin{figure}[H]
    \centering
    \includegraphics[width = 0.33\textwidth]{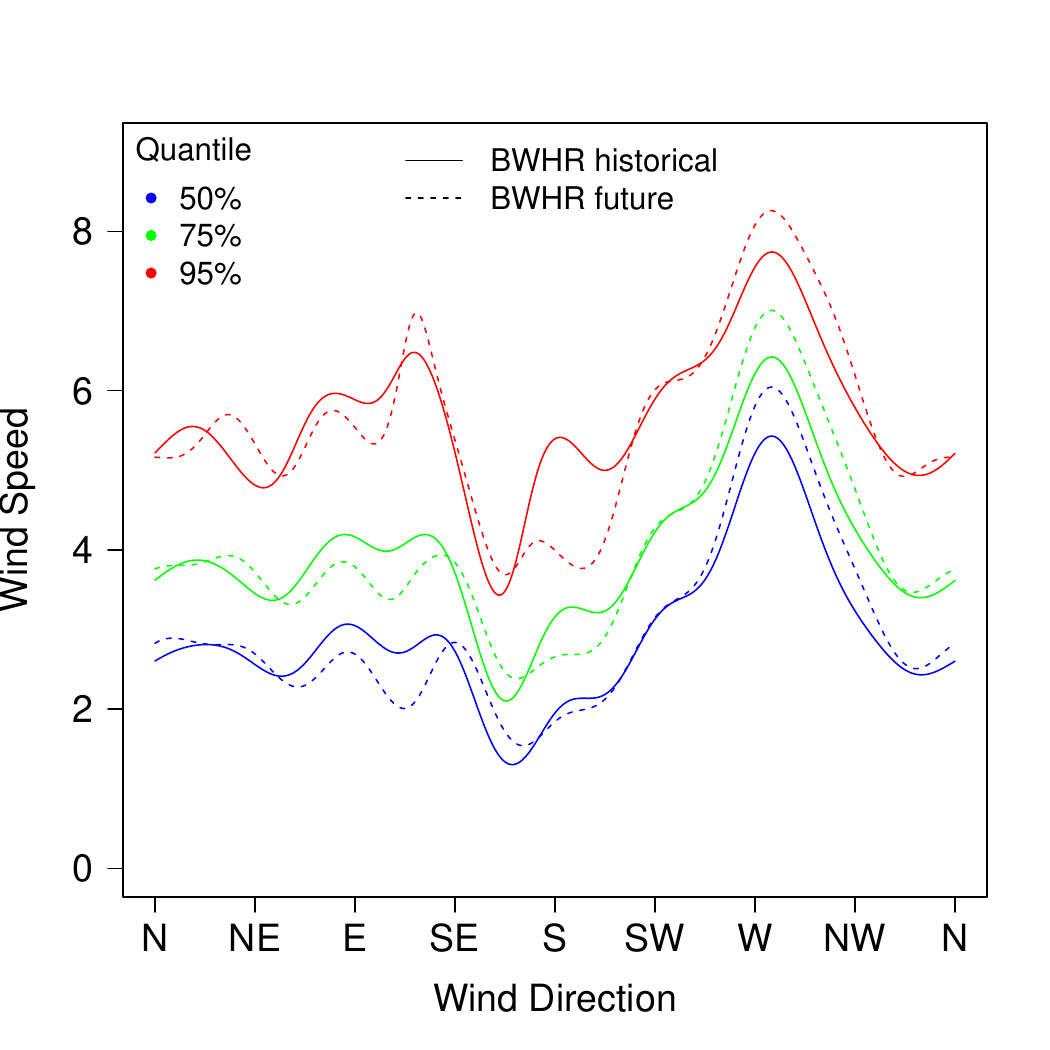}%
    \includegraphics[width = 0.33\textwidth]{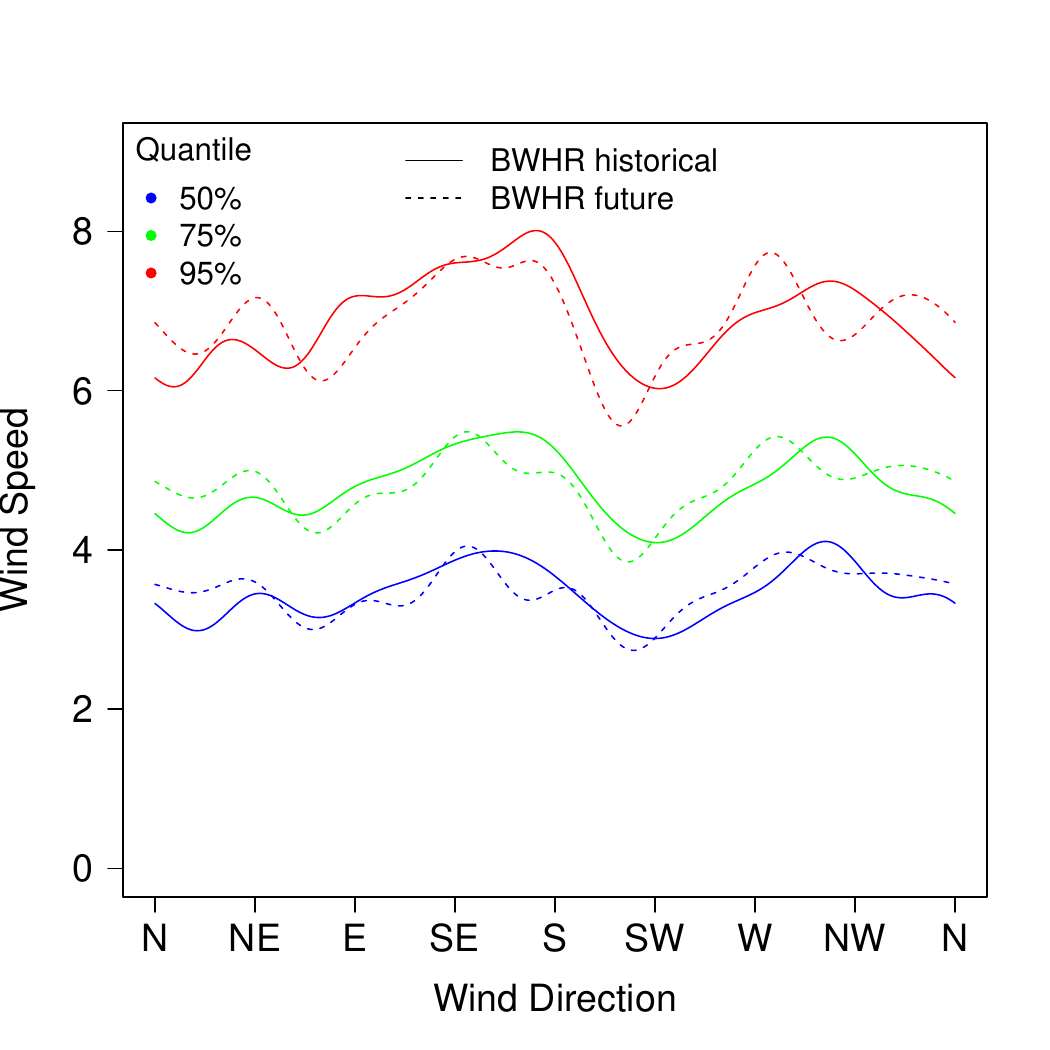}%
    \includegraphics[width = 0.33\textwidth]{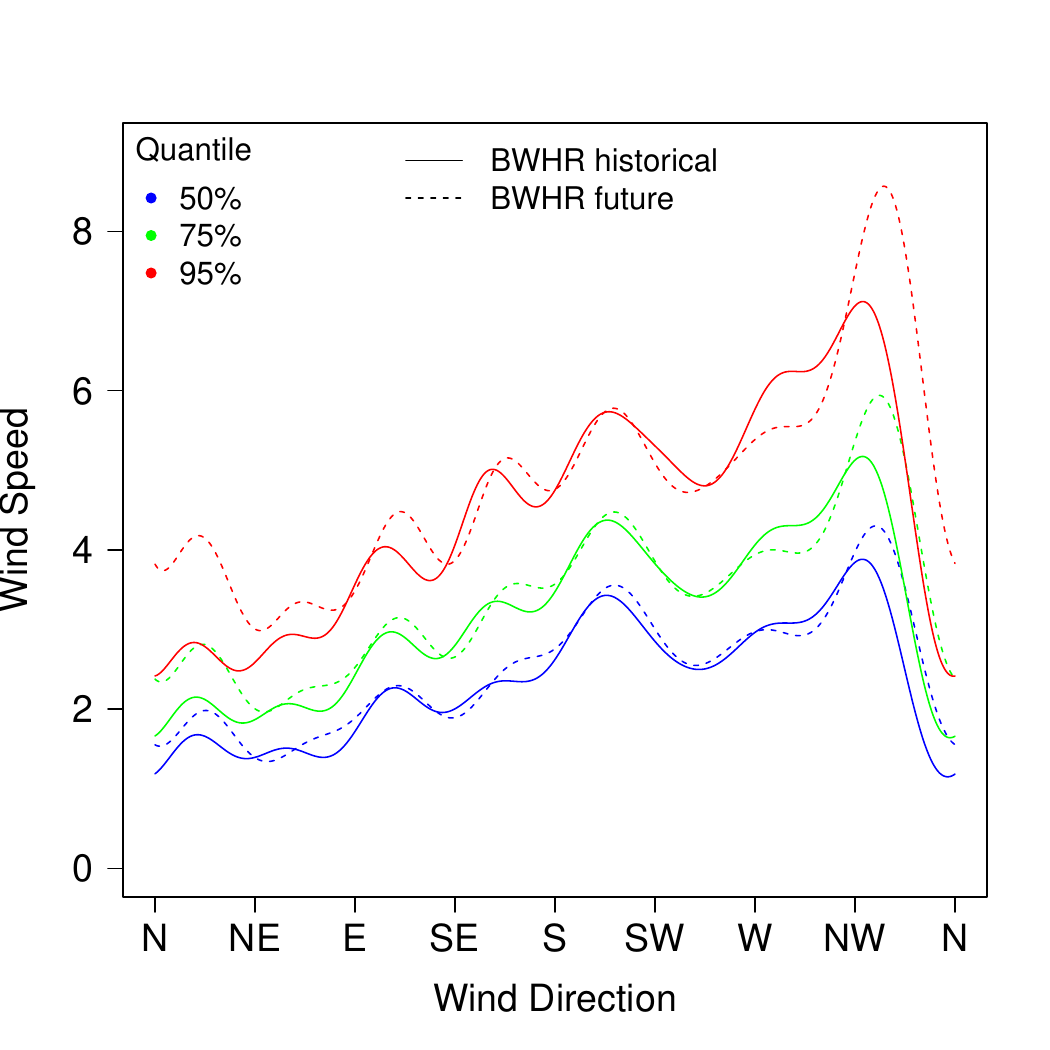}\\
    \includegraphics[width = 0.33\textwidth]{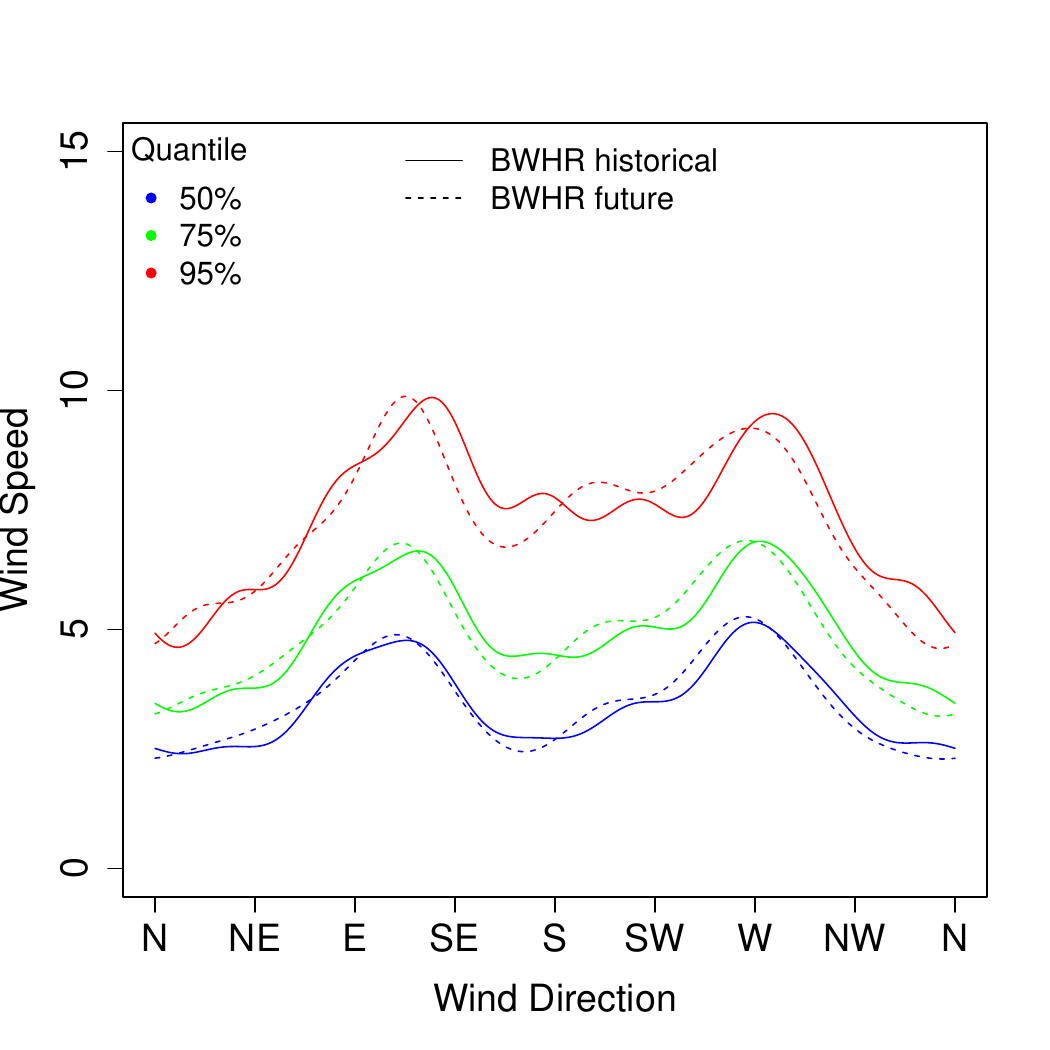}%
    \includegraphics[width = 0.33\textwidth]{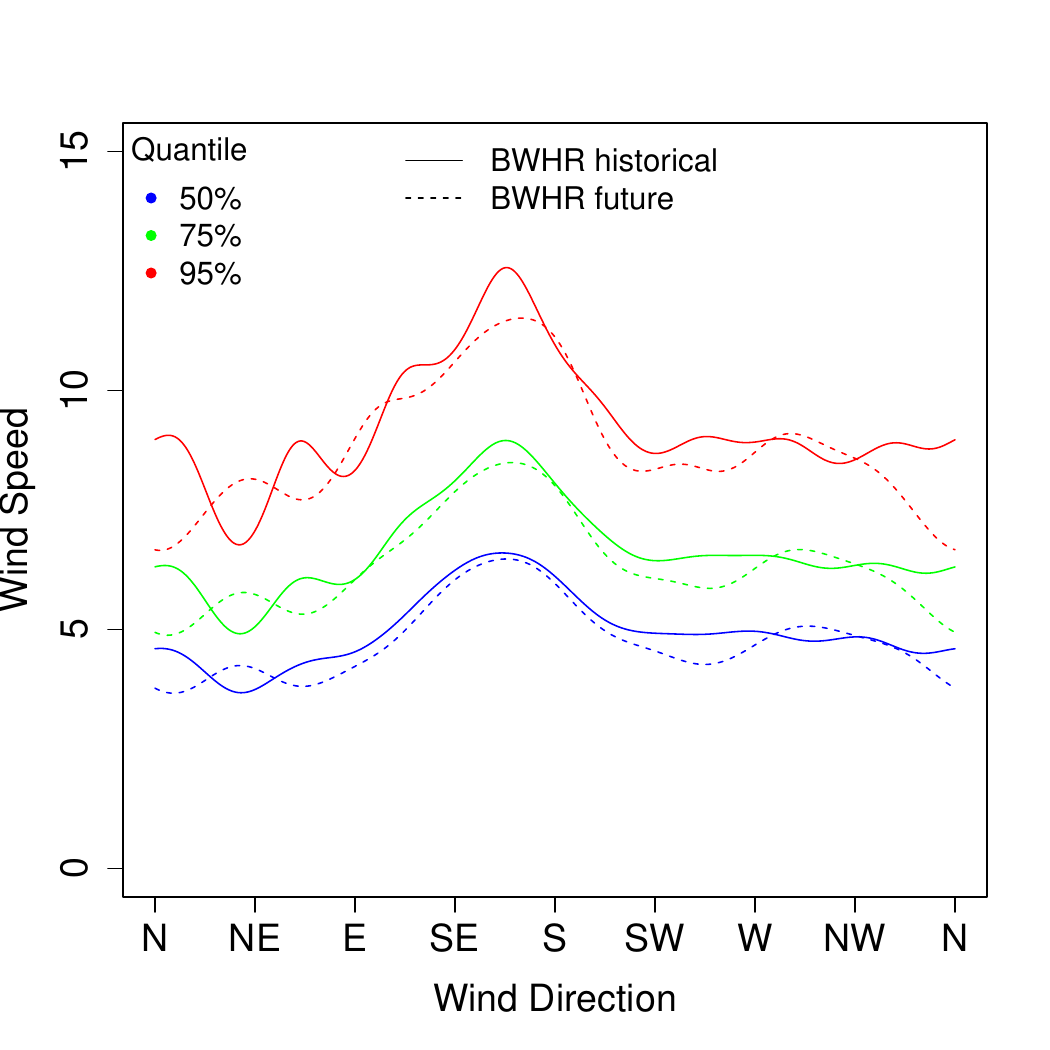}%
    \includegraphics[width = 0.33\textwidth]{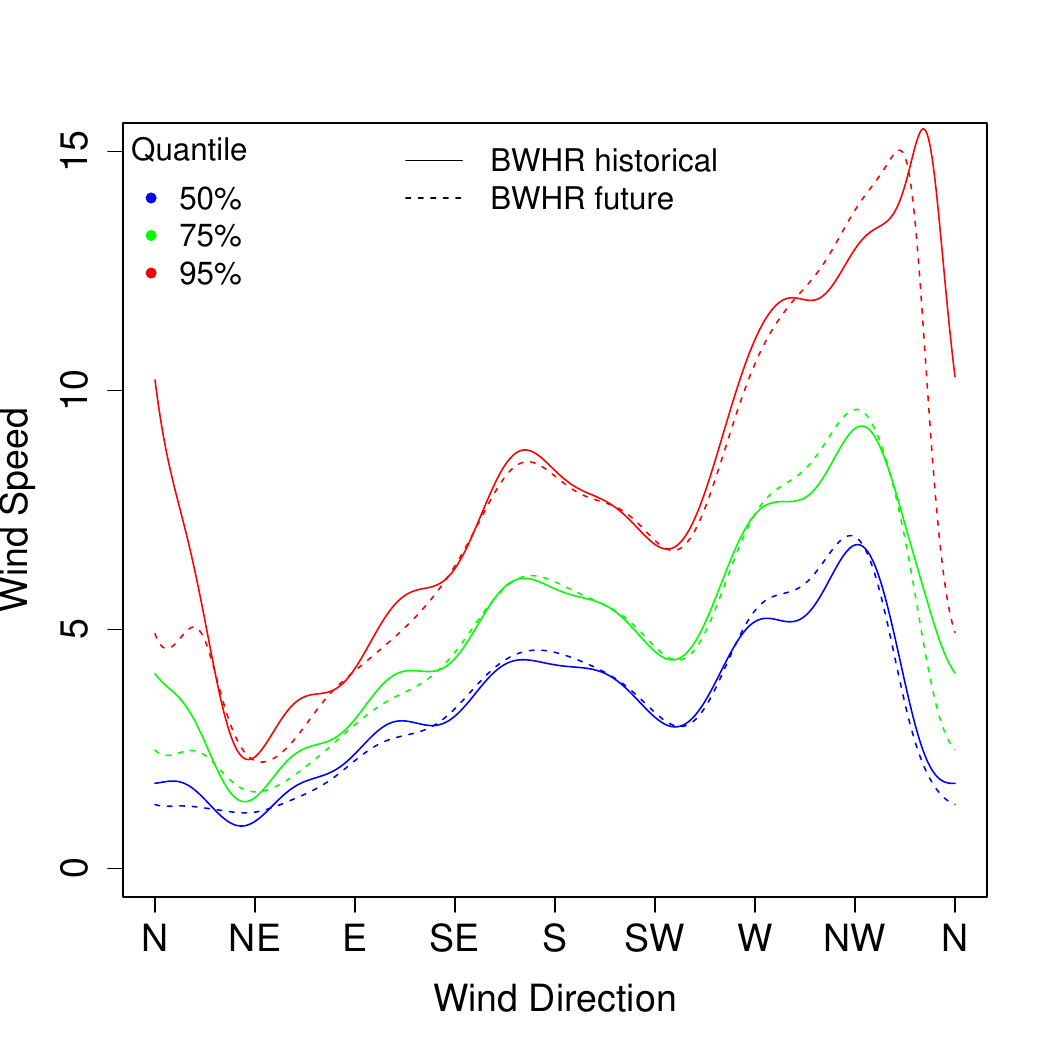}
    \caption{\small{ $50\%$ (blue), $75\%$ (green) and $95\%$ (red) conditional quantiles curves computed using the \texttt{BWHR} method of historical (solid lines) and future (dashed lines) summer (\textbf{top row}) and winter (\textbf{bottom row}) directional wind speed data. Each column represents a different location.}}
    \label{PresvsFuture}
\end{figure}

To study the change in the directional wind speed distribution, we apply the \texttt{BWHR} method to $\{r_{i}, \phi_{i}\}_{i=1}^{n}$ following the procedure described in Section \ref{method}.  Fig. \ref{PresvsFuture} presents the estimates of the $50\%$, $75\%$ and $95\%$ quantile curves at each of the locations. 

\begin{figure}[H]
    \centering
    \includegraphics[width = .33\textwidth, page = 1]{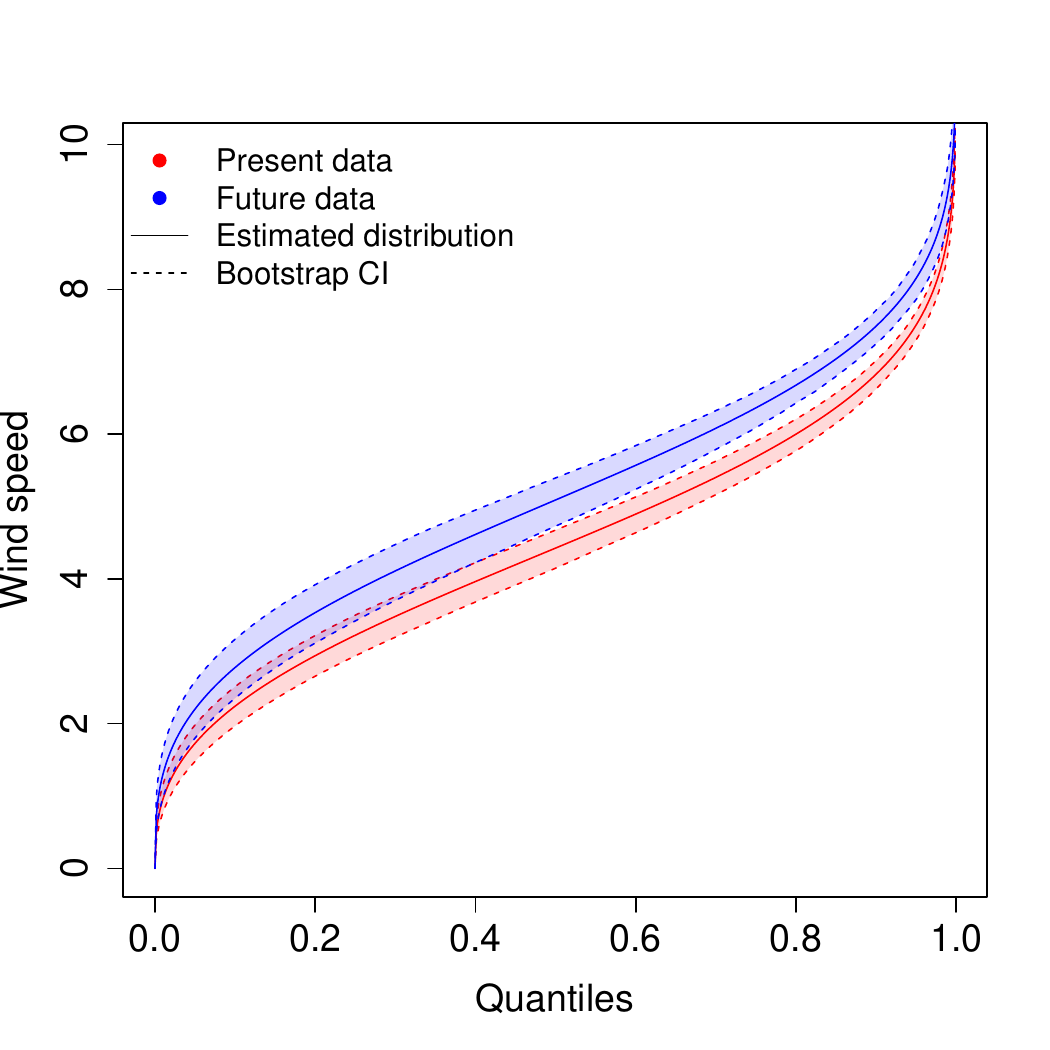}%
    \includegraphics[width = .33\textwidth, page = 2]{Figure/WS_quantile.pdf}%
    \includegraphics[width = .33\textwidth, page = 3]{Figure/WS_quantile.pdf}\\
    \includegraphics[width = 0.33\textwidth]{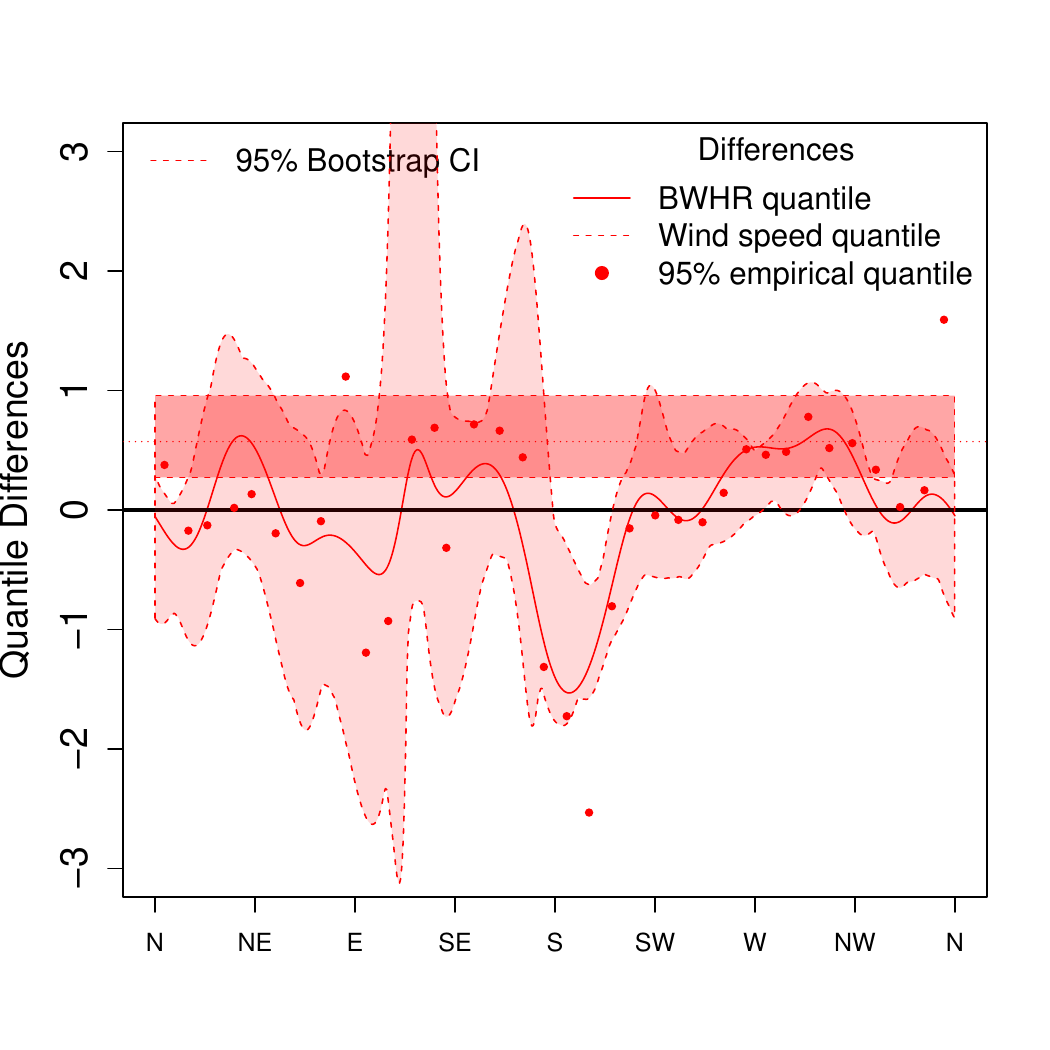}%
    \includegraphics[width = 0.33\textwidth]{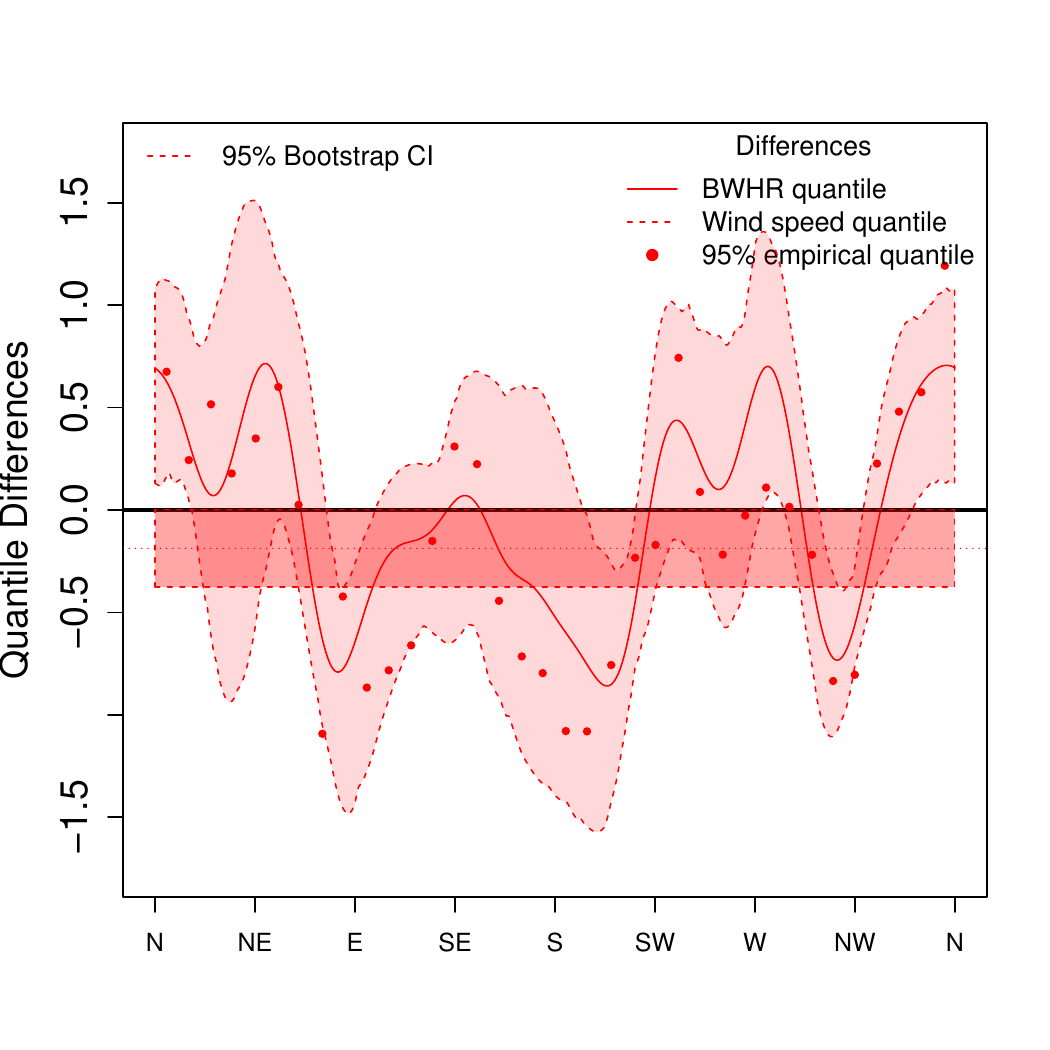}%
    \includegraphics[width = 0.33\textwidth]{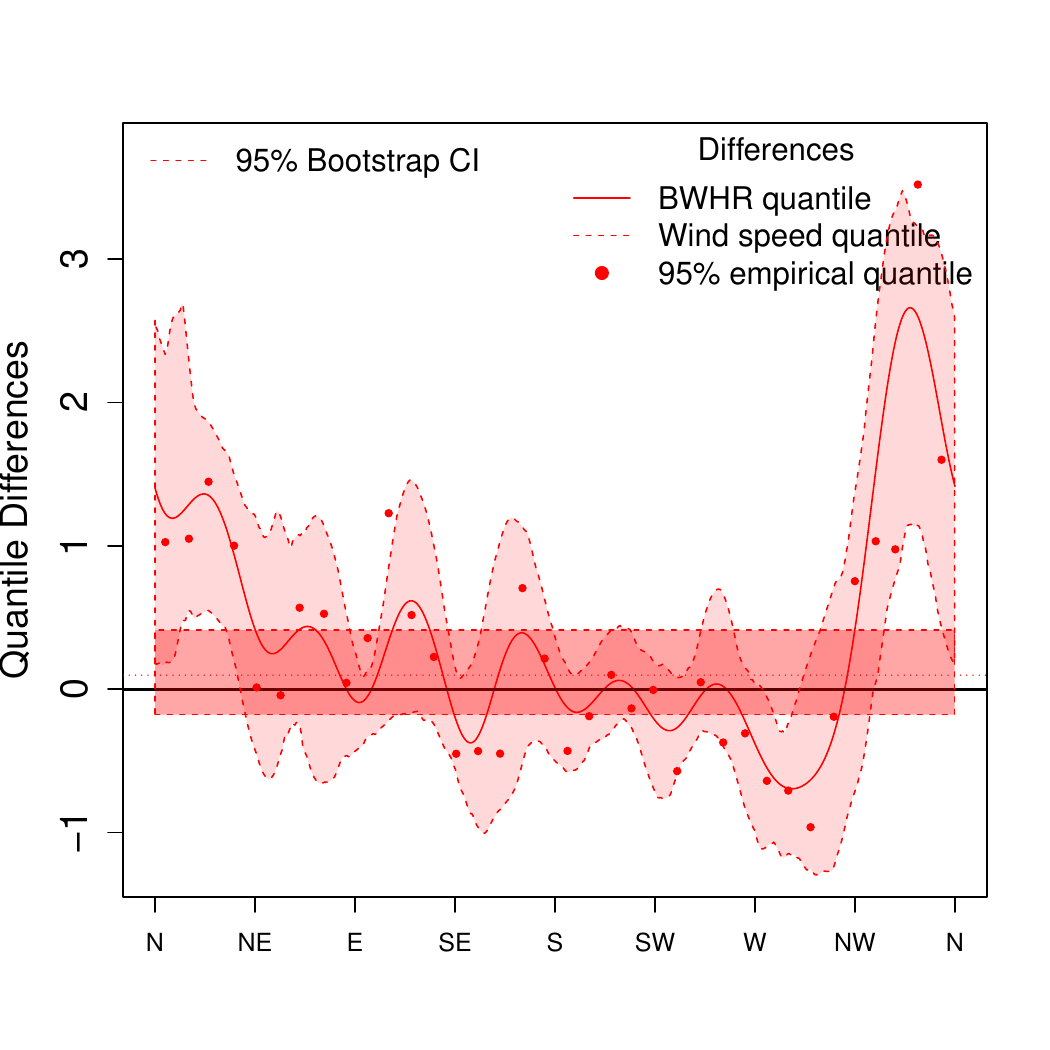}\\
    \includegraphics[width = 0.33\textwidth]{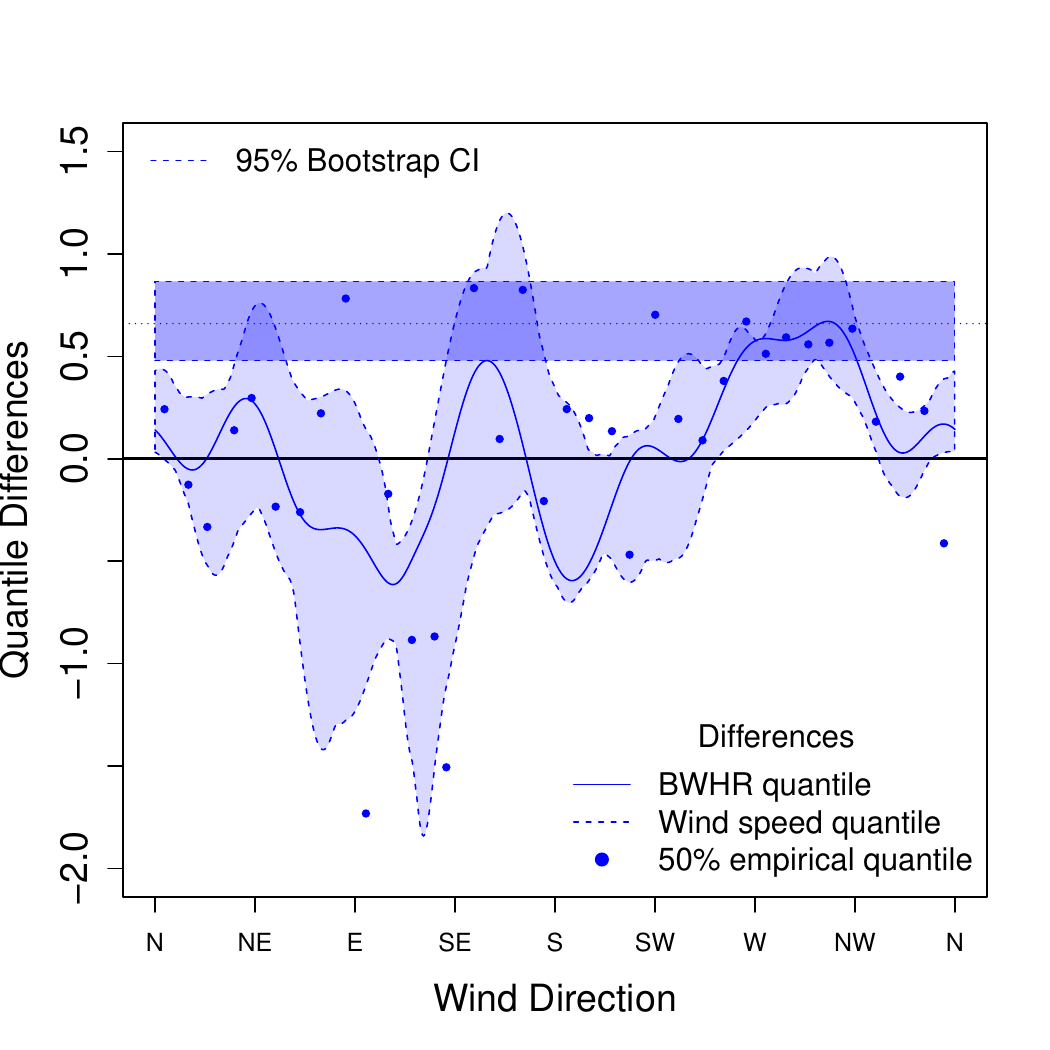}%
    \includegraphics[width = 0.33\textwidth]{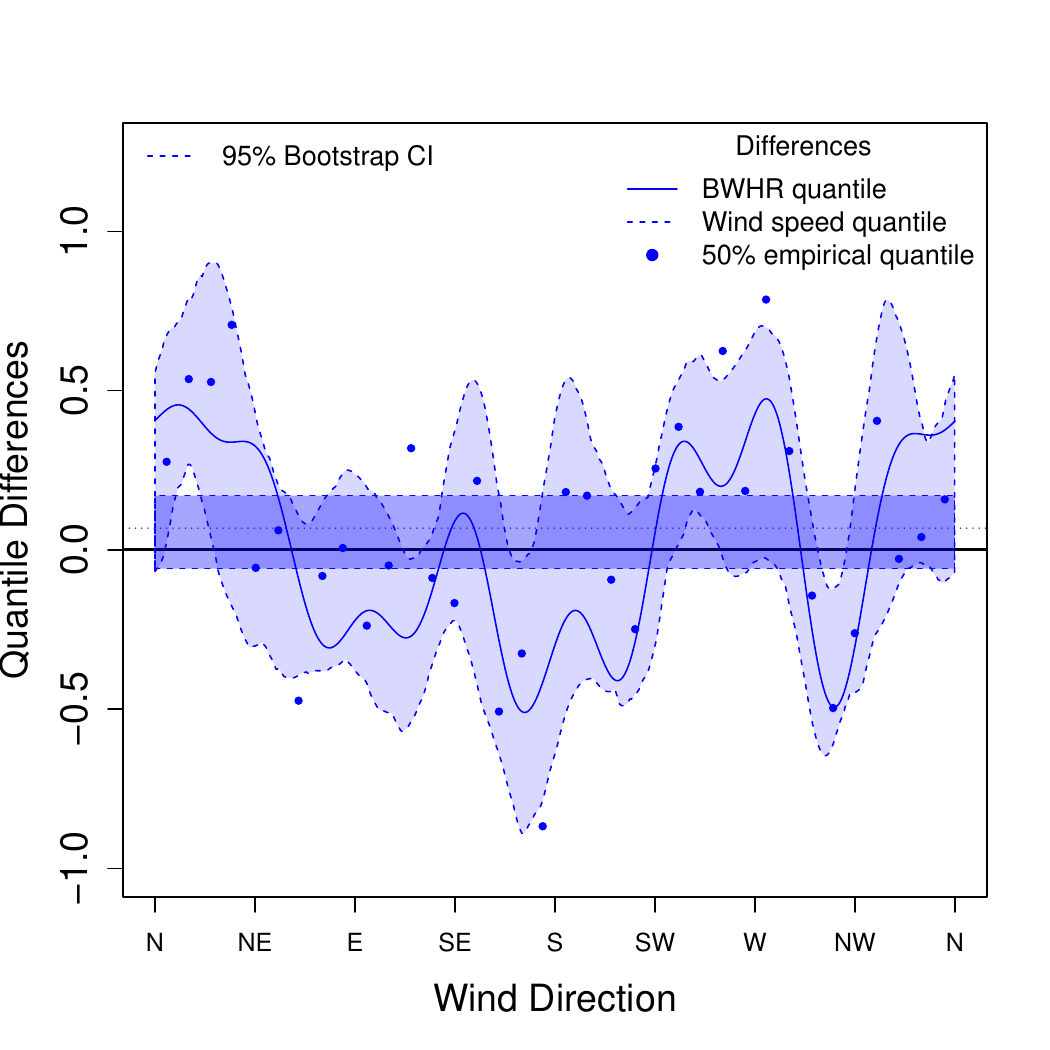}%
    \includegraphics[width = 0.33\textwidth]{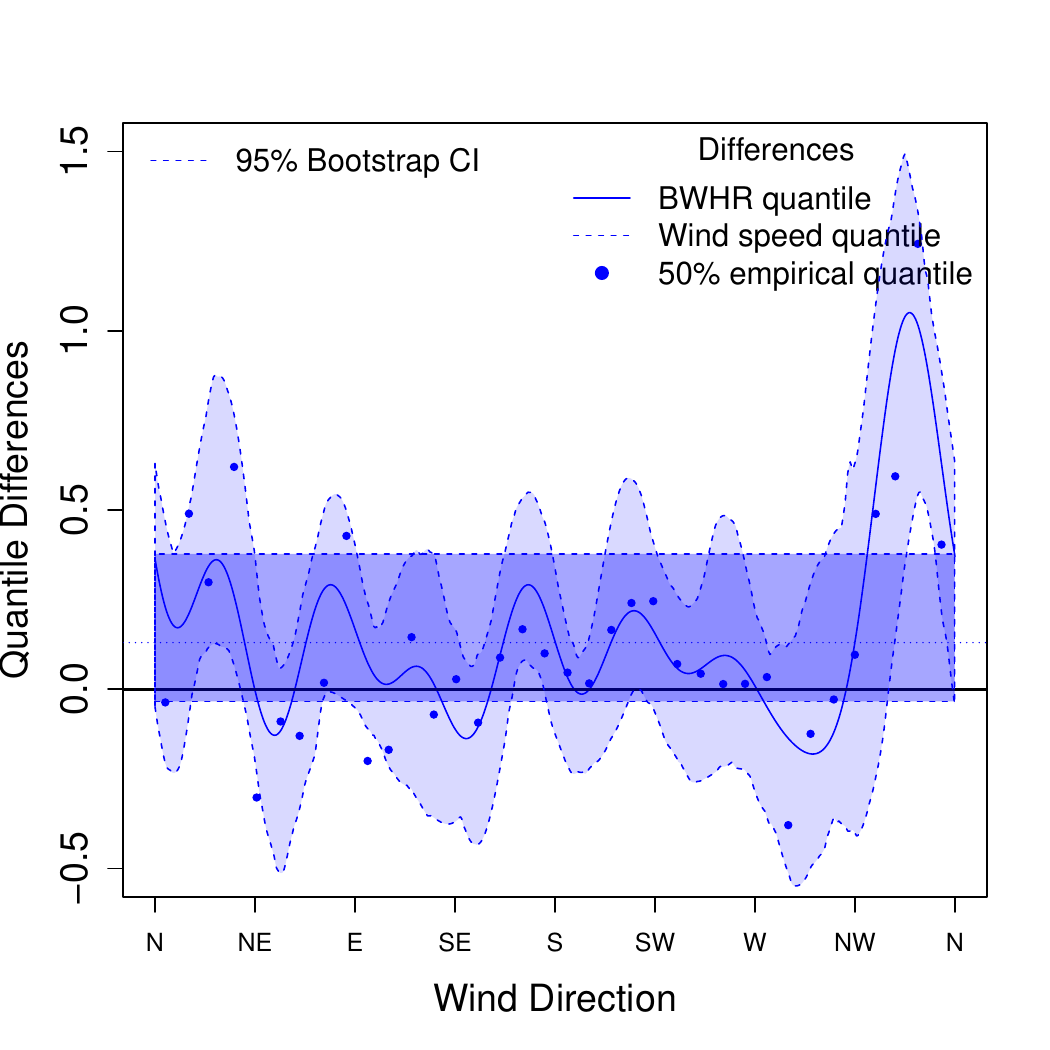}%
    \caption{\small{\textbf{Summer wind data}: Present (\textbf{red}) and future (\textbf{blue}) wind speed quantile curve (\textbf{top row}); Present and future quantile differences of the $95\%$ (\textbf{middle row}) and $50\%$ (\textbf{bottom row}) quantiles with the $95\%$ bootstrap confidence interval (\textbf{shaded polygon}); \textbf{dotted horizontal lines} in middle and bottom row, represent the quantile differences of present and future wind speed data and the \textbf{darker shaded polygons} represent the corresponding $95\%$ bootstrap confidence interval.} }
    \label{diff}
\end{figure}

\begin{figure}[H]
    \centering
    \includegraphics[width = .33\textwidth, page = 1]{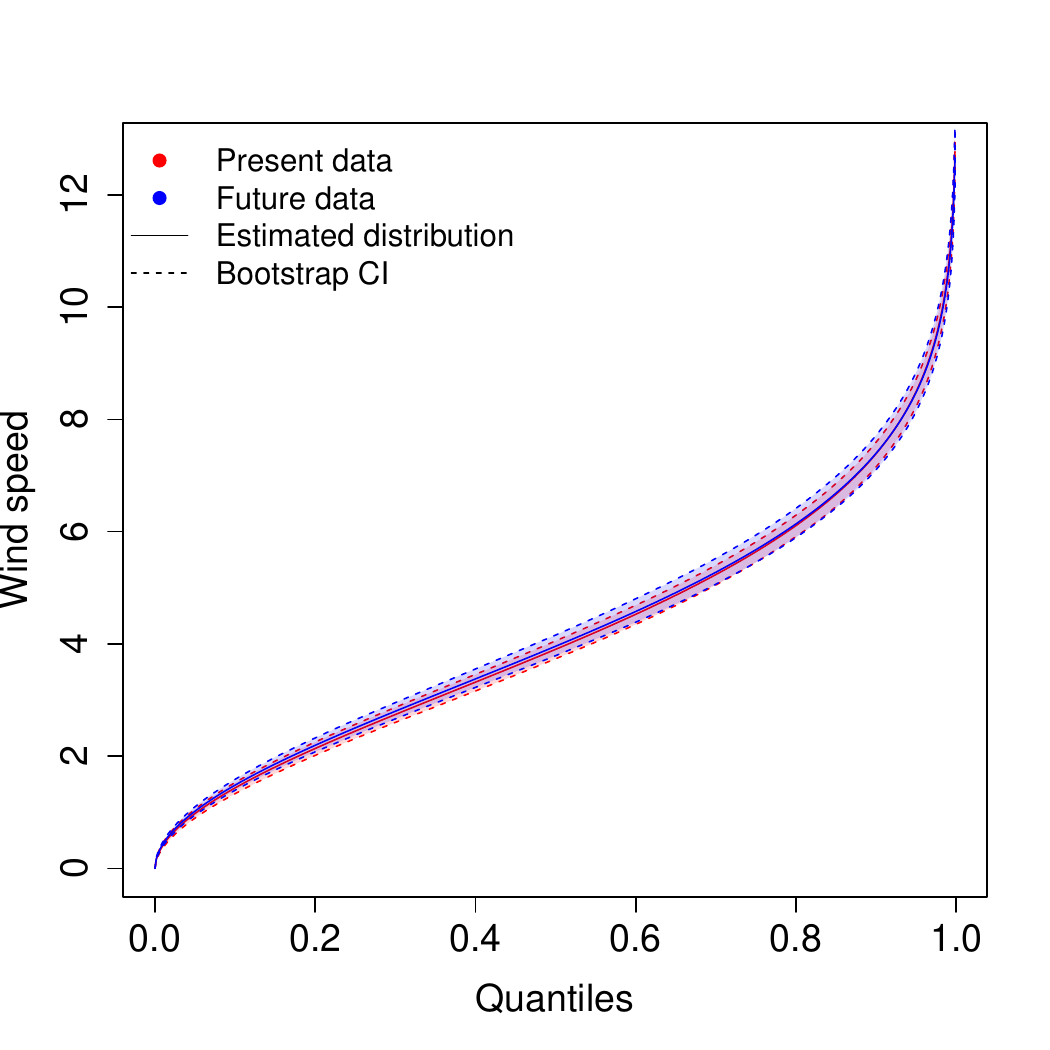}%
    \includegraphics[width = .33\textwidth, page = 2]{Figure/WS_quantile_winter.pdf}%
    \includegraphics[width = .33\textwidth, page = 3]{Figure/WS_quantile_winter.pdf}\\
    \includegraphics[width = 0.33\textwidth, page =1]{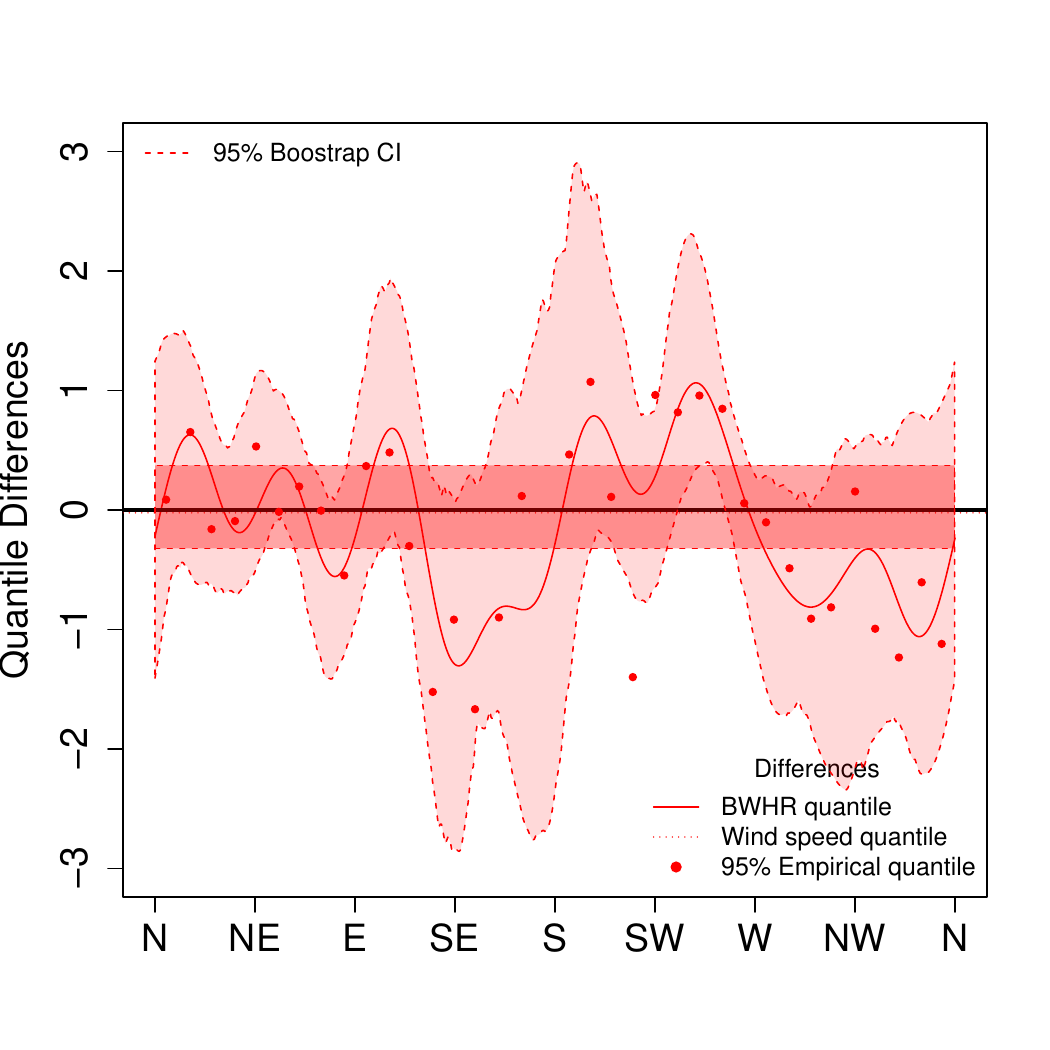}%
    \includegraphics[width = 0.33\textwidth, page = 2]{Figure/Quant_95_diff_win.pdf}%
    \includegraphics[width = 0.33\textwidth, page = 3]{Figure/Quant_95_diff_win.pdf}\\
    \includegraphics[width = 0.33\textwidth, page = 1]{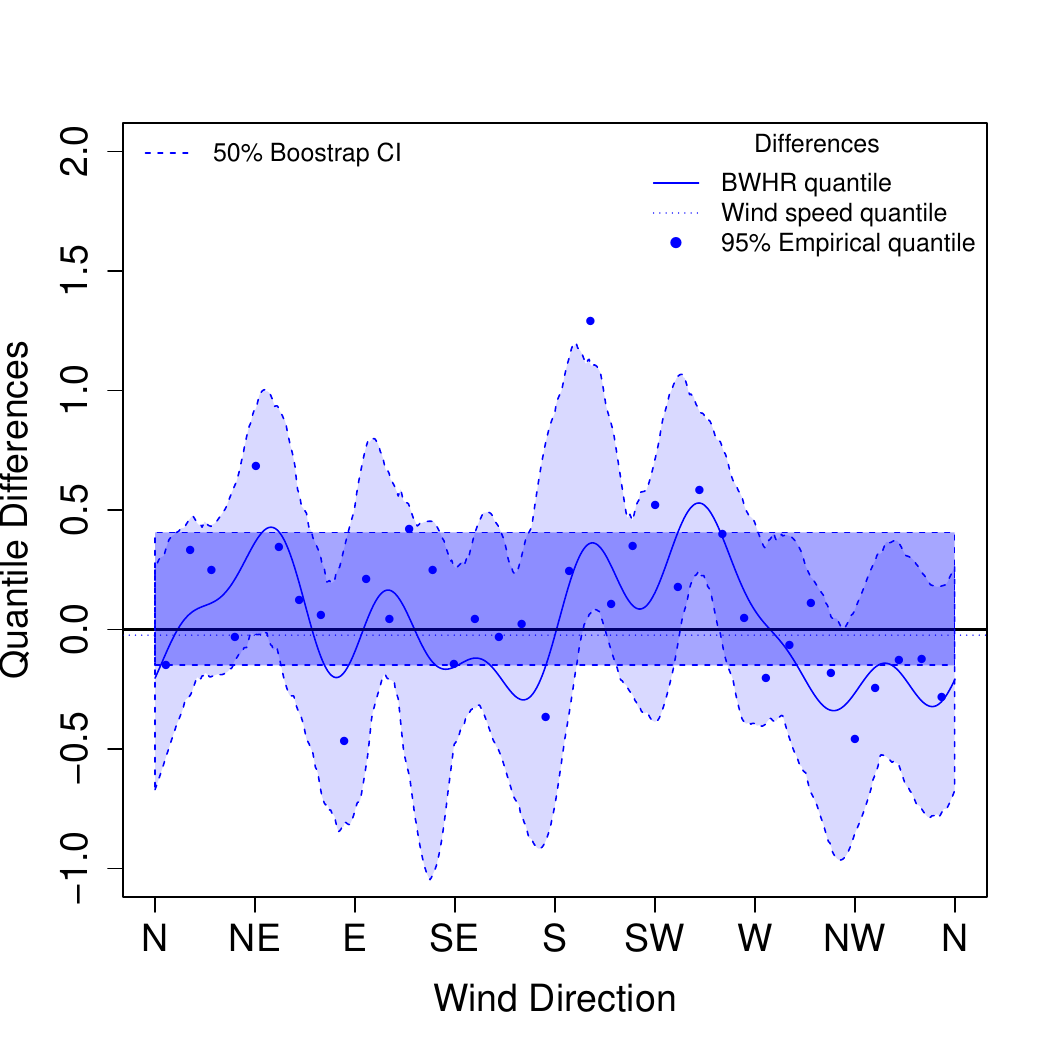}%
    \includegraphics[width = 0.33\textwidth, page = 2]{Figure/Quant_50_diff_win.pdf}%
    \includegraphics[width = 0.33\textwidth, page = 3]{Figure/Quant_50_diff_win.pdf}
    \caption{As in Fig.~\ref{diff} but for winter season. }
    \label{diffwin}
\end{figure}

In order to assess any potential changes between the future and present directional wind speed distributions, we calculate the differences between the estimated future and present directional quantiles. To evaluate the added insights provided by our method in comparison to changes in wind speed distribution, we also calculate the estimated differences for the respective wind speed quantiles averaged across directions. We present the differences in the $95\%$ and $50\%$ of directional quantiles and wind speed quantiles along with their $95\%$ bootstrap confidence interval in Fig. \ref{diff} and Fig.~\ref{diffwin}. For the TX\_GP location, our method reveals potentially greater changes in the $95\%$ directional quantile, particularly towards the South wind direction during summer and the Southeast wind direction during winter. These changes appear more pronounced than the $95\%$ quantile of the wind speed distribution. Similarly, in the case of ND\_GP and NC\_mtn, our method suggests considerably greater changes in the $95\%$ quantiles of the Northerly wind speeds than the $95\%$ quantile of wind speed.

We conclude this section by stating that the \texttt{BWHR} method allows for identifying changes in the wind speed distribution with respect to wind direction. If we had solely analyzed the differences in the quantiles of the wind speed distribution, we would have overlooked these differences.

\section{Summary and Discussion} \label{discussion}

In this work we develop a methodology for estimating the joint distribution of wind speed and wind direction through a conditional framework. The estimation is carried out by decomposing the bivariate distribution into the product of the marginal distribution of wind direction  and the directional wind speed distribution, for which each component can be handle independently. For the marginal modeling of wind direction a von Mises mixture model is used, whereas for the modeling of the directional wind speed distribution we develop the \texttt{BWHR} method to estimate the directional dependent Weibull parameters, which can accommodate fairly flexible conditional distribution while maintaining the circular constraint.

A Monte Carlo simulation study is conducted to compare the model performance of the \texttt{BWHR} method to a periodic B-spline QR, \texttt{BPQR},and a cylindrical distribution, the \texttt{AL} method. The results indicate that, for the three locations considered, \texttt{BWHR} performs better than \texttt{BPQR} and\texttt{AL} method in terms of the directional density weighted integrated mean relative error (\texttt{MIRE}). We illustrate this framework by applying the \texttt{BWHR} to estimate changes in the joint distribution of wind vector from present to future climate scenarios using the output from a regional climate model. Relative to the locations under consideration, our proposed methodology enables the detection of potential changes in wind speed given wind direction, in particular in the higher quantiles. Moreover, by taking into account the wind direction, our approach enables us to capture changes in wind speed that may have been overlooked if the wind direction was not taken into account.
The proposed \texttt{BWHR} method is constructed using equally spaced bins. Nevertheless, one can also choose to bin wind direction that each bin includes the same amount of data points. We observe, however, that in this case, the model fails to appropriately capture the conditional quantile curves of the \texttt{BWHR} method when the wind direction data is sparse. This issue relates back to the conclusions made in Section~\ref{sec:BWHR} regarding the tuning parameter selection, that is, the bins are too large and estimating the wind speed distribution by Weibull distribution might be inappropriate. Furthermore, the \texttt{BWHR} method uses Fourier series to model the dependence structure between the Weibull parameters and wind direction. This dependence can also be modeled using periodic B-splines. In this case the degrees of freedom have to be determined. We find that this change gives a higher MIRE value compared to the method using Fourier series (see supplementary materials).

The method proposed in this study offers a procedure for modeling wind speed and direction measured at one location during one season. However, wind speed and wind direction exhibit a coherent spatial and temporal structure. Therefore, our intention is to extend this model to a spatio-temporal framework, considering the dependence on both space and time. To achieve this, we can incorporate spatio-temporal methods when modeling the parameters of the Weibull distribution, as suggested in previous studies \citep{stWeibull1, stWeibull2}. In the case of wind direction, it is necessary to account for both space and time in addition to allowing for spatial and temporal variation in the marginal distribution. This can be achieved by building upon the methodologies discussed in \cite{linCircST, STcircGelfand}, and \cite{stVonMises}. Considering the similarity of our work to a hierarchical model, a Bayesian hierarchical modeling approach, such as that proposed by \cite{mastrantonio2015}, could be used to incorporate the temporal aspect. However, this approach presents significant estimation challenges. It would likely increase computational complexity and inference would become more complicated, as it requires estimating the parameters of the circular and linear components and integrating these into the already complex Bayesian estimation procedures.

\textbf{Supplementary information}\\
Additional supplementary information can be found  \href{https://github.com/evamurphy100/Joint-modeling-of-wind-speed-and-wind-direction/blob/main/Supplementary%20Material/Supplementary_Material.pdf}{here}. 
\newpage
\singlespacing
\bibliography{manuscript_clean.bib}

\begin{thebibliography}{80}
\providecommand{\natexlab}[1]{#1}
\providecommand{\url}[1]{\texttt{#1}}
\expandafter\ifx\csname urlstyle\endcsname\relax
  \providecommand{\doi}[1]{doi: #1}\else
  \providecommand{\doi}{doi: \begingroup \urlstyle{rm}\Url}\fi

\bibitem[Abatzoglou et~al.(2020)Abatzoglou, Hatchett, Fox-Hughes, Gershunov, and Nauslar]{fire}
J.~T. Abatzoglou, B.~J. Hatchett, P.~Fox-Hughes, A.~Gershunov, and N.~J. Nauslar.
\newblock Global climatology of synoptically-forced downslope winds.
\newblock \emph{International Journal of Climatology}, 2020.

\bibitem[Abe and Ley(2017)]{abe2017tractable}
T.~Abe and C.~Ley.
\newblock A tractable, parsimonious and flexible model for cylindrical data, with applications.
\newblock \emph{Econometrics and Statistics}, 4:\penalty0 91--104, 2017.

\bibitem[Ailliot et~al.(2015)Ailliot, Bessac, Monbet, and Pene]{regimBess}
P.~Ailliot, J.~Bessac, V.~Monbet, and F.~Pene.
\newblock Non-homogeneous hidden {Markov-switching} models for wind time series.
\newblock \emph{Journal of Statistical Planning and Inference}, pages 75--88, 2015.
\newblock \doi{10.1016/j.jspi.2014.12.005}.

\bibitem[Andrade et~al.(2023)Andrade, Pereira, and Artes]{circquant}
A.~C. Andrade, G.~H. Pereira, and R.~Artes.
\newblock The circular quantile residual.
\newblock \emph{Computational Statistics \& Data Analysis}, 178:\penalty0 107612, 2023.
\newblock ISSN 0167-9473.
\newblock \doi{https://doi.org/10.1016/j.csda.2022.107612}.

\bibitem[Banerjee et~al.(2005)Banerjee, Dhillon, Ghosh, and Sra]{banerjee}
A.~Banerjee, I.~Dhillon, J.~Ghosh, and S.~Sra.
\newblock Clustering on the unit hypersphere using von {Mises-Fisher} distributions.
\newblock \emph{Journal of Machine Learning Research}, 6, 2005.

\bibitem[Bessac et~al.(2016)Bessac, Ailliot, Cattiaux, and Monbet]{uvhidden}
J.~Bessac, P.~Ailliot, J.~Cattiaux, and V.~Monbet.
\newblock Comparison of hidden and observed regime-switching autoregressive models for ($u,v$)-components of wind fields in the northeastern atlantic.
\newblock \emph{Advances in Statistical Climatology, Meteorology and Oceanography}, 2\penalty0 (1):\penalty0 1--16, 2016.
\newblock \doi{10.5194/ascmo-2-1-2016}.
\newblock URL \url{https://ascmo.copernicus.org/articles/2/1/2016/}.

\bibitem[Breckling(1989)]{berck}
J.~Breckling.
\newblock \emph{The analysis of directional time series: applications to wind speed and direction}, volume~61.
\newblock Springer Science \& Business Media, 1989.

\bibitem[Brown and R.W.~Katz(1984)]{weibull}
B.~Brown and A.~M. R.W.~Katz.
\newblock Time series models to simulate and forecast wind speed and wind power.
\newblock \emph{Journal of Applied Meteorology and Climatology}, 23, 1984.

\bibitem[Carta et~al.(2009)Carta, Ram\'{i}rez, and Vel\'{a}zquez]{wsreview}
J.~Carta, P.~Ram\'{i}rez, and S.~Vel\'{a}zquez.
\newblock A review of wind speed distributions used in wind energy analysis: {Case} studies in the {Canary Islands}.
\newblock \emph{Renewable and Sustainable Energy Reviews}, 13, 2009.

\bibitem[Carta et~al.(2008)Carta, Ram\'{i}rez, and Bueno]{windenergy}
J.~A. Carta, P.~Ram\'{i}rez, and C.~Bueno.
\newblock A joint probability density function of wind speed and direction for wind energy analysis.
\newblock \emph{Energy Conversion and Management}, 49, 2008.

\bibitem[Chen(2006)]{building3}
Q.~Chen.
\newblock \emph{Chapter 6: Wind in building environment design, Sustainable Urban Housing in China}.
\newblock Springer, 2006.

\bibitem[Coles and Walshaw(1994)]{colsaw}
S.~G. Coles and D.~Walshaw.
\newblock Directional modeling of extreme wind speeds.
\newblock \emph{Journal of the Royal Statistical Society. Series C (Applied Statistics)}, 43, 1994.

\bibitem[Dempster et~al.(1977)Dempster, Laird, and D.B.Rubin]{em}
A.~Dempster, N.~Laird, and D.B.Rubin.
\newblock Maximum likelihood for incomplete data via the em algorithm.
\newblock \emph{Journal of the Royal Statistical Society. Series B (Methodological)}, 39, 1977.

\bibitem[Ding(2020)]{Ding}
Y.~Ding.
\newblock \emph{Data Science for Wind Energy}.
\newblock Taylor \& Francis Group, LLC, 2020.

\bibitem[Fahrmeir et~al.(2021)Fahrmeir, Kneib, Lang, and Marx]{distrreg}
L.~Fahrmeir, T.~Kneib, S.~Lang, and B.~D. Marx.
\newblock \emph{Distributional regression models}.
\newblock Springer Berlin Heidelberg, 2021.

\bibitem[Farlie(1960)]{copula2}
D.~J.~G. Farlie.
\newblock The performance of some correlation coefficients for a general bivariate distribution.
\newblock \emph{Biometrika}, 47, 1960.

\bibitem[Frumento and Bottai(2016)]{Frumento2016}
P.~Frumento and M.~Bottai.
\newblock Parametric modeling of quantile regression coefficient functions.
\newblock \emph{Biometrics}, 72\penalty0 (1):\penalty0 74--84, March 2016.
\newblock \doi{10.1111/biom.12410}.
\newblock URL \url{https://doi.org/10.1111/biom.12410}.

\bibitem[Gent et~al.(2011)Gent, Danabasoglu, Donner, Holland, Hunke, Jayne, Lawrence, Neale, Rasch, Vertenstein, and et.al]{data}
R.~Gent, G.~Danabasoglu, L.~Donner, M.~Holland, E.~Hunke, S.~Jayne, D.~Lawrence, R.~Neale, P.~Rasch, M.~Vertenstein, and et.al.
\newblock The community climate system model version 4.
\newblock \emph{Journal of climate}, 24(19): 4973-4991, 2011.

\bibitem[Greco et~al.(2021)Greco, Saraceno, and Agostinelli]{Greco2021}
L.~Greco, G.~Saraceno, and C.~Agostinelli.
\newblock Robust fitting of a wrapped normal model to multivariate circular data and outlier detection.
\newblock \emph{Stats}, 4\penalty0 (2):\penalty0 454--471, 2021.
\newblock \doi{10.3390/stats4020028}.
\newblock URL \url{https://doi.org/10.3390/stats4020028}.

\bibitem[Hering and Genton(2010)]{stwf}
A.~S. Hering and M.~G. Genton.
\newblock Powering up with space-time wind forecasting.
\newblock \emph{Journal of the American Statistical Association}, 105, 2010.

\bibitem[Hill(1977)]{bessel}
G.~Hill.
\newblock Algorithm 518: Incomplete bessel function {$I_0$}. {The} {von Misses} distribution {[S14]}.
\newblock \emph{{ACM} Transition of Mathematical Software}, 3, 1977.

\bibitem[Holmes and Bekele(2013)]{building1}
J.~D. Holmes and S.~Bekele.
\newblock \emph{Wind Loading of Structures}.
\newblock Springer Science \& Business Media, 2013.

\bibitem[Hornik and Gr\"un(2014)]{movFMR}
K.~Hornik and B.~Gr\"un.
\newblock {movMF}: An {R} package for fitting mixtures of von {Mises-Fisher} distributions.
\newblock \emph{Journal of Statistical Software}, 58\penalty0 (10):\penalty0 1--31, 2014.
\newblock \doi{10.18637/jss.v058.i10}.

\bibitem[IPCC(2021)]{ipcc}
IPCC.
\newblock \emph{Climate Change 2021: The physical science basis. Contribution of working group {I} to the Sixth Assessment Report of the Intergovernmental Panel on Climate Change}.
\newblock Cambridge University Press, Cambridge, United Kingdom and New York, NY, USA, In Press, 2021.
\newblock \doi{10.1017/9781009157896}.

\bibitem[Irish et~al.(2013)Irish, Resion, and Ratcliff]{irish}
J.~Irish, D.~Resion, and J.~Ratcliff.
\newblock The influence of storm size on hurricane surge.
\newblock \emph{Journal of Physical Oceanography}, 38(9), 2013.

\bibitem[Joe(1997)]{copulaJoe}
H.~Joe.
\newblock \emph{Multivariate models and dependence concepts}.
\newblock Chapman and Hall/CRC, 1997.

\bibitem[Johnson and Wehrly(1978)]{copula3}
R.~A. Johnson and T.~E. Wehrly.
\newblock Some angular-linear distributions and related regression models.
\newblock \emph{Journal of the American Statistical Association}, 73, 1978.

\bibitem[KNMI(2013)]{cesar}
KNMI.
\newblock Cesar database.
\newblock \emph{Royal Netherlands Meteorological Institute (KNMI, the Netherlands), available at: http://www. cesar-database.nl/}, 2013.

\bibitem[Koenker(2005)]{Koenker2005}
R.~Koenker.
\newblock \emph{Quantile Regression}.
\newblock Cambridge University Press, 2005.

\bibitem[Koenker(2019)]{quantregR}
R.~Koenker.
\newblock \emph{quantreg: Quantile regression}, 2019.
\newblock URL \url{https://CRAN.R-project.org/package=quantreg}.
\newblock R package version 5.54.

\bibitem[Koenker and Bassett~Jr(1978)]{quantreg}
R.~Koenker and G.~Bassett~Jr.
\newblock Regression quantiles.
\newblock \emph{Econometrica: journal of the Econometrics Society, pages 33-50}, 1978.

\bibitem[Kunsch(1989{\natexlab{a}})]{blockboot1}
H.~R. Kunsch.
\newblock The jackknife and the bootstrap for general stationary observations.
\newblock \emph{The Annals of Statistics}, 17\penalty0 (3):\penalty0 1217--1241, 1989{\natexlab{a}}.

\bibitem[Kunsch(1989{\natexlab{b}})]{kunsch1989}
H.~R. Kunsch.
\newblock The jackknife and the bootstrap for general stationary observations.
\newblock \emph{The annals of Statistics}, pages 1217--1241, 1989{\natexlab{b}}.

\bibitem[Lagona(2022)]{stVonMises}
F.~Lagona.
\newblock \emph{Spatial Autoregressive Models for Circular Data}, pages 297--313.
\newblock Springer Nature Singapore, Singapore, 2022.
\newblock \doi{10.1007/978-981-19-1044-9_16}.
\newblock URL \url{https://doi.org/10.1007/978-981-19-1044-9_16}.

\bibitem[Lagona et~al.(2015)Lagona, Picone, and Maruotti]{Lagona2015}
F.~Lagona, M.~Picone, and A.~Maruotti.
\newblock A hidden markov model for the analysis of cylindrical time series.
\newblock \emph{Environmetrics}, 26\penalty0 (8):\penalty0 534--544, 2015.
\newblock \doi{https://doi.org/10.1002/env.2355}.

\bibitem[Lahiri(2003)]{blockboot}
S.~Lahiri.
\newblock \emph{Resampling methods for dependent data}.
\newblock Springer - Verlag, 2003.

\bibitem[Ley et~al.(2014)Ley, Sabbah, and Verdebout]{quantdir}
C.~Ley, C.~Sabbah, and T.~Verdebout.
\newblock {A new concept of quantiles for directional data and the angular Mahalanobis depth}.
\newblock \emph{Electronic Journal of Statistics}, 8\penalty0 (1):\penalty0 795 -- 816, 2014.
\newblock \doi{10.1214/14-EJS904}.

\bibitem[Lie and Eidsvik(2021)]{Eidsvik2021}
H.~S. Lie and J.~Eidsvik.
\newblock Inference in cylindrical models having latent markovian classes—with an application to ocean current data.
\newblock \emph{Spatial Statistics}, 41:\penalty0 100497, 2021.
\newblock ISSN 2211-6753.
\newblock \doi{https://doi.org/10.1016/j.spasta.2021.100497}.

\bibitem[Lindsay(1995{\natexlab{a}})]{lindsay1995}
B.~G. Lindsay.
\newblock Mixture models: theory, geometry, and applications.
\newblock Ims, 1995{\natexlab{a}}.

\bibitem[Lindsay(1995{\natexlab{b}})]{mixture}
B.~G. Lindsay.
\newblock Mixture models: theory, geometry and applications.
\newblock \emph{NSF-CBMS Regional Conference Series in Probability and Statistics}, 5:\penalty0 i--163, 1995{\natexlab{b}}.

\bibitem[Mao and Rychlik(2016)]{stWeibull2}
W.~Mao and I.~Rychlik.
\newblock Estimation of weibull distribution for wind speeds along ship routes.
\newblock \emph{Proceedings of the Institution of Mechanical Engineers, Part M: Journal of Engineering for the Maritime Environment}, 231, 07 2016.
\newblock \doi{10.1177/1475090216653495}.

\bibitem[Mardia(1972)]{vonMises}
K.~V. Mardia.
\newblock \emph{Statistics of directional data}.
\newblock Academic Press, 1972.

\bibitem[Mardia(1975)]{vonmis2}
K.~V. Mardia.
\newblock Statistics of directional data.
\newblock \emph{Journal of the Royal Statistical Society. Series B (Methodological)}, 37, 1975.

\bibitem[Mastrantonio(2018)]{Mastrantonio2018}
G.~Mastrantonio.
\newblock The joint projected normal and skew-normal: A distribution for poly-cylindrical data.
\newblock \emph{Journal of Multivariate Analysis}, 165:\penalty0 14--26, 2018.
\newblock ISSN 0047-259X.
\newblock \doi{https://doi.org/10.1016/j.jmva.2017.11.006}.

\bibitem[Mastrantonio et~al.(2015)Mastrantonio, Maruotti, and Jona-Lasinio]{mastrantonio2015}
G.~Mastrantonio, A.~Maruotti, and G.~Jona-Lasinio.
\newblock Bayesian hidden markov modelling using circular-linear general projected normal distribution.
\newblock \emph{Environmetrics}, 26\penalty0 (2):\penalty0 145--158, 2015.
\newblock \doi{https://doi.org/10.1002/env.2326}.

\bibitem[Mastrantonio et~al.(2016)Mastrantonio, Jona~Lasinio, and Gelfand]{STcircGelfand}
G.~Mastrantonio, G.~Jona~Lasinio, and A.~E. Gelfand.
\newblock Spatio-temporal circular models with non-separable covariance structure.
\newblock \emph{TEST}, 25\penalty0 (3):\penalty0 331--350, 2016.
\newblock \doi{10.1007/s11749-015-0458-y}.

\bibitem[McLachlan and Peel(2000)]{bic}
G.~McLachlan and D.~Peel.
\newblock \emph{Finite mixture models}.
\newblock John Wiley and Sons, 2000.

\bibitem[McLachlan et~al.(2019)McLachlan, Lee, and Rathnayake]{mclachlan2019}
G.~J. McLachlan, S.~X. Lee, and S.~I. Rathnayake.
\newblock Finite mixture models.
\newblock \emph{Annual review of statistics and its application}, 6:\penalty0 355--378, 2019.

\bibitem[McWilliams and Sprevak(1979)]{uvmodel2}
B.~McWilliams and D.~Sprevak.
\newblock The probability distribution of wind velocity and direction.
\newblock \emph{Wind Engineering}, 3, No 4, 1979.

\bibitem[Mendis et~al.(2007)Mendis, Ngo, Haritos, and Samali]{building2}
P.~Mendis, T.~Ngo, N.~Haritos, and J.~Samali, B.and~Cheung.
\newblock Wind loading on tall buildings.
\newblock \emph{Electronic Journal of Structural Engineering}, 7:41-54, 2007.

\bibitem[Moen(1982)]{windchar}
A.~N. Moen.
\newblock \emph{The biology and management of wild ruminants, Chapter Fourteen}.
\newblock CornerBrook Press, 1982.

\bibitem[Monahan(2014)]{monahan1}
A.~Monahan.
\newblock \emph{Wind Speed Probability distribution, Encyclopedia of Natural Resources - Water and Air}, volume~2.
\newblock CRC Press, 2014.

\bibitem[Mosteller and Tukey(1977)]{quantreg2}
F.~Mosteller and J.~Tukey.
\newblock \emph{Data analysis and regression: A Second Course in Statistics}.
\newblock Addison-Wesley Publishing Company, 1977.

\bibitem[Mullen et~al.(2011)Mullen, Ardia, Gil, Windover, and Cline]{DEoptim}
K.~M. Mullen, D.~Ardia, D.~L. Gil, D.~Windover, and J.~Cline.
\newblock Deoptim: An r package for global optimization by differential evolution.
\newblock \emph{Journal of Statistical Software}, 40\penalty0 (6):\penalty0 1–26, 2011.
\newblock \doi{10.18637/jss.v040.i06}.

\bibitem[Nelsen(2006)]{copulaNelsen}
R.~B. Nelsen.
\newblock \emph{An introduction to Copulas}.
\newblock Springer, 2006.

\bibitem[Paolo~Frumento and Fernández-Val(2021)]{Frumento2021}
M.~B. Paolo~Frumento and I.~Fernández-Val.
\newblock Parametric modeling of quantile regression coefficient functions with longitudinal data.
\newblock \emph{Journal of the American Statistical Association}, 116\penalty0 (534):\penalty0 783--797, 2021.
\newblock \doi{10.1080/01621459.2021.1892702}.

\bibitem[Plackett(1965)]{copula1}
R.~L. Plackett.
\newblock A class of bivariate distributions.
\newblock \emph{Journal of the American Statistical Association}, 60\penalty0 (310):\penalty0 516--522, 1965.

\bibitem[Politis and Romano(1991)]{blockboot2}
D.~Politis and J.~Romano.
\newblock A circular block-resampling procedure for stationary data.
\newblock \emph{Technical Report No. 370, tanford University}, 1991.

\bibitem[{R Core Team}(2021)]{R}
{R Core Team}.
\newblock \emph{R: A Language and environment for statistical computing}.
\newblock R Foundation for Statistical Computing, Vienna, Austria, 2021.
\newblock URL \url{https://www.R-project.org/}.

\bibitem[Ranalli and Maruotti(2020)]{ranalli2020model}
M.~Ranalli and A.~Maruotti.
\newblock Model-based clustering for noisy longitudinal circular data, with application to animal movement.
\newblock \emph{Environmetrics}, 31\penalty0 (2):\penalty0 e2572, 2020.

\bibitem[Reich and Fuentes(2007)]{uvrep}
B.~Reich and M.~Fuentes.
\newblock A multivariate semiparametric {Bayesian} spatial modeling framework for hurricane surface wind fields.
\newblock \emph{The Annals of Applied Statistics}, 1, 2007.

\bibitem[Rychlik(2015)]{stWeibull1}
I.~Rychlik.
\newblock Spatio-temporal model for wind speed variability.
\newblock \emph{Annales de l’ISUP}, 59\penalty0 (1-2):\penalty0 25--56, 2015.
\newblock URL \url{https://hal.archives-ouvertes.fr/ffhal-03604750f}.

\bibitem[Sadeghianpourhamami et~al.(2019)Sadeghianpourhamami, Benoit, Deschrijver, and Develder]{Sadeghianpourhamami2019}
N.~Sadeghianpourhamami, D.~Benoit, D.~Deschrijver, and C.~Develder.
\newblock Bayesian cylindrical data modeling using abe–ley mixtures.
\newblock \emph{Applied Mathematical Modelling}, 68:\penalty0 629--642, 2019.
\newblock ISSN 0307-904X.
\newblock \doi{https://doi.org/10.1016/j.apm.2018.11.039}.

\bibitem[S.C.Pryor et~al.(2009)S.C.Pryor, Barthelmie, Young, Takle, Arritt, D.Flory, Jr., Nunes, and J.Roads]{windclimate}
S.C.Pryor, R.~Barthelmie, D.~Young, E.~Takle, R.~Arritt, D.Flory, W.~G. Jr., A.~Nunes, and J.Roads.
\newblock Wind speed trends over the contiguous {United States}.
\newblock \emph{Journal of Geophysical Research}, 114, 2009.

\bibitem[Scrucca et~al.(2016)Scrucca, Fop, Murphy, and Raftery]{mclust}
L.~Scrucca, M.~Fop, T.~B. Murphy, and A.~E. Raftery.
\newblock {mclust} 5: clustering, classification and density estimation using {G}aussian finite mixture models.
\newblock \emph{The {R} Journal}, 8\penalty0 (1):\penalty0 289--317, 2016.
\newblock URL \url{https://doi.org/10.32614/RJ-2016-021}.

\bibitem[Skamarock et~al.(2008)Skamarock, Klemp, Dudhia, Barker, and Duda]{WRF}
W.~C. Skamarock, J.~B. Klemp, D.~O. Dudhia, J.and~Gill, D.~Barker, and J.~G. Duda, M. G. …~Powers.
\newblock A description of the advanced research wrf version 3. ncar tech. note ncar/tn-475+str, 2008.

\bibitem[Smith(1971)]{uvmodel1}
O.~Smith.
\newblock An application of distributions derived from the bivariate normal density function.
\newblock \emph{Bulletin of the American Meteorological Society}, 52, No. 3, 1971.

\bibitem[Solari and Losada(2016)]{solari}
S.~Solari and M.~A. Losada.
\newblock Simulation of non-stationary wind speed and direction time series.
\newblock \emph{Wind Energy and Industrial Aerodynamics}, 149, 2016.

\bibitem[Stanley et~al.(2020)Stanley, King, and Ning]{cleanenergy}
A.~P.~J. Stanley, J.~King, and A.~Ning.
\newblock Wind farm layout optimization with loads considerations.
\newblock \emph{Journal of Physics: Conference Series}, 1452, 2020.

\bibitem[Titterington et~al.(1985)Titterington, Afm, Smith, Makov, et~al.]{titterington1985}
D.~M. Titterington, S.~Afm, A.~F. Smith, U.~Makov, et~al.
\newblock \emph{Statistical analysis of finite mixture distributions}, volume 198.
\newblock John Wiley \& Sons Incorporated, 1985.

\bibitem[Wang et~al.(2015)Wang, Gelfand, and Jona-Lasinio]{linCircST}
F.~Wang, A.~E. Gelfand, and G.~Jona-Lasinio.
\newblock Joint spatio-temporal analysis of a linear and a directional variable: Space-time modeling of wave heights and wave directions in the adriatic sea.
\newblock \emph{Statistica Sinica}, 25\penalty0 (1):\penalty0 25--39, 2015.
\newblock URL \url{http://www.jstor.org/stable/24311002}.

\bibitem[Wang and Kotamarthi(2015)]{wang2015}
J.~Wang and V.~R. Kotamarthi.
\newblock High-resolution dynamically downscaled projections of precipitation in the mid and late 21st century over {North America}.
\newblock \emph{Earth's Future}, 3\penalty0 (7):\penalty0 268--288, 2015.

\bibitem[Wang(2013)]{pbsR}
S.~Wang.
\newblock \emph{pbs: Periodic B splines}, 2013.
\newblock URL \url{https://CRAN.R-project.org/package=pbs}.
\newblock R package version 1.1.

\bibitem[Weber(1997)]{uvmodel3}
R.~O. Weber.
\newblock Estimators for the standard deviation of horizontal wind direction.
\newblock \emph{Journal of Applied Meteorology and Climatology}, 36, 1997.

\bibitem[Westerling et~al.(2004)Westerling, Cayan, Brown, Hall, and Riddle]{wildfire}
A.~L. Westerling, D.~R. Cayan, T.~J. Brown, B.~L. Hall, and L.~G. Riddle.
\newblock Climate, {Santa Ana} winds and autumn wildfires in {Southern California}.
\newblock \emph{Eos, Transactions American Geophysical Union}, 85(31):289–296, 2004.

\bibitem[Wohland et~al.(2019)Wohland, Omrani, Witthaut, and Keenlyside]{windclimate2}
J.~Wohland, N.~Omrani, D.~Witthaut, and N.~Keenlyside.
\newblock Inconsistent wind speed trends in current twentieth century reanalyses.
\newblock \emph{Journal of Geophysical Research: Atmosphere}, 124, 2019.

\bibitem[Wu et~al.(2022)Wu, Bessac, Huang, Wang, and Kotamarthi]{wu2022}
Q.~Wu, J.~Bessac, W.~Huang, J.~Wang, and R.~Kotamarthi.
\newblock A conditional approach for joint estimation of wind speed and direction under future climates.
\newblock \emph{Advances in Statistical Climatology, Meteorology and Oceanography}, 2022.

\bibitem[Zannetti(2013)]{airpolution}
P.~Zannetti.
\newblock \emph{Air pollution modeling: theories, computational methods and available software}.
\newblock Springer Science \& Business Media., 2013.

\bibitem[Zobel et~al.(2018{\natexlab{a}})Zobel, Wang, Wuebbles, and Kotamarthi]{zobel2018a}
Z.~Zobel, J.~Wang, D.~J. Wuebbles, and V.~R. Kotamarthi.
\newblock Analyses for high-resolution projections through the end of the 21st century for precipitation extremes over the {United States}.
\newblock \emph{Earth's Future}, 6\penalty0 (10):\penalty0 1471--1490, 2018{\natexlab{a}}.

\bibitem[Zobel et~al.(2018{\natexlab{b}})Zobel, Wang, Wuebbles, and Kotamarthi]{zobel2018b}
Z.~Zobel, J.~Wang, D.~J. Wuebbles, and V.~R. Kotamarthi.
\newblock Evaluations of high-resolution dynamically downscaled ensembles over the contiguous {United States}.
\newblock \emph{Climate Dynamics}, 50\penalty0 (3):\penalty0 863--884, 2018{\natexlab{b}}.

\end{thebibliography}

\newpage
 \appendix

\end{document}